\documentclass[a4paper, onecolumn, 11pt, accepted=2024-05-03]{quantumarticle}
\pdfoutput=1
\usepackage{breakurl}
\usepackage{stmaryrd}
\usepackage{amsthm,amstext,amssymb}
\usepackage{epsfig}
\usepackage{slashed}
\usepackage{fancyhdr}
\usepackage{graphicx}
\usepackage{pdfpages}
\usepackage{float}
\usepackage{subfig}
\usepackage{simplewick}
\usepackage{graphics,psfrag}
\usepackage{pstricks}
\usepackage{latexsym}
\usepackage{changepage}
\usepackage{caption}
\usepackage{paralist}
\usepackage{tikzit}

\tikzstyle{boundary vertex}=[inner sep=0mm, minimum size=1mm, shape=circle, draw=black, fill=black]
\tikzstyle{grey_dot}=[fill={rgb,255: red,191; green,191; blue,191}, draw={rgb,255: red,191; green,191; blue,191}, shape=circle, minimum size=1mm, inner sep=0mm]
\tikzstyle{blue_dot}=[fill={rgb,255: red,202; green,251; blue,255}, draw=black, shape=circle, minimum size=1mm, inner sep=0mm]
\tikzstyle{white_dot}=[fill=white, draw=black, shape=circle, minimum size=1.5mm, inner sep=0mm]

\tikzstyle{arrow}=[->]
\tikzstyle{red_arrow}=[->, draw=red]
\tikzstyle{cyan_arrow}=[->, draw=cyan]
\tikzstyle{red_dash}=[-, dashed, draw=red]
\tikzstyle{grey dash}=[-, fill=none, draw={rgb,255: red,191; green,191; blue,191}, dashed]
\tikzstyle{dashed arrow}=[->, dashed]
\tikzstyle{blue_edge}=[-, draw={rgb,255: red,46; green,126; blue,255}]
\tikzstyle{red_edge}=[-, draw=red]

\tikzstyle{gate}=[shape=rectangle, text height=1.5ex, text depth=0.25ex, yshift=0.5mm, fill=white, draw=black, minimum height=5mm, yshift=-0.5mm, minimum width=5mm, font={\small}, tikzit category=circuit]
\tikzstyle{big gate}=[shape=rectangle, text height=1.5ex, text depth=0.25ex, yshift=0.5mm, fill=white, draw=black, minimum height=18mm, yshift=-0.5mm, minimum width=5mm, font={\small}, tikzit category=circuit]
\tikzstyle{Z dot}=[inner sep=0mm, minimum size=2mm, shape=circle, draw=black, fill={rgb,255: red,221; green,255; blue,221}, tikzit category=zx]
\tikzstyle{Z phase dot}=[minimum size=5mm, font={\footnotesize\boldmath}, shape=rectangle, rounded corners=2mm, inner sep=0.2mm, outer sep=-2mm, scale=0.8, tikzit shape=circle, draw=black, fill={rgb,255: red,221; green,255; blue,221}, tikzit draw=blue, tikzit category=zx]
\tikzstyle{X dot}=[Z dot, shape=circle, draw=black, fill={rgb,255: red,255; green,136; blue,136}, tikzit category=zx]
\tikzstyle{X phase dot}=[Z phase dot, tikzit shape=circle, tikzit draw=blue, fill={rgb,255: red,255; green,136; blue,136}, font={\footnotesize\boldmath}, tikzit category=zx]
\tikzstyle{hadamard}=[fill=yellow, draw=black, shape=rectangle, inner sep=0.6mm, minimum height=1.5mm, minimum width=1.5mm, tikzit category=zx]
\tikzstyle{paulibox}=[fill={rgb,255: red,221; green,221; blue,255}, draw=black, shape=rectangle, inner sep=0.6mm, minimum height=5mm, minimum width=5mm, font={\footnotesize}, text height=1.5ex, text depth=0.25ex, tikzit category=zx]
\tikzstyle{vertex}=[inner sep=0mm, minimum size=1mm, shape=circle, draw=black, fill=black, tikzit category=misc]
\tikzstyle{vertex set}=[inner sep=0mm, minimum size=1mm, shape=circle, draw=black, fill=white, font={\footnotesize\boldmath}, tikzit category=misc]
\tikzstyle{small black dot}=[fill=black, draw=black, shape=circle, inner sep=0pt, minimum width=1.2mm, tikzit category=circuit]
\tikzstyle{cnot ctrl}=[fill=black, draw=black, shape=circle, inner sep=0pt, minimum width=1.2mm, tikzit category=circuit]
\tikzstyle{cnot targ}=[fill=white, draw=white, shape=circle, tikzit category=circuit, label={center:$\oplus$}, inner sep=0pt, minimum width=2.1mm, tikzit fill={rgb,255: red,102; green,204; blue,255}, tikzit draw=black]
\tikzstyle{ket}=[fill=white, draw=black, shape=regular polygon, regular polygon sides=3, regular polygon rotate=-30, scale=0.7, inner sep=1pt, tikzit category=circuit, tikzit shape=rectangle, tikzit fill=green]
\tikzstyle{bra}=[fill=white, draw=black, shape=regular polygon, regular polygon sides=3, regular polygon rotate=30, scale=0.7, inner sep=1pt, tikzit category=circuit, tikzit shape=rectangle, tikzit fill=red]
\tikzstyle{scalar}=[shape=rectangle, text height=1.5ex, text depth=0.25ex, yshift=0.5mm, fill=white, draw=black, minimum height=5mm, yshift=-0.5mm, minimum width=5mm, font={\small}]
\tikzstyle{clabel}=[fill=white, draw=none, shape=rectangle, tikzit fill={rgb,255: red,56; green,255; blue,242}, font={\footnotesize}, inner sep=1pt, tikzit category=labels]
\tikzstyle{empty diagram}=[draw={gray!40!white}, dashed, shape=rectangle, minimum width=1cm, minimum height=1cm, tikzit category=misc]

\tikzstyle{hadamard edge}=[-, dashed, dash pattern=on 2pt off 0.5pt, thick, draw={rgb,255: red,68; green,136; blue,255}]
\tikzstyle{box edge}=[-, dashed, dash pattern=on 2pt off 0.5pt, thick, draw={rgb,255: red,203; green,192; blue,225}]
\tikzstyle{brace edge}=[-, tikzit draw=blue, decorate, decoration={brace,amplitude=1mm,raise=-1mm}]
\tikzstyle{diredge}=[->]
\tikzstyle{double edge}=[-, double, shorten <=-1mm, shorten >=-1mm, double distance=2pt]
\tikzstyle{gray edge}=[-, {gray!60!white}]
\tikzstyle{pointer edge}=[->, very thick, gray]
\tikzstyle{boldedge}=[-, line width=1.6pt, shorten <=-0.17mm, shorten >=-0.17mm]

\usepackage{graphicx}
\usepackage{array}
\usepackage{hyperref}
\usepackage{tikz-cd}
\usepackage{mathrsfs}
\usepackage{braket}
\usepackage{soul}

\usepackage{scalefnt}
\usepackage[all]{xypic}
\usepackage{slashed}
\usepackage{extarrows}
\usepackage{lettrine}
\usepackage{booktabs}
\usepackage{paralist}
\usepackage{subfig}
\usepackage{wrapfig}
\usepackage{xcolor}

\DeclareMathAlphabet{\mathcal}{OMS}{cmsy}{m}{n}
\newtheorem{lemma}{Lemma}[section] 
\newtheorem{proposition}[lemma]{Proposition}
\newtheorem{corollary}[lemma]{Corollary}
\newtheorem{example}[lemma]{Example}

\newtheorem{definition}[lemma]{Definition}
\newtheorem{remark}[lemma]{Remark}
\newtheorem{conjec}[lemma]{Conjecture}

\renewcommand{\imath}{\mathrm{i}}

\newcommand{\CC}{\hbox{{$\mathcal C$}}}

\newcommand{\CF}{\hbox{{$\mathcal F$}}}
\newcommand{\CG}{\hbox{{$\mathcal G$}}}

\newcommand{\CT}{\hbox{{$\mathcal T$}}}

\newcommand{\CP}{\hbox{{$\mathcal P$}}}

\newcommand{\CE}{\hbox{{$\mathcal E$}}}

\newcommand{\C}{\mathbb{C}}

\newcommand{\F}{\mathbb{F}}
\newcommand{\Z}{\mathbb{Z}}
\newcommand{\N}{\mathbb{N}}

\newcommand{\Hom}{\mathrm{Hom}}

\newcommand{\del}{\partial}

\newcommand{\tens}{\mathop{{\otimes}}}

\newcommand{\id}{\mathrm{id}}
\newcommand{\im}{\mathrm{im}}

\def\lcross{{>\!\!\!\triangleleft}}

\newcommand{\MatF}{\mathtt{Mat}_{\F_2}}
\newcommand{\Chains}{\mathtt{Ch}(\MatF)}
\newcommand{\Coch}{\mathtt{Coch}(\MatF)}
\newcommand{\Grph}{\mathtt{Grph}}
\newcommand{\OGrph}{\mathtt{OGrph}}
\newcommand{\ACC}{\mathtt{ACC}}
\newcommand{\OACC}{\mathtt{OACC}}
\newcommand{\coeq}{{\textit coeq}}
\newcommand{\FHilb}{\mathtt{FHilb}}
\newcommand{\CatSet}{\mathtt{Set}}

\newcommand{\supp}{{\rm supp}}

\newsavebox{\pullback}
\sbox\pullback{%
\begin{tikzpicture}%
\draw (0,0) -- (2ex,0ex);%
\draw (2ex,0ex) -- (2ex,2ex);%
\end{tikzpicture}}

\newsavebox{\pushout}
\sbox\pushout{%
\begin{tikzpicture}%
\draw (0,0) -- (0,2ex);%
\draw (0,2ex) -- (2ex,2ex);%
\end{tikzpicture}}

\setstcolor{red}

\begin{document}
\title{CSS code surgery as a universal construction}

\author{Alexander Cowtan}
\affiliation{Dept. of Computer Science,  University of Oxford,
Wolfson Building, Parks Road, Oxford OX1 3QD, UK}
\affiliation{Quantinuum,
Terrington House, 13-15 Hills Road, Cambridge CB2 1NL, UK}
\email{akcowtan@gmail.com}
\author{Simon Burton}
\affiliation{Quantinuum,
Terrington House, 13-15 Hills Road, Cambridge CB2 1NL, UK}
\email{simon.burton@quantinuum.com}

\maketitle

\begin{abstract}
We define code maps between Calderbank-Shor-Steane (CSS) codes using maps between chain complexes, and describe code surgery between such codes using a specific colimit in the category of chain complexes. As well as describing a surgery operation, this gives a general recipe for new codes. As an application we describe how to `merge' and `split' along a shared $\overline{X}$ or $\overline{Z}$ operator between arbitrary CSS codes in an error-corrected manner, so long as certain technical conditions concerning gauge fixing and code distance are satisfied. We prove that such merges and splits on LDPC codes yield codes which are themselves LDPC.
\end{abstract}

\section{Introduction}
Quantum computers have become larger and more sophisticated in recent years \cite{AAB,Colcode}, but fault-tolerance is necessary to perform many practically relevant quantum algorithms. Qubit stabiliser error-correction codes are a well-studied approach to fault-tolerant quantum computing \cite{Got} and are favourable both for their practicality and theoretical simplicity. Such codes store logical data using entangled states of physical qubits and repeated many-body measurements, and so long as the physical errors on the qubits stay below a certain threshold the logical data is protected.

The most well-known example of a qubit stabiliser code
is the toric code, in which qubits are embedded on the
surface of a torus, and properties of the logical space
are determined by the topology of the surface \cite{DKLP, Kit}.
This is a basic example of a qubit Calderbank-Shor-Steane
(CSS) code; there are several equivalent ways of defining
CSS codes, but for our purposes we shall describe them
as codes which are all \textit{homological} in a suitable
sense \cite{BE2,BM}.

This means that we can study CSS codes using the tools of homological algebra \cite{Weib}. This approach has recently seen much success, for example in the construction of so-called good quantum low-density parity check (LDPC) code families using a lifted product of chain complexes \cite{PK1}. Such code families have an encoding rate $k/n$ of logical to physical qubits which is constant in the code size, while maintaining a linear code distance $d$, a substantial asymptotic improvement over simpler examples such as the toric code. The main caveat is, informally, that the connectivity between physical qubits is non-local. This complicates the architecture of the system, and also complicates the protocols for performing logical gates.

There have been several recent works on protocols for logical gates in CSS codes \cite{Kris,Coh,Burt1,QWV,HJY}, of varying generality. Here, we build on this work by defining surgery, in the abstract, using arbitrary CSS codes which form a categorical span, although the practical implementation of such surgery has several important caveats. The idea is that merging two codes works by identifying a common structure in each code and quotienting it out. CSS code surgery is particularly convenient when the CSS codes are \textit{compatible}, in the sense that they have at least one identical $\overline{Z}$ or $\overline{X}$ logical operator. In this case, the common structure being quotiented out is the logical operator. In order to formalise this, we take a step back and look at the category of chain complexes $\Chains$.

We start by giving a recap of the relevant categorical
background of chain complexes, and the view of classical
linear binary codes and qubit CSS codes using chain complexes.
We then define code maps between CSS codes using morphisms
between chain complexes. These are maps which send $X$-checks
to $X$-checks and $Z$-checks to $Z$-checks in a coherent
way, and have a convenient presentation as phase-free
ZX diagrams, which we prove in Proposition~\ref{prop:CNOT_circuit}.

We believe that code maps crop up throughout the CSS
code literature. We see 3 primary use-cases for code maps:
\begin{enumerate}
\item Encoders/decoders \cite{DCP,Del,Hig}.
\item Constructing new codes.
\item Designing fault-tolerant logical operations.
\end{enumerate}
We intend to expound on code maps in future work, but
presently we focus on items 2 and 3.

We define CSS code merges as a colimit -- specifically, a coequaliser/pushout -- in the category of chain complexes. Not only does the construction describe a surgery operation, but it also gives a general recipe for new codes. An application of our treatment is the description of certain classes of code surgery whereby the codes are merged or split along a $\overline{Z}$ or $\overline{X}$ operator. This is closely related to the notion of `welding' in \cite{Mich}, and generalises the cases for 2D topological codes given in \cite{HFDM,LRA,NFB}. We prove that merging two LDPC codes in such a manner still yields an LDPC code. We give a series of examples, including the specific case of lattice surgery between surface codes. Lastly, we discuss how to apply such protocols in practice. We prove that when two technical conditions are satisfied then code surgery can be performed while maintaining the error-correcting properties of the code, allowing us to perform logical parity measurements on codes.

\subsection{Guide to reading the paper}
Section~\ref{sec:universal} gives a bird's eye view of category theory and universal constructions, which will be useful later on. Section~\ref{sec:chain_comp} describes the category of chain complexes with morphisms as matrices over $\F_2$. Category theorists may wish to skip past these sections.

We then give a rundown of CSS codes viewed as chain complexes in Section~\ref{sec:codes}. Readers familiar with basic category theory and this perspective of CSS codes can safely skip to Section~\ref{sec:code_maps}, where we introduce the notion of \textit{code maps}, that is coherent transforms between codes.

We introduce surgery of codes as a colimit in Section~\ref{sec:CSS_surgery}. This is when the notion of `gluing' codes together comes in, and we prove several results about these codes when the colimit uses logical $\overline{Z}$ or $\overline{X}$-operators.
Lastly, we introduce a protocol for performing logical $\overline{Z}\tens\overline{Z}$ and $\overline{X}\tens\overline{X}$ measurements in Section~\ref{sec:practical}.

\section{Universal constructions}\label{sec:universal}

In this section 
we provide a cartoon introduction to category theory
and universal constructions. See \cite{Lein} for a more in-depth introduction.

A \emph{category} is a collection of \emph{objects} and \emph{morphisms}.
We will begin by drawing an object as a box
with a decoration, such as
\begin{center}
\includegraphics{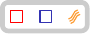}\ .
\end{center}
Morphisms are arrows between objects, like this
\begin{center}
\includegraphics{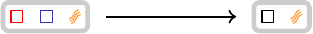}\ .
\end{center}
The arrow notation suggests that we can \emph{compose} these.
\begin{center}
\includegraphics{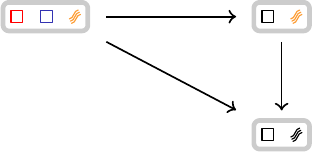}\ .
\end{center}

The \emph{product} of two objects in a category is an object,
together with two arrows,
\begin{center}
\includegraphics{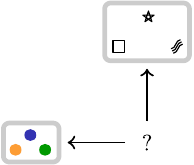}\ .
\end{center}
The product decoration combines the two decorations,
\begin{center}
\includegraphics{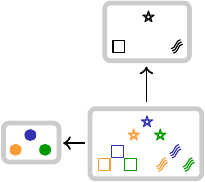}\ .
\end{center}
The product also must satisfy a \emph{universal property}.
This states that any other object that also combines the
two decorations is already compatible with the product object in
a unique way.
In other words, for all test objects 
there exists a unique \emph{comparison} morphism:
\begin{center}
\includegraphics{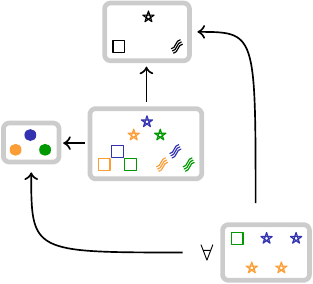}\ \ \ \ 
\includegraphics{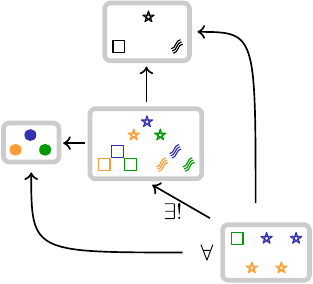}\ .
\end{center}
The real product is the minimal object
that projects down to the factors. 
Any other test object lives over the real product.

This universal property has the immediate consequence
that any other object that satisfies all these requirements,
will be isomorphic via a unique isomorphism that commutes
with the other morphisms,
\begin{center}
\includegraphics{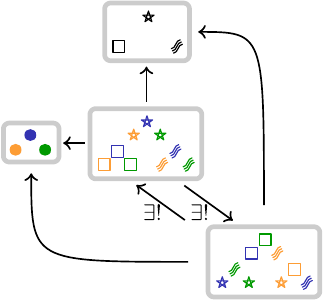}\ .
\end{center}

A \emph{pullback} is a product with constraints:
\begin{center}
\includegraphics{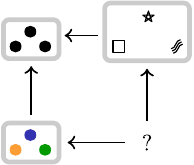}\ .
\end{center}
The resulting square should \emph{commute}: 
if we compose any two paths of arrows with the same
source object and the same target object then these paths should
be equal.
\begin{center}
\includegraphics{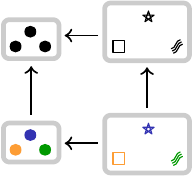}\ .
\end{center}
As with products, we also require the pullback to
satisfy a universal property.
\begin{center}
\includegraphics{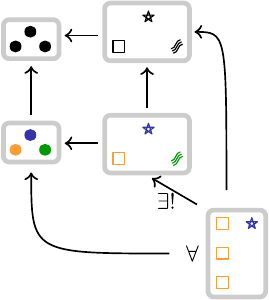}\ .
\end{center}

All of these statements have \emph{dual} statements,
which we get by reversing all the arrows.
When we do this we sometimes put a \emph{co-} prefix on the terminology.
For example, a \emph{coproduct}, which would normally be called a sum, looks like this
\begin{center}
\includegraphics{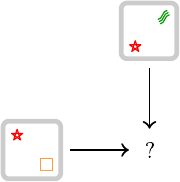}\ .
\end{center}
Once again, we require any such candidate coproduct
\begin{center}
\includegraphics{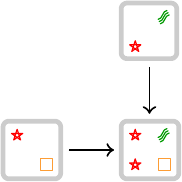}
\end{center}
to satisfy a universal property
\begin{center}
\includegraphics{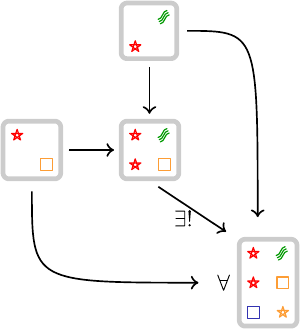}\ .
\end{center}
We think of a coproduct as a way of gluing together
objects.
By adding constraints we can express where we wish to glue
\begin{center}
\includegraphics{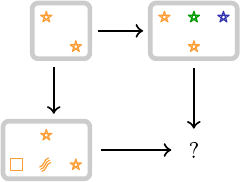}
\end{center}
The answer to this question is called a \emph{pushout}:
it is an object together
with two morphisms,
\begin{center}
\includegraphics{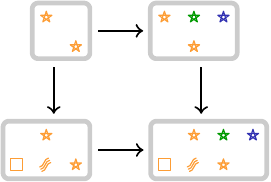}
\end{center}
that satisfies a universal property,
\begin{center}
\includegraphics{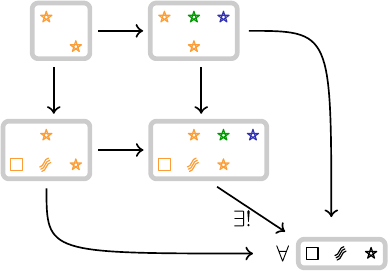}
\end{center}

We have purposefully avoided describing the
decorations in these diagrams: how they work, what they mean.
A more in-depth introduction to category theory would
describe these systematically, possibly mentioning the
category of \emph{finite sets and functions}.
In this case, objects are sets, with \emph{elements},
and we can combine these in various ways to make
other sets.
Instead of telling this story, we skip to the
punchline, which is that there are no elements,
or rather, an element of an object
is really a morphism into that object:
\begin{center}
\includegraphics{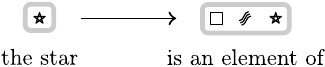}
\end{center}

To push this idea home, and also move toward the goals
of this paper, we consider the category $\MatF$ of finite dimensional
matrices over the field $\F_2.$
This has as objects the natural numbers and
matrices over $\F_2$ as morphisms. 
Composition of morphisms is matrix multiplication.
We will show each object as a box with dots.
For example, here is a composition of two morphisms in
this category
\begin{center}
\includegraphics{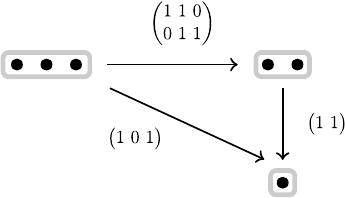}\ .
\end{center}
The objects serve only as anchors for the morphisms (matrices) where all the
action is taking place.
The vector elements of a vector space have vanished into the
morphisms:
\begin{center}
\includegraphics{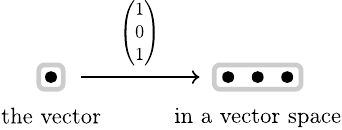}
\end{center}
A coproduct (sum) in this category is
an object together with two morphisms 
\begin{center}
\includegraphics{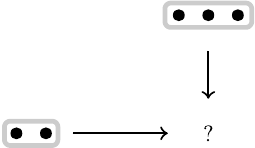}
\end{center}
satisfying the universal property of coproducts.
Here is one candidate:
\begin{center}
\includegraphics{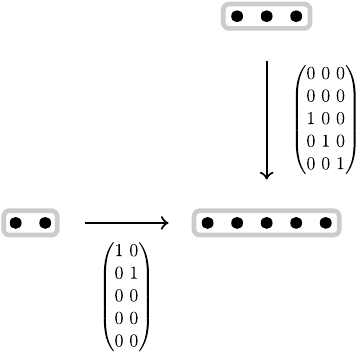}\ .
\end{center}
This coproduct will not be unique (except for some degenerate cases),
but the universal property of the coproduct
guarantees it is unique up to unique isomorphism.
We have reinvented the \emph{direct sum} of vector spaces.

For a pushout of vector spaces
\begin{center}
\includegraphics{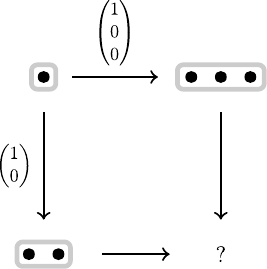}\ .
\end{center}
we get
\begin{center}
\includegraphics{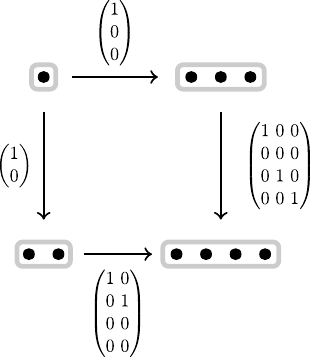}\ .
\end{center}
This is gluing of a two dimensional vector space
and a three dimensional vector space along a one dimensional
vector space.

But what about products? 
A curious thing happens in the category $\MatF$;
we can get the dual universal construction by
transposing matrices. For example, the above coproduct
becomes the product:
\begin{center}
\includegraphics{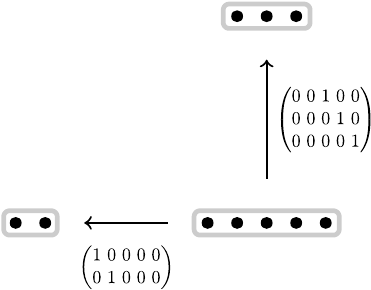}
\end{center}
and similarly with pullbacks.
The transpose duality of $\MatF$
will follow us throughout the rest of this paper.

Here we have been taking the objects of $\MatF$ to be just
natural numbers. In the rest of the paper we will use a slightly
different definition for the objects: each natural number $n$
is replaced by a basis set of size $n$ for an $n$-dimensional
vector space.

\section{Chain complexes}\label{sec:chain_comp}
We now recap some elementary homological algebra. All of this section is known \cite{Weib}, but we fix notation and look explicitly at the particular category of interest.

Let $\MatF$ be the category which has as objects linearised finite sets over $\F_2$, so each object is a vector space $V$ equipped with a specified basis $\tilde{V}$ such that $V = \F_2 \tilde{V}$. Each element of $\tilde{V}$ corresponds to an entry for vectors in $V$. Importantly, we require the vector spaces to have this form on the nose. Throughout, we will call these \textit{based vector spaces}. A morphism $f: V\rightarrow W$ in $\MatF$ is a $\dim W\times \dim V$ matrix valued in $\F_2$.

Let $\Chains$ be the category of bounded chain complexes in $\MatF$. We now recap some of the basic properties of this category. A chain complex $C_\bullet$ looks like this:
\[\begin{tikzcd}
 \cdots \arrow[r] & C_{n+1}\arrow[r, "\del_{n+1}"] & C_{n}\arrow[r, "\del_{n}"] & C_{n-1}\arrow[r] & \cdots
 \end{tikzcd}\]
where each component $C_i$ is a based vector space and $n\in \Z$ is called the degree of the component in $C_\bullet$. $C_\bullet$ has $\F_2$-matrices as differentials $\del_{n+1}:C_{n+1}\rightarrow C_{n}$ such that $\del_n\circ\del_{n+1}=0 \pmod 2$, $\forall n \in \Z$. To disambiguate differentials between chain complexes we will use $\del^{C_\bullet}_n := \del_n \in C_\bullet$ when necessary.

All our chain complexes are bounded, meaning there is some $k\in \Z$ such that $C_{n > k} = 0$ and $l\in\Z$ such that $C_{n < l} =0$, i.e. it is bounded above and below. We call $k-l$ the length of $C_\bullet$ for $k$ and $l$ the smallest and largest possible values respectively.

\begin{definition}
Given a chain complex $C_\bullet$ we let
\[Z_n(C_\bullet) = \ker (\del_{n});\quad B_n(C_\bullet) = \im(\del_{n+1}) \]
and call $Z_n, B_n$ the $n$-cycles and $n$-boundaries. We also define a quotient $H_n(C_\bullet) = Z_n(C_\bullet)/B_n(C_\bullet)$, and call $H_n$ the $n$th homology space of $C_\bullet$.
\end{definition}

Recall that $\dim (\ker (\del_{n})) = {\rm null}(\del_{n}) = \dim C_n - {\rm rank}(\del_{n})$. Throughout we sometimes use $\ker (f)$ of a matrix $f$ to mean the kernel object, i.e. subspace, and sometimes the kernel morphism, i.e. inclusion map. It should be clear from context which is meant.

\begin{example}\label{ex:incidence}
Let $\Gamma = (V,E)$ be a finite simple undirected graph. We can form the incidence chain complex $C_\bullet$ of $\Gamma$, which has $C_{0} = \F_2 V$, $C_1 = \F_2E$. All other components are zero. The sole nonzero differential $\del_{1}$ is the incidence matrix of $\Gamma$, with $(\del_{1})_{ij} = 1$ if the $j$th edge is attached to the $i$th vertex, and 0 otherwise.
$H_1(C_\bullet)$ is determined by the graph homology of $\Gamma$ \cite{Weib}.
\end{example}

\begin{definition}\label{def:chain_map}
A morphism $f_\bullet: C_\bullet\rightarrow D_\bullet$ in $\Chains$ is called a chain map, and consists of a collection of matrices $\{f_i: C_i\rightarrow D_i\}_{i\in \Z}$ such that each resultant square of maps commutes:
\[\begin{tikzcd}\cdots \arrow[r] & C_{n+1}\arrow[r, "\del^{C_\bullet}_{n+1}"]\arrow[d, "f_{n+1}"] & C_{n}\arrow[r, "\del^{C_\bullet}_{n}"]\arrow[d, "f_{n}"] & C_{n-1}\arrow[r]\arrow[d,"f_{n-1}"] & \cdots\\
\cdots \arrow[r] & D_{n+1}\arrow[r, "\del^{D_\bullet}_{n+1}"] & D_{n}\arrow[r, "\del^{D_\bullet}_{n}"] & D_{n-1}\arrow[r] & \cdots\end{tikzcd}
\]
\end{definition}

As we specified \textit{bounded} chain complexes only a finite number of the $f_i$ matrices will be non-zero. A chain map $f_\bullet$ is an isomorphism in $\Chains$ iff all $f_i$ are invertible, in which case one can think of the isomorphism as being a `change of basis' for all components, which thus transforms the differential matrices appropriately. Every pair of chain complexes has at least two chain maps, the zero chain maps, between them, given by a collection of entirely zero matrices either way.

\begin{definition}\label{def:basis_preserving}
We say that a chain map is {\rm basis-preserving} when every matrix at each component maps basis elements to basis elements.
\end{definition}
This does not require that the map is either monic or epic, and is a property associated only with based vector spaces, as it evidently does not arise with abstract vector spaces. We can equivalently see this property as the case when the chain map is a collection of functions between the underlying sets $\{\tilde{f_i}: \tilde{C_i} \rightarrow \tilde{D_i}\}_{i\in\Z}$.

\begin{lemma}\label{lem:chain_map_rest}
A chain map at a component $f_n : C_n \rightarrow D_n$ lifts to a matrix $H_n(f_\bullet) : H_n(C_\bullet)\rightarrow H_n(D_\bullet)$.
\end{lemma}
\proof
It is easy to check that $f_n$ induces matrices from $Z_n(C_\bullet)\rightarrow Z_n(D_\bullet)$ and the same for $B_n$.
\endproof

This lemma is equivalent to saying that $H_n(-)$ is a functor from $\Chains \rightarrow \MatF$.

$\Chains$ has several known categorical properties which will be useful to us. One way to see a chain complex $C_\bullet$ in $\MatF$ is as a $\Z$-graded $\F_2$-vector space, with a specified basis and a distinguished map $\del : C_\bullet \rightarrow C_\bullet$ with components $\del_i : C_{i+1} \rightarrow C_i$, such that $\del\circ\del = 0$. Many of the properties of $\Chains$ are inherited directly from those of $\Z$-graded $\F_2$-vector spaces.

$\Chains$ is an additive category, i.e. it has all finite biproducts, i.e. direct sums. These have components
\[(C\oplus D)_n = C_n \oplus D_n\]
and the same for differentials. This is both a categorical product and coproduct. Homology preserves direct sums: given chain complexes $C_\bullet$ and $D_\bullet$,
\[H_n((C\oplus D)_\bullet) \cong H_n(C_\bullet)\oplus H_n(D_\bullet)\]
This is obvious, considering the blocks of each differential in $(C\oplus D)_\bullet$.

A co-chain complex is similar to a chain complex, with some notational differences.
\begin{definition}
Given a chain complex $C_\bullet$, the co-chain complex $C^\bullet$ has components
\[C^n = C_n\]
and differentials
\[\delta^n = \del_{n-1}^\intercal.\]
\end{definition}
We also have co-cycles $Z^n = \ker(\delta^n)$, co-boundaries $B^n = \im(\delta^{n-1})$ and co-homology spaces $H^n = Z^n/B^n$. It is easy to show that $H_n(C_\bullet) \cong H^n(C^{\bullet})$.

A co-chain map is defined similarly to a chain map:
\[\begin{tikzcd}\cdots \arrow[r] & C^{n-1}\arrow[r, "\delta_{C^\bullet}^{n-1}"]\arrow[d, "f^{n-1}"] & C^{n}\arrow[r, "\delta_{C^\bullet}^{n}"]\arrow[d, "f^{n}"] & C^{n+1}\arrow[r]\arrow[d,"f^{n+1}"] & \cdots\\
\cdots \arrow[r] & D^{n-1}\arrow[r, "\delta_{D^\bullet}^{n-1}"] & D^{n}\arrow[r, "\delta_{D^\bullet}^{n}"] & D^{n+1}\arrow[r] & \cdots\end{tikzcd}
\]
and these are morphisms in the category $\Coch$. There is a functor $H^n(-) : \Coch \rightarrow \MatF$. We can also ask for co-chain maps to be basis-preserving. Given a chain map $f_\bullet$, we automatically also have the cochain map $f^\bullet$, with components $f^n = f_n^\intercal$.

\section{Quantum codes}\label{sec:codes}

Here we introduce classes of both classical and quantum codes as chain complexes. We give easy examples such as the surface and toric codes. Up until Section~\ref{sec:code_maps}, this part is also well-known, although we describe the relationship between $Z$ and $X$ operators in greater detail than we have found elsewhere.

\subsection{Codes as chain complexes}

Binary linear classical codes which encode $k$ bits using $n$ bits can be described by a $m\times n$ parity check $\F_2$-matrix $P$. The parity check matrix $P$, when applied to any codeword of length $n$, gives $Pc = 0$, and thus $k = \dim \ker(P)$; if the result is non-zero then an error has been detected, and under certain assumptions can be corrected. The distance $d$ of a binary linear classical code is the minimum Hamming weight of its nonzero codewords, and one characterisation of codes is by the parameters $[n,k,d]$. We may trivially view a binary linear classical code as a length 1 chain complex, with indices chosen for convenience:
\[\begin{tikzcd}C_\bullet = C_1\arrow[r, "\del_{1}"]&C_{0}\end{tikzcd}\]
where $C_1 = \F_2^n$, $C_{0} = \F_2^m$, and $\del_{1} = P$, the chosen $m\times n$ parity check matrix. Then we have $k = \dim H_{1}(C_\bullet) =\dim Z_{1}(C_\bullet)$, where $Z_{1}(C_\bullet)$ is the codespace.

\begin{example}\label{ex:rep_code}
Let $C_\bullet$ be a $[3,1,3]$ repetition code, encoding 1 bit into 3 bits. In this case, let
\[P = \begin{pmatrix} 1&1&0\\1&0&1\end{pmatrix}\]
\end{example}
\begin{example}
Let $C_\bullet$ be the $[7,4,3]$ Hamming code. Then let
\[P = \begin{pmatrix} 1&1&0&1&1&0&0\\1&0&1&1&0&1&0\\0&1&1&1&0&0&1 \end{pmatrix}\]
\end{example}

We now move on to quantum codes. Qubit Calderbank-Shor-Steane (CSS) codes are a type of stabiliser quantum code. Let $\mathscr{P}_n = \mathscr{P}^{\tens n}$ be the Pauli group over $n$ qubits. Stabiliser codes start by specifying an Abelian subgroup $\mathscr{S} \subset \mathscr{P}_n$, called a stabiliser subgroup, such that the codespace $\mathscr{H}$ is the mutual +1 eigenspace of all operators in $\mathscr{S}$. That is,
\[U\ket{\psi} = \ket{\psi} \quad \forall U \in \mathscr{S}, \ket{\psi}\in \mathscr{H}\]
We then specify a generating set of $\mathscr{S}$, of size $m$. For CSS codes, this generating set has as elements tensor product strings of either $\{I,X\}$ or $\{I,Z\}$ Pauli terms, with no scalars other than 1. One can define two parity check $\F_2$-matrices $P_X, P_Z$, for the $X$s and $Z$s, which together define a particular code. Each column in $P_X$ and $P_Z$ represents a physical qubit, and each row a measurement/stabiliser generator. $P_X$ and $P_Z$ thus map $Z$ and $X$ operators on physical qubits respectively to sets of measurement outcomes, with a $1$ outcome if the operators anticommute with a given stabiliser generator, and $0$ otherwise; these outcomes are also called \textit{syndromes}. $P_X$ is a $m_X \times n$ matrix, and $P_Z$ is $m_Z\times n$, with $m_X,m_Z$ marking the division of the generating set into $X$s and $Z$s respectively, satisfying $m=m_X+m_Z$. We do not require the generating set to be minimal, and hence $P_X$ and $P_Z$ need not be full rank.

\begin{definition}\label{def:weights}
We say that $w^Z$ is the maximal weight of all $Z$-type generators and $w^X$ the same for the $X$-type generators. These are the highest weight rows of $P_Z$ and $P_X$ respectively. Similarly, we say that $q^Z$, $q^X$ is the maximal number of $Z$, $X$ generators sharing a single qubit. These are the highest weight columns of $P_Z$ and $P_X$.
\end{definition}

CSS codes are described by parameters $\llbracket n,k,d \rrbracket$, with $k$ the number of encoded qubits and $d$ the code distance, which we define presently.

That the stabilisers must commute is equivalent to the requirement that $P_X P_Z^\intercal = P_Z P_X^\intercal = 0$. We may therefore view these matrices as differentials in a length 2 chain complex:
\[\begin{tikzcd}C_\bullet = C_{2}\arrow[r, "\del_2"]& C_1\arrow[r, "\del_{1}"]& C_{0}\end{tikzcd}\]
where $\del_2 = P_Z^\intercal$ and $\del_{1} = P_X$, or the other way round ($\del_2 = P_X^\intercal, \del_{1} = P_Z$) if desired, but we use the former for consistency with the literature. The quantum code then has $C_1 = \F_2^n$, and thus:
\[\begin{tikzcd}C_\bullet = \F_2^{m_Z}\arrow[r, "P_Z^\intercal"]& \F_2^n\arrow[r, "P_X"]& \F_2^{m_X}\end{tikzcd}\]
The code also has $k = \dim\ H_1(C_\bullet)$. To see this, observe first that $C_1$ represents the space of $Z$ Paulis on the set of physical qubits, with a vector being a Pauli string e.g. $v=\begin{pmatrix} 1&0&1\end{pmatrix}^\intercal \leadsto Z\tens I \tens Z$. Each vector in $H_1(C_\bullet)$ can be interpreted as an equivalence class $[v]$ of $Z$ operators on the set of physical qubits, modulo $Z$ operators which arise as $Z$ stabilisers. That this vector is in $Z_1(C_\bullet)$ means that the $Z$ operators commute with all $X$ stabilisers, and when the vector is not in $[0] = B_1(C_\bullet)$ it means that the $Z$ operators act nontrivially on the logical space. A basis of $H_1(C_\bullet)$ constitutes a choice of individual logical Paulis $\overline{Z}$, that is a tensor product decomposition of the space of logical $Z$ operators, and we set $\overline{Z}_1 = \overline{Z} \tens \overline{I}\cdots \tens \overline{I}$ on \textit{logical} qubits, $\overline{Z}_2 = \overline{I} \tens \overline{Z}\cdots \tens \overline{I}$ etc. There is a logical qubit for every logical $Z$, hence $k =\dim H_1(C_\bullet)$.

To get the logical $X$ operators, consider the cochain complex $C^\bullet$. The vectors in $H^{1}(C^\bullet)$ then correspond to $\overline{X}$ operators in the same manner. As $H_i(C_\bullet) \cong H^i(C^{\bullet})$ there must be an $\overline{X}$ operator for every $\overline{Z}$ operator and vice versa.

\begin{lemma}\label{lem:duality_basis}
A choice of basis $\{[v]_i\}_{i\leq k}$ for $H_1(C_\bullet)$ implies a choice of basis $\{[w]_j\}_{j\leq k}$ for $H^1(C^\bullet)$.
\end{lemma}
\proof
First, recall that we have the nondegenerate bilinear form 
\[\cdot : \F_2^n \times \F_2^n \rightarrow \F_2;\quad u\cdot v = u^\intercal v\]
which is equivalent to $\cdot : C_1 \times C^1 \rightarrow \F_2$; computationally, this tells us whether a $Z$ operator commutes or anticommutes with an $X$ operator. Now, let $u \in Z_1(C_\bullet)$ be a (possibly trivial) logical $Z$ operator, and $v \in B^1(C^\bullet)$ be a product of $X$ stabilisers. Then $P_X u = 0$, and $v = P_X^\intercal w$ for some $w \in C^2$. Thus $u \cdot v = u^\intercal v = u^\intercal P_X^\intercal w = (P_X u)^\intercal w = 0$, and so products of $X$ stabilisers commute with logical $Z$ operators. The same applies for $Z$ stabilisers and logical $X$ operators.

As a consequence, $v\cdot w = (v + s)\cdot (w+t)$ for any $v \in Z_1(C_\bullet)$, $w\in Z^1(C^\bullet)$, $s \in B_1(C_\bullet)$, $t\in B^1(C^\bullet)$, and so we may define $[v]\cdot [w] = v\cdot w$ for any $[v] \in H_1(C_\bullet)$, $[w]\in H^1(C^\bullet)$ with representatives $v, w$. The duality pairing of $C_1, C^1$ thus lifts to $H_1(C_\bullet), H^1(C^\bullet)$, and a choice of basis $\{[v]_i\}_{i\leq k}$ for $H_1(C_\bullet)$ implies a choice of basis of $H^1(C^\bullet)$, determined uniquely by $[v]_i\cdot [w]_j = \delta_{i,j}$.
\endproof

The above lemma ensures that picking a tensor product decomposition of logical $Z$ operators also entails the same tensor product decomposition of logical $X$ operators, so that $\overline{X}_i\overline{Z}_j = (-1)^{\delta_{i,j}}\overline{Z}_j\overline{X}_i$, for operators on the $i$th and $j$th logical qubits.

Let 
\[d^Z = \min_{v \in Z_1(C_\bullet)\backslash B_1(C_\bullet)} |v| ;\quad d^X = \min_{w \in Z^1(C^\bullet)\backslash B^1(C^\bullet)} |w|\]
where $|\cdot |$ is the Hamming weight of a vector, then the code distance $d = \min(d^Z,d^X)$. $d^Z$ and $d^X$ are called the systolic and cosystolic distances, and represent the lowest weight nontrivial $Z$ and $X$ logical operators respectively.

As all the data required for a CSS code is contained within the chain complex $C_\bullet$ -- and potentially a choice of basis of $H_1(C_\bullet)$ -- then we could define a CSS code as just the single chain complex, but it will be convenient to have direct access to the cochain complex as well.
\begin{definition}\label{def:CSS_complex}
A CSS code is a pair $(C_\bullet, C^\bullet)$, with $C_\bullet$ a length 2 chain complex centred at degree 1, so we have:
\[\begin{tikzcd}C_\bullet = C_{2}\arrow[r, "P_Z^\intercal"]& C_1\arrow[r, "P_X"]& C_{0}\end{tikzcd};\quad \begin{tikzcd}C^\bullet = C^{0}\arrow[r, "P_X^\intercal"]& C^1\arrow[r, "P_Z"]& C^{2}\end{tikzcd}\]
We call the first of the pair the $Z$-type complex, as vectors in $C_1$ correspond to $Z$-operators, and the second the $X$-type complex.
A {\rm based} CSS code additionally has a choice of basis for $H_1(C_\bullet)$, and hence for $H^1(C^\bullet)$. 
\end{definition}

Employing the direct sum $(C \oplus D)_\bullet$ of chain complexes we have the CSS code $((C \oplus D)_\bullet, (C \oplus D)^\bullet)$, which means the CSS codes $(C_\bullet, C^\bullet)$ and $(D_\bullet, D^\bullet)$ perform in parallel on disjoint sets of qubits, without any interaction. The $Z$ and $X$ operators will then be the tensor product of operators in each.

In summary, there is a bijection between length 1 chain complexes in $\Chains$ and binary linear classical codes, and between length 2 chain complexes in $\Chains$ and CSS codes. There are CSS codes for higher dimensional qudits, but for simplicity we stick to qubits.

Rather than just individual codes we tend to be interested in families of codes, where $n,k,d$ scale with the size of code in the family. Of particular practical interest are quantum \textit{low density parity check} (LDPC) CSS codes, which are families of codes where all $w^Z$, $w^X$, $q^Z$ and $q^X$ in the family are bounded from above by a constant. Equivalently, this means the Hamming weight of each column and row in each differential is bounded by a constant.

\subsection{Basic quantum codes}

\begin{example}\label{ex:shor}
Let $(C_\bullet, C^\bullet)$ be the $\llbracket 9,1,3\rrbracket $ Shor code, so we have $C_{2} = \F_2^2$, $C_1 = \F_2^9$, $C_0 = \F_2^6$. The parity check matrices are given by
\[
P_X = \begin{pmatrix} 
    1&1&1&1&1&1&0&0&0\\
    1&1&1&0&0&0&1&1&1
    \end{pmatrix}; \quad
P_Z = \begin{pmatrix}
    1&1&0&0&0&0&0&0&0\\
    1&0&1&0&0&0&0&0&0\\
    0&0&0&1&1&0&0&0&0\\
    0&0&0&1&0&1&0&0&0\\
    0&0&0&0&0&0&1&1&0\\
    0&0&0&0&0&0&1&0&1
    \end{pmatrix}
\]
We then have $\dim Z_1(C_\bullet) =\dim  C_1-{\rm rank}(P_X) = 9-2 = 7$ and $\dim  B_1(C_\bullet) = {\rm rank}(P_Z^\intercal) = 6$. Thus $k =\dim  H_1(C_\bullet) = 1$. There is a single nonzero equivalence class $[v] \in H_1(C_\bullet)$, with a representative $v = \begin{pmatrix}1&1&1&1&1&1&1&1&1\end{pmatrix}^\intercal$. Similarly there is the nonzero vector $w = \begin{pmatrix}1&1&1&1&1&1&1&1&1\end{pmatrix}^\intercal$, which is a representative of $[w]\in H_1(C_\bullet)$. Hence, we have two logical operators $\overline{Z} = \bigotimes_i^9 Z_i$, $\overline{X} = \bigotimes_i^9 X_i$ with $Z_i$ on the $i$th qubit and the same for $X_i$. We equally have, say, $\overline{Z} = Z_1\tens Z_4\tens Z_7$ and $\overline{X} = X_1\tens X_2\tens X_3$ in the same equivalence classes as those above, $[v]$ and $[w]$.
\end{example}

We now consider two examples which come from square lattices. This can be done much more generally. In Appendix~\ref{app:cells} we formalise categorically the procedure of acquiring chain complexes -- and therefore CSS codes -- from square lattices, which are a certain type of cell complex.
\begin{example}\label{ex:homological}
Consider the following square lattice:
\[\begin{tikzpicture}
	\begin{pgfonlayer}{nodelayer}
		\node [style=boundary vertex] (5) at (-3, 3) {};
		\node [style=boundary vertex] (6) at (-1, 3) {};
		\node [style=boundary vertex] (7) at (1, 3) {};
		\node [style=boundary vertex] (8) at (-3, 1) {};
		\node [style=boundary vertex] (9) at (-1, 1) {};
		\node [style=boundary vertex] (10) at (1, 1) {};
		\node [style=boundary vertex] (11) at (-3, -1) {};
		\node [style=boundary vertex] (12) at (-1, -1) {};
		\node [style=boundary vertex] (13) at (1, -1) {};
		\node [style=none] (14) at (3, 3) {};
		\node [style=none] (15) at (3, 1) {};
		\node [style=none] (16) at (3, -1) {};
		\node [style=none] (17) at (1, -3) {};
		\node [style=none] (18) at (-1, -3) {};
		\node [style=none] (19) at (-3, -3) {};
		\node [style={grey_dot}] (20) at (-3, -3) {};
		\node [style={grey_dot}] (21) at (-1, -3) {};
		\node [style={grey_dot}] (22) at (1, -3) {};
		\node [style={grey_dot}] (23) at (3, -1) {};
		\node [style={grey_dot}] (24) at (3, 1) {};
		\node [style={grey_dot}] (25) at (3, 3) {};
		\node [style={white_dot}] (36) at (-2, 2) {};
		\node [style={white_dot}] (37) at (0, 2) {};
		\node [style={white_dot}] (38) at (2, 2) {};
		\node [style={white_dot}] (39) at (2, 0) {};
		\node [style={white_dot}] (40) at (0, 0) {};
		\node [style={white_dot}] (41) at (-2, 0) {};
		\node [style={white_dot}] (42) at (-2, -2) {};
		\node [style={white_dot}] (43) at (0, -2) {};
		\node [style={white_dot}] (44) at (2, -2) {};
	\end{pgfonlayer}
	\begin{pgfonlayer}{edgelayer}
		\draw (5) to (8);
		\draw (8) to (9);
		\draw (9) to (10);
		\draw (10) to (7);
		\draw (7) to (6);
		\draw (6) to (9);
		\draw (5) to (6);
		\draw (8) to (11);
		\draw (11) to (12);
		\draw (12) to (9);
		\draw (12) to (13);
		\draw (13) to (10);
		\draw (7) to (14.center);
		\draw (10) to (15.center);
		\draw (13) to (16.center);
		\draw (13) to (17.center);
		\draw (12) to (18.center);
		\draw (11) to (19.center);
	\end{pgfonlayer}
\end{tikzpicture}\]
Edges in the lattice are qubits, so $n = 18$, the 9 $X$-checks are associated with vertices and the 9 $Z$-checks are associated with faces, which are indicated by white circles. Grey vertices indicate periodic boundary conditions, so the lattice can be embedded on a torus. This is an instance of the standard toric code \cite{Kit}.

The abstracted categorical homology from before is now the homology of the tessellated torus, with cycles, boundaries etc. having their usual meanings. $k = \dim H_1(C_\bullet) = 2$, and (co)systolic distances are the lengths of the essential cycles of the torus.
\end{example}

\begin{example}\label{ex:in_hom_code}
Now consider a different square lattice:
\[\begin{tikzpicture}
	\begin{pgfonlayer}{nodelayer}
		\node [style=none] (15) at (-2, 3) {};
		\node [style=none] (16) at (-2, -3) {};
		\node [style=none] (17) at (0, -3) {};
		\node [style=none] (18) at (0, 3) {};
		\node [style=boundary vertex] (21) at (-2, 1) {};
		\node [style=boundary vertex] (22) at (0, 1) {};
		\node [style=boundary vertex] (23) at (2, 1) {};
		\node [style=boundary vertex] (24) at (-2, -1) {};
		\node [style=boundary vertex] (25) at (0, -1) {};
		\node [style=boundary vertex] (26) at (2, -1) {};
		\node [style=boundary vertex] (27) at (-2, 1) {};
		\node [style=boundary vertex] (28) at (-2, -1) {};
		\node [style=boundary vertex] (29) at (0, -1) {};
		\node [style=boundary vertex] (30) at (0, 1) {};
		\node [style=none] (31) at (2, 3) {};
		\node [style=none] (32) at (2, -3) {};
		\node [style=boundary vertex] (33) at (2, 1) {};
		\node [style=boundary vertex] (34) at (2, -1) {};
		\node [style={white_dot}] (35) at (-1, 2) {};
		\node [style={white_dot}] (36) at (1, 2) {};
		\node [style={white_dot}] (37) at (1, 0) {};
		\node [style={white_dot}] (38) at (-1, 0) {};
		\node [style={white_dot}] (39) at (-1, -2) {};
		\node [style={white_dot}] (40) at (1, -2) {};
	\end{pgfonlayer}
	\begin{pgfonlayer}{edgelayer}
		\draw (15.center) to (27);
		\draw (27) to (28);
		\draw (28) to (16.center);
		\draw (18.center) to (30);
		\draw (30) to (29);
		\draw (29) to (17.center);
		\draw (24) to (25);
		\draw (25) to (26);
		\draw (21) to (22);
		\draw (22) to (23);
		\draw (31.center) to (33);
		\draw (33) to (34);
		\draw (34) to (32.center);
	\end{pgfonlayer}
\end{tikzpicture}\]
This represents a patch of surface code $(D_\bullet, D^\bullet)$, where we have two smooth sides, on the left and right, and two rough sides to the patch, on the top and bottom. There are `dangling' edges at the top and bottom, which do not terminate at vertices. We have
\[\dim D_2 =  \dim D_{0} = 6;\quad n = \dim D_1 = 13;\quad k=\dim H_1(D_\bullet) = 1 \]
The systolic distance is $3$, the length of the shortest path from the top to bottom boundary, and the cosystolic distance $3$, the same but from left to right.
\end{example}
\subsection{Code maps}\label{sec:code_maps}

One may wish to convert one code into another, making a series of changes to the set of stabiliser generators to be measured, and potentially also to the physical qubits. The motivation behind such protocols is typically to perform logical operations which are not available natively to the code; not only might the target code have other logical operations, but the protocol is itself a map between logical spaces when chosen carefully. An example of a change to the measurements and qubits is code deformation. We do not formalise code deformation here, as that has some specific connotations \cite{VLCABT}. Instead we define a related notion, called a \textit{code map}, which has some overlap. A code map is also related to, but not the same as, the `homomorphic gadgets' from \cite{HJY}.

\begin{definition}\label{def:code_def}
A $\overline{Z}$-preserving code map $\CF_{\overline{Z}}$ from a CSS code $(C_\bullet, C^\bullet)$ to $(D_\bullet, D^\bullet)$ is a paired chain map and cochain map $(f_\bullet, f^\bullet)$, for $f_\bullet: C_\bullet \rightarrow D_\bullet$ and $f^\bullet: D^\bullet \rightarrow C^\bullet$.
\[\begin{tikzcd}
\ \arrow[d, "\CF_{\overline{Z}}"', Rightarrow] & (C_\bullet, C^\bullet)\arrow[d, "f_\bullet"', shift right=2ex]\\
\ & (D_\bullet, D^\bullet)\arrow[u, "f^\bullet"', shift right=2ex]
\end{tikzcd}\]
\end{definition}

Note that the cochain map is strictly speaking obsolete, as all the data is contained in a single chain map $f_\bullet$, but as with CSS codes it will be handy to keep both around.

Let us unpack this definition. $\CF_{\overline{Z}}$ first maps $Z$-operators in $C_1$ to $Z$-operators in $D_1$, using $f_1$. It may map a single $Z$ on a qubit to a tensor product of $Z$s, or to $I$. It then has a map $f_2$ on $Z$ generators, and another $f_{0}$ on $X$ checks. Recalling Definition~\ref{def:chain_map}, we have:
\[\begin{tikzcd}C_{2}\arrow[r, "\del^{C_\bullet}_{2}"]\arrow[d, "f_{2}"'] & C_{1}\arrow[r, "\del^{C_\bullet}_{1}"]\arrow[d, "f_{1}"] & C_{0}\arrow[d,"f_{0}"]\\
D_{2}\arrow[r, "\del^{D_\bullet}_{2}"'] & D_{1}\arrow[r, "\del^{D_\bullet}_{1}"']\arrow[ul, phantom, "\textrm{I}"] & D_{0}\arrow[ul, phantom, "\textrm{II}"]\end{tikzcd}
\]
With two commuting squares labelled I and II. I stipulates that applying products of $Z$ stabiliser generators on the code and then performing the code map should be equivalent to performing the code map and then applying products of $Z$ stabiliser generators, i.e. $f_1\circ\del^{C_\bullet}_{2} =\del^{D_\bullet}_{2}\circ f_2 $. II stipulates that performing the $X$ measurements and then mapping the code should be equivalent to mapping the code and then performing $X$ measurements, so there is a consistent mapping between all measurement outcomes, i.e. $f_{0}\circ\del^{C_\bullet}_{1} =\del^{D_\bullet}_{1}\circ f_1 $.

Then there is the cochain map $f^\bullet$. This has the component $f^1 = f^\intercal_1 : D^1 \rightarrow C^1$, which maps an $X$-operator in $D^1$ back to an $X$-operator in $C^1$. Similarly for $f^0$ and $f^\intercal_2$, each of which come with commuting squares which are just the transposed conditions of those in $f_\bullet$, so they say nothing new. This is not surprising, as all the data for $f^\bullet$ is given by $f_\bullet$ already.

We now show that this definition entails some elementary properties. For a start, Lemma~\ref{lem:chain_map_rest} implies that a code map gives a map from a $\overline{Z}$ operator in $H_1(C_\bullet)$ to $\overline{Z}$s in $H_1(D_\bullet)$; this can also map to a tensor product of logical $\overline{Z}$s, and in particular map $\overline{Z}$ to zero i.e. $\overline{I}$, but it must not map a $\overline{Z}$ to an operator which can be detected by the $X$ stabiliser measurements. Hence $(f_\bullet, f^\bullet)$ preserves the fact that any $\overline{Z}$ is an undetectable operator on the codespace. A similar requirement holds for $\overline{X}$ operators, but this time the condition is inverted. Every $\overline{X}$ in $H^1(D^\bullet)$ must have a map only to logical operators in $H^1(C^\bullet)$, but the other way is not guaranteed.

Let $n_C$ and $n_D$ be the number of physical qubits in codes $(C_\bullet, C^\bullet)$ and $(D_\bullet, D^\bullet)$ respectively. We may interpret $\CF_{\overline{Z}}$ as a $\C$-linear map $M$ in $\FHilb$, the category of Hilbert spaces. This $\C$-linear map has the property that $M U_Z = U'_Z M$, where $U_Z$ is a tensor product of $Z$ Paulis on $n_C$ qubits and $U'_Z$ is a tensor product of $Z$ Paulis on $n_D$ qubits. In particular, given any $U_Z$ we have a specified $U'_Z$. The same is not true the other way round, as the map $f_1$ is not necessarily injective or surjective. Similarly, $M U_X = U'_X M$. This time, however, given any unique $U'_X$ on $n_D$ qubits we have a specified $U_X$ but vice versa is not guaranteed, depending on $f_1^\intercal$.

As a consequence, the linear map $M$ is \textit{stabiliser}, in the sense that it maps Paulis to Paulis, but not \textit{unitary} in general. $M$ is unitary iff $f_1$ is invertible.

If $M$ is not even an isometry, it cannot be performed deterministically, and the code map must include measurements on physical qubits. There will in general be Kraus operators corresponding to different measurement outcomes which will determine whether the code map has been implemented as desired; for now we assume that $M$ is performed deterministically, and leave this complication for Section~\ref{sec:practical}. Similarly, while the code map can be interpreted as a circuit between two codes, we do not claim that such a circuit can be performed fault-tolerantly in general.

\begin{remark}
For the following proposition, and at various points throughout the rest of the paper, we will use the ZX-calculus, a formal graphical language for reasoning about computation with qubits. We do not give a proper introduction to this calculus for brevity, but Sections 1-3 of \cite{JvdW} are sufficient for the interested reader. Our use of ZX diagrams is unsophisticated, and primarily for convenience. 
\end{remark}

\begin{proposition}\label{prop:CNOT_circuit}
Let $\CF_{\overline{Z}}$ be a $\overline{Z}$-preserving code map between codes $(C_\bullet,  C^\bullet)$ and $(D_\bullet,  D^\bullet)$ with qubit counts $n_C$ and $n_D$. The interpretation of $\CF_{\overline{Z}}$ as a $\C$-linear map $M$ in $\FHilb$ has a presentation as a circuit with gates drawn from $\{{\rm CNOT},\ket{+},\bra{0}\}$.
\end{proposition}
\proof
We start with the linear map $M: (\C^2)^{\tens n_C}\rightarrow (\C^2)^{\tens n_D}$:
\[\begin{tikzpicture}
	\begin{pgfonlayer}{nodelayer}
		\node [style=none] (0) at (-2, 0) {};
		\node [style=none] (1) at (2, 0) {};
		\node [style=gate] (2) at (0, 0) {$M$};
		\node [style=none] (3) at (-1.25, 0.5) {$n_C$};
		\node [style=none] (4) at (1.25, 0.5) {$n_D$};
		\node [style=none] (5) at (-1.5, -0.25) {};
		\node [style=none] (6) at (-1, 0.25) {};
		\node [style=none] (7) at (1, -0.25) {};
		\node [style=none] (8) at (1.5, 0.25) {};
	\end{pgfonlayer}
	\begin{pgfonlayer}{edgelayer}
		\draw (0.center) to (2);
		\draw (2) to (1.center);
		\draw (5.center) to (6.center);
		\draw (7.center) to (8.center);
	\end{pgfonlayer}
\end{tikzpicture}\]
By employing the partial transpose in the computational basis we convert it into the state
\[\begin{tikzpicture}
	\begin{pgfonlayer}{nodelayer}
		\node [style=none] (0) at (0.5, 1) {};
		\node [style=none] (1) at (3, -1) {};
		\node [style=none] (4) at (2.25, -0.5) {$n_D$};
		\node [style=none] (7) at (2, -1.25) {};
		\node [style=none] (8) at (2.5, -0.75) {};
		\node [style=none] (9) at (2.25, 1.5) {$n_C$};
		\node [style=none] (10) at (2, 0.75) {};
		\node [style=none] (11) at (2.5, 1.25) {};
		\node [style=none] (12) at (3, 1) {};
		\node [style=none] (13) at (-2.25, 0) {$\ket{\psi}=$};
		\node [style=none] (14) at (0.5, -1) {};
		\node [style=gate] (15) at (1, -1) {$M$};
	\end{pgfonlayer}
	\begin{pgfonlayer}{edgelayer}
		\draw (7.center) to (8.center);
		\draw (10.center) to (11.center);
		\draw (0.center) to (12.center);
		\draw (14.center) to (15);
		\draw (15) to (1.center);
		\draw [bend right=270, looseness=2.50] (14.center) to (0.center);
	\end{pgfonlayer}
\end{tikzpicture}\]
i.e. inserting $n_C$ Bell pairs. By the definition of $f_1$ we know that this has an independent stabiliser, with one $Z$ and $n_C-1$ $I$s followed by some $n_D$-fold tensor product of $Z$ and $I$, for each of the $n_C$ qubits. From $f_1^\intercal$ it also has an independent stabiliser, with some $n_C$-fold tensor product of $X$ and $I$ followed by $n_D-1$ Is and one $X$, for each of the $n_D$ qubits. $\ket{\psi}$ is therefore a stabiliser state. Further, from Theorem 5.1 of \cite{Kis} it has a presentation as a `phase-free ZX diagram', of the form
\[\begin{tikzpicture}
	\begin{pgfonlayer}{nodelayer}
		\node [style=Z dot] (0) at (-2, 2.75) {};
		\node [style=Z dot] (1) at (-2, 1.25) {};
		\node [style=Z dot] (2) at (-2, -0.25) {};
		\node [style=X dot] (3) at (0, 0.5) {};
		\node [style=X dot] (4) at (0, -1) {};
		\node [style=none] (5) at (0.5, 3.5) {};
		\node [style=none] (6) at (0.5, 5) {};
		\node [style=none] (7) at (-2.25, 2.25) {$\vdots$};
		\node [style=none] (8) at (-2.25, 0.75) {$\vdots$};
		\node [style=none] (9) at (0.25, 4.25) {$\vdots$};
		\node [style=none] (10) at (0.25, 0) {$\vdots$};
		\node [style=none] (11) at (0.5, 0.5) {};
		\node [style=none] (12) at (0.5, -1) {};
		\node [style=none] (13) at (0.5, 2) {};
		\node [style=none] (14) at (0.25, 2.75) {$\vdots$};
		\node [style=none] (15) at (1.5, 3.5) {$n_C$};
		\node [style=none] (16) at (1.5, -0.25) {$n_D$};
		\node [style=none] (17) at (-3.75, 1.25) {$\ket{\psi}=$};
		\node [style=X dot] (18) at (-2, 2.75) {};
		\node [style=X dot] (19) at (-2, 1.25) {};
		\node [style=X dot] (20) at (-2, -0.25) {};
		\node [style=Z dot] (21) at (0, 0.5) {};
		\node [style=Z dot] (22) at (0, -1) {};
	\end{pgfonlayer}
	\begin{pgfonlayer}{edgelayer}
		\draw (2) to (4);
		\draw (0) to (4);
		\draw (2) to (3);
		\draw (1) to (4);
		\draw (0) to (3);
		\draw (3) to (11.center);
		\draw (4) to (12.center);
		\draw (2) to (13.center);
		\draw (0) to (6.center);
		\draw (1) to (5.center);
	\end{pgfonlayer}
\end{tikzpicture}\]
where the top $n_C$ qubits do not have a green spider. We perform the partial transpose again to convert the state $\ket{\psi}$ back into the map $M$, which has the form
\[\begin{tikzpicture}
	\begin{pgfonlayer}{nodelayer}
		\node [style=Z dot] (0) at (0, 2.75) {};
		\node [style=Z dot] (1) at (0, 1.25) {};
		\node [style=Z dot] (2) at (0, -0.25) {};
		\node [style=X dot] (3) at (2, 2) {};
		\node [style=X dot] (4) at (2, 0.5) {};
		\node [style=none] (5) at (-0.5, 1.25) {};
		\node [style=none] (6) at (-0.5, 2.75) {};
		\node [style=none] (8) at (-0.25, 0.75) {$\vdots$};
		\node [style=none] (9) at (-0.25, 2.25) {$\vdots$};
		\node [style=none] (10) at (2.25, 1.5) {$\vdots$};
		\node [style=none] (11) at (2.5, 2) {};
		\node [style=none] (12) at (2.5, 0.5) {};
		\node [style=none] (13) at (-0.5, -0.25) {};
		\node [style=none] (17) at (-6.5, 1.25) {};
		\node [style=none] (18) at (-2.5, 1.25) {};
		\node [style=gate] (19) at (-4.5, 1.25) {$M$};
		\node [style=none] (20) at (-6, 1.75) {$n_C$};
		\node [style=none] (21) at (-3, 1.75) {$n_D$};
		\node [style=none] (22) at (-6, 1) {};
		\node [style=none] (23) at (-5.5, 1.5) {};
		\node [style=none] (24) at (-3.5, 1) {};
		\node [style=none] (25) at (-3, 1.5) {};
		\node [style=none] (26) at (-1.5, 1.25) {$=$};
		\node [style=X dot] (27) at (0, 2.75) {};
		\node [style=X dot] (28) at (0, 1.25) {};
		\node [style=X dot] (29) at (0, -0.25) {};
		\node [style=Z dot] (30) at (2, 2) {};
		\node [style=Z dot] (31) at (2, 0.5) {};
	\end{pgfonlayer}
	\begin{pgfonlayer}{edgelayer}
		\draw (2) to (4);
		\draw (0) to (4);
		\draw (2) to (3);
		\draw (1) to (4);
		\draw (0) to (3);
		\draw (3) to (11.center);
		\draw (4) to (12.center);
		\draw (2) to (13.center);
		\draw (0) to (6.center);
		\draw (1) to (5.center);
		\draw (17.center) to (19);
		\draw (19) to (18.center);
		\draw (22.center) to (23.center);
		\draw (24.center) to (25.center);
	\end{pgfonlayer}
\end{tikzpicture}\]
Any ZX diagram of this form can be expressed as a matrix over $\F_2$, mapping $X$-basis states from $(\C^2)^{\tens n_C}$ to $(\C^2)^{\tens n_D}$. The example above, ignoring the ellipses, has the matrix
\[\begin{pmatrix}
1&0&1\\1&1&1
\end{pmatrix}\]
which is equal to $f_1$; the point of the above rigmarole is thus to say that $f_1$ is precisely a linear map between $X$-basis states, which one can check easily. One can explicitly calculate $M$ as a matrix in the $X$-basis in $\FHilb$. For the first column, we compute $f_1 \begin{pmatrix}0 \\ 0 \\ 0\end{pmatrix} = \begin{pmatrix}0 \\ 0\end{pmatrix}$, so $M \ket{+}\ket{+}\ket{+} = \ket{+}\ket{+}$. Overall, we have
\[M = \begin{pmatrix}
1&0&0&0&0&1&0&0\\
0&0&1&0&0&0&0&1\\
0&0&0&1&0&0&1&0\\
0&1&0&0&1&0&0&0
\end{pmatrix}\]
which again is in the $X$-basis, not the (computational) $Z$-basis.

Returning to $f_1$, we can perform Gaussian elimination, performing row operations, which produce CNOTs on the r.h.s. of the diagram in the manner of \cite{KisM}, until the matrix is in reduced row echelon form. We then perform column operations producing CNOTs on the l.h.s. of the diagram, until the matrix has at most one 1 in each row and column. This can be performed using the leading coefficients to remove all other 1s in that row. The final matrix just represents a permutation of qubits with some states and effects. An empty column corresponds to a $\bra{0}$ effect, and an empty row a $\ket{+}$ state. We thus end up with a presentation of $M$ in the form
\[\begin{tikzpicture}
	\begin{pgfonlayer}{nodelayer}
		\node [style=none] (11) at (0.25, -0.25) {};
		\node [style=big gate] (13) at (-1.5, 1) {${\rm CNOT}$};
		\node [style=none] (16) at (-1.25, -0.25) {};
		\node [style=none] (20) at (-3.25, -0.25) {};
		\node [style=none] (21) at (-1.75, -0.25) {};
		\node [style=Z dot] (23) at (0.25, -0.25) {};
		\node [style=X dot] (24) at (-3.25, -0.25) {};
		\node [style=none] (27) at (1, 2) {};
		\node [style=none] (28) at (-1.25, 2) {};
		\node [style=none] (29) at (-4, 2) {};
		\node [style=none] (30) at (-1.75, 2) {};
		\node [style=none] (32) at (-3.5, 2.5) {$n_C$};
		\node [style=none] (33) at (-3.5, 1.75) {};
		\node [style=none] (34) at (-3, 2.25) {};
		\node [style=none] (36) at (0.5, 2.5) {$n_D$};
		\node [style=none] (37) at (0, 1.75) {};
		\node [style=none] (38) at (0.5, 2.25) {};
		\node [style=none] (39) at (0.25, 1) {};
		\node [style=none] (40) at (-1.25, 1) {};
		\node [style=none] (41) at (-3.25, 1) {};
		\node [style=none] (42) at (-1.75, 1) {};
		\node [style=Z dot] (43) at (0.25, 1) {};
		\node [style=X dot] (44) at (-3.25, 1) {};
		\node [style=none] (45) at (-3, 0.5) {$\vdots$};
		\node [style=none] (46) at (0, 0.5) {$\vdots$};
		\node [style=Z dot] (47) at (-3.25, 1) {};
		\node [style=Z dot] (48) at (-3.25, -0.25) {};
		\node [style=X dot] (49) at (0.25, 1) {};
		\node [style=X dot] (50) at (0.25, -0.25) {};
	\end{pgfonlayer}
	\begin{pgfonlayer}{edgelayer}
		\draw (16.center) to (11.center);
		\draw (21.center) to (20.center);
		\draw (28.center) to (27.center);
		\draw (30.center) to (29.center);
		\draw (33.center) to (34.center);
		\draw (37.center) to (38.center);
		\draw (40.center) to (39.center);
		\draw (42.center) to (41.center);
	\end{pgfonlayer}
\end{tikzpicture}\]
On our example, this is then
\[\begin{tikzpicture}
	\begin{pgfonlayer}{nodelayer}
		\node [style=Z dot] (0) at (-3.5, 2.25) {};
		\node [style=Z dot] (1) at (-3.5, 0.75) {};
		\node [style=Z dot] (2) at (-3.5, -0.75) {};
		\node [style=X dot] (3) at (-1.5, 1.5) {};
		\node [style=X dot] (4) at (-1.5, 0) {};
		\node [style=none] (5) at (-4, 0.75) {};
		\node [style=none] (6) at (-4, 2.25) {};
		\node [style=none] (10) at (-1, 1.5) {};
		\node [style=none] (11) at (-1, 0) {};
		\node [style=none] (12) at (-4, -0.75) {};
		\node [style=none] (13) at (-0.25, 0.75) {$=$};
		\node [style=none] (14) at (0.5, 2.25) {};
		\node [style=none] (15) at (0.5, 0.75) {};
		\node [style=none] (16) at (0.5, -0.75) {};
		\node [style=X dot] (17) at (1.5, 2.25) {};
		\node [style=X dot] (18) at (2.5, 0.75) {};
		\node [style=Z dot] (19) at (2.5, 2.25) {};
		\node [style=Z dot] (20) at (1.5, -0.75) {};
		\node [style=none] (21) at (3.5, 2.25) {};
		\node [style=none] (22) at (3.5, 0.75) {};
		\node [style=Z dot] (23) at (2.5, -0.75) {};
		\node [style=Z dot] (24) at (-1.5, 1.5) {};
		\node [style=Z dot] (25) at (-1.5, 0) {};
		\node [style=Z dot] (26) at (1.5, 2.25) {};
		\node [style=Z dot] (27) at (2.5, 0.75) {};
		\node [style=X dot] (28) at (2.5, 2.25) {};
		\node [style=X dot] (29) at (1.5, -0.75) {};
		\node [style=X dot] (30) at (2.5, -0.75) {};
		\node [style=X dot] (31) at (-3.5, 2.25) {};
		\node [style=X dot] (32) at (-3.5, 0.75) {};
		\node [style=X dot] (33) at (-3.5, -0.75) {};
	\end{pgfonlayer}
	\begin{pgfonlayer}{edgelayer}
		\draw (2) to (4);
		\draw (0) to (4);
		\draw (2) to (3);
		\draw (1) to (4);
		\draw (0) to (3);
		\draw (3) to (10.center);
		\draw (4) to (11.center);
		\draw (2) to (12.center);
		\draw (0) to (6.center);
		\draw (1) to (5.center);
		\draw (14.center) to (17);
		\draw (17) to (19);
		\draw (19) to (21.center);
		\draw (22.center) to (18);
		\draw (18) to (19);
		\draw (17) to (20);
		\draw (16.center) to (20);
		\draw (20) to (23);
		\draw (15.center) to (18);
	\end{pgfonlayer}
\end{tikzpicture}\]
which one can check maps $Z\tens I\tens I \mapsto Z\tens Z$ etc.
\endproof
As a consequence $\bar{M} = M$, i.e. the conjugate of $M$ is just $M$.

\begin{corollary}\label{cor:zero_inputs}
If $n_C =0$ then the map $M$ is actually a stabiliser state of the form $M = \ket{+}^{\otimes n_D}$. When $n_D = 0$ then $M = \bra{0}^{\otimes n_C}$.
\end{corollary}
\proof
When $n_C =0$ we see that $M$ has exactly $n_D$ independent stabilisers with 1 $X$ and $n_D-1$ $I$s, for each qubit to put $X$ on. The flipped argument applies when $n_D=0$.
\endproof

\begin{definition}
An $\overline{X}$-preserving code map $\CF_{\overline{X}}$ from a CSS code $(D_\bullet,  D^\bullet)$ to $(C_\bullet,  C^\bullet)$ is a paired chain map and cochain map $(f_\bullet, f^\bullet)$, for $f_\bullet: C_\bullet \rightarrow D_\bullet$ and $f^\bullet:  D^\bullet \rightarrow  C^\bullet$.
\[\begin{tikzcd}
\ \arrow[d, "\CF_{\overline{X}}"', Leftarrow] & (C_\bullet,  C^\bullet)\arrow[d, "f_\bullet"', shift right=2ex]\\
\ & (D_\bullet,  D^\bullet)\arrow[u, "f^\bullet"', shift right=2ex]
\end{tikzcd}\]
\end{definition}

So $\CF_{\overline{X}}$ is just mapping in the other direction to $\CF_{\overline{Z}}$ from before, and we say that $\CF_{\overline{X}}$ is \textit{opposite} to $\CF_{\overline{Z}}$. In this case, when we interpret $\CF_{\overline{X}}$ as a $\C$-linear map $L$, it has the property that $L U_X = U'_XL$ and that any $U_X$ gives a specified $U'_X$, and $L U_Z = U'_ZL$, but that any $U'_Z$ gives a specified $U_Z$ but not vice versa.

By inspecting the stabilisers we see that, for $\CF_{\overline{Z}}$ with interpretation $M$ and $\CF_{\overline{X}}$ with interpretation $L$, $L = M^\dagger = M^\intercal$.

\begin{corollary}\label{cor:CNOTs}
Let $\CF_{\overline{X}}$ be an $\overline{X}$-preserving code map between codes $(D_\bullet,  D^\bullet)$ and $(C_\bullet,  C^\bullet)$ with qubit counts $n_D$ and $n_C$. The interpretation of $\CF_{\overline{X}}$ as a $\C$-linear map $L$ in $\FHilb$ has a presentation as a circuit with gates drawn from $\{{\rm CNOT},\ket{0}, \bra{+}\}$.
\end{corollary}

\begin{corollary}\label{cor:zero_outputs}
If $n_D = 0$ then $L = \ket{0}^{\otimes n_C}$, and if $n_C = 0$ then $L = \bra{+}^{\otimes n_D}$.
\end{corollary}

\begin{corollary}\label{cor:restriction_maps}
The restrictions of $\CF_{\overline{Z}}$ and $\CF_{\overline{X}}$ to use only $H_1(f_\bullet)$ and $H^1(f^\bullet)$ also have interpretations as $\C$-linear maps on logical qubits in the same way, and Proposition~\ref{prop:CNOT_circuit} and Corollary~\ref{cor:CNOTs} also apply to such interpretations.
\end{corollary}

While our definitions are for chain complexes of length 2, in principle one can map between any two codes with an arbitrary number of meta-checks, or between a classical code and quantum code, which could be interpreted as `switching on/off' either $X$ or $Z$ stabiliser measurements.

Code maps are related to code deformations, but we are aware of code deformation protocols which do not appear to fit in the model of chain maps described. For example, when moving defects around on the surface code for the purpose of, say, defect braiding \cite{FMMC}, neither $\overline{Z}$ nor $\overline{X}$ operators are preserved in the sense we give here.

\section{CSS code surgery}\label{sec:CSS_surgery}
\subsection{Colimits in $\Chains$}
To understand code surgery we require some additional chain complex technology, namely colimits. Coproducts, pushouts and coequalisers are directly relevant for our applications. Coproducts are just direct sums, so we describe pushouts and coequalisers here.

\begin{definition}\label{def:pushout}
The pushout of chain maps $f_\bullet: A_\bullet \rightarrow C_\bullet$ and $g_\bullet:A_\bullet \rightarrow D_\bullet$ gives the chain complex $Q_\bullet$, where each component is the pushout $Q_n$ of $f_n$ and $g_n$. The differentials $\del^{Q_\bullet}_n$ are given by the unique mediating map from each component's pushout. Specifically, if we have the pushout
\[\begin{tikzcd}
A_\bullet \arrow[r, "g_\bullet"]\arrow[d, "f_\bullet"'] & D_\bullet \arrow[d,"l_\bullet"]\\
C_\bullet \arrow[r,"k_\bullet"'] & Q_\bullet\arrow[ul, phantom, "\usebox\pushout", very near start]
\end{tikzcd}\]
then for degrees $n, n+1$ we have
\[\begin{tikzcd}
A_n\arrow[rrr,"g_n"]\arrow[ddd,"f_n"'] & & & D_n\arrow[ddd, "l_n"]\\
& A_{n+1}\arrow[ul, "\del^{A_\bullet}_{n+1}"]\arrow[r, "g_{n+1}"]\arrow[d, "f_{n+1}"'] & D_{n+1}\arrow[ur, "\del^{D_\bullet}_{n+1}"]\arrow[d, "l_{n+1}"] &\\
& C_{n+1}\arrow[dl,"\del^{C_\bullet}_{n+1}"]\arrow[r,"k_{n+1}"'] & Q_{n+1}\arrow[dr, dotted, "\del^{Q_\bullet}_{n+1}"]\arrow[ul, phantom, "\usebox\pushout", very near start] &\\
C_n\arrow[rrr, "k_n"'] & & & Q_n
\end{tikzcd}\]
where 
\[Q_n = (C \oplus D)_n/f_n\sim g_n; \quad k_n(c)=[c] \in Q_n;\quad l_n(d)=[d]\in Q_n.\] 
with $[c]$ being the equivalence class in $Q_n$ having $c$ as a representative, and the same for $[d]$. As $k_n\circ \del^{C_\bullet}_{n+1}\circ f_{n+1} = l_n\circ \del^{D_\bullet}_{n+1}\circ g_{n+1}$ and the inner square is a pushout in $\MatF$, there is a unique matrix $\del^{Q_\bullet}_{n+1}$. The differentials satisfy $\del^{Q_\bullet}_{n}\circ \del^{Q_\bullet}_{n+1} = 0$, and one can additionally check that this is indeed a pushout in $\Chains$ by considering the universal property at each component.
\end{definition}

\begin{definition}\label{def:coequaliser}
The coequaliser of chain maps $\begin{tikzcd} C_\bullet \arrow[r, "f", shift left=1.5ex]\arrow[r, "g"', shift right=1.5ex]& D_\bullet \end{tikzcd}$ is a chain complex $E_\bullet$ and chain map $\coeq(f,g)_\bullet : D_\bullet \rightarrow E_\bullet$, which we will just call $\coeq_\bullet$. We have $E_n = D_n/f_n\sim g_n$ and $\coeq_n(d) = [d]$.
\end{definition}

Doing some minor diagram chasing one can check that this is indeed a coequaliser in $\Chains$. 

\begin{remark}\label{rem:push_coeq}
We can view the pushout
\[\begin{tikzcd}
A_\bullet \arrow[r, "g_\bullet"]\arrow[d, "f_\bullet"'] & D_\bullet \arrow[d,"l_\bullet"]\\
C_\bullet \arrow[r,"k_\bullet"'] & Q_\bullet\arrow[ul, phantom, "\usebox\pushout", very near start]
\end{tikzcd}\]
as the coequaliser of $\begin{tikzcd}A_\bullet \arrow[r, "\tau_\bullet \circ f_\bullet",shift left=1.5ex]\arrow[r, "\omega_\bullet \circ g_\bullet"',shift right=1.5ex]& (C\oplus D)_\bullet\end{tikzcd}$  for the inclusion maps $\begin{tikzcd}C_\bullet \arrow[r, "\tau_\bullet", hookrightarrow] & (C\oplus D)_\bullet\end{tikzcd}$, $\begin{tikzcd}D_\bullet \arrow[r, "\omega_\bullet", hookrightarrow] & (C\oplus D)_\bullet\end{tikzcd}$.
The difference is that the pair of chain maps $k_\bullet, l_\bullet$ have been replaced with the single map $\coeq_\bullet$, so we have
\[\begin{tikzcd}A_\bullet \arrow[r, "\tau_\bullet\circ f_\bullet",shift left=1.5ex]\arrow[r, "\omega_\bullet\circ g_\bullet"',shift right=1.5ex]& (C\oplus D)_\bullet \arrow[r, "\coeq_\bullet"] &Q_\bullet
\end{tikzcd}\]
We can view coequalisers as instances of pushouts as well, doing a sort of reverse of the procedure above.
\end{remark}

As with all colimits, those above are defined by the category theory only up to isomorphism.
Because we are working over a field, the isomorphism class of a chain complex $Q_\bullet$ is completely determined by the dimensions of the underlying vector spaces $\{\dim Q_i\}_i$ and its \emph{Betti numbers}, which is the set $\{\dim H_i(Q_\bullet)\}_i$ of dimensions of the homology spaces. This is a homological version of the rank-nullity theorem. These are very large iso-classes, and we require more fine-grained control over which chain complexes are chosen by the colimits.

One way to choose a specific pushout of chain maps
is via an explicit definition of the coequaliser of two based linear maps. 
For this we need not just a basis for our vector spaces, but an ordered basis.
Using these coequalisers we can then construct pushouts of linear maps and their universal arrows,
which is used in turn to define the pushout of chain maps.
For the coequaliser of linear maps $r,s:V\to W$, we may take $s=0$ by linearity.
Then we let $r^{+}$ be the reflexive generalised inverse of $r$, which always exists by \cite{WD},
and see that $P = I - r r^{+}$ is a projector $P:W\to W$ that coequalizes $r,0:V\to W.$
To make this a universal projector we need to row-reduce the matrix of $P$, i.e. put $P$ into row echelon form and remove all-zero rows, which is
where we use the order on the basis of $W$. 
This row-reduced matrix will then have full rank and will be a universal coequaliser.

Alternatively, there is a straightforward way to choose the representatives we want from these iso-classes when the chain maps $f_\bullet, g_\bullet$ in the span are basis-preserving in the sense of Definition~\ref{def:basis_preserving}.

\begin{lemma}\label{lem:basis_preserve}
Let the chain maps $f_\bullet, g_\bullet$ be basis-preserving. Then there is always a choice of pushout such that the chain maps $k_\bullet, l_\bullet$ are also basis-preserving.
\end{lemma}
\proof
Let $Q_n = (C \oplus D)_n /f_n\sim g_n$. Then, for any basis element $a \in A_n$, the elements $f_n(a)$ and $g_n(a)$ are mapped by $k_n$ and $l_n$ respectively to the same basis element in $Q_n$. For basis elements in $C_n$, $D_n$ which are not in the images of $f_n$ and $g_n$, $k_n$ and $l_n$ will map them to distinct basis elements in $Q_n$. So this choice of representative of the isomorphism class of pushouts has $k_\bullet$, $l_\bullet$ being basis-preserving.
\endproof

We can think of the pushout at each component as being a pushout in the category $\CatSet$, freely promoted to $\MatF$. The differentials are then defined by the universal property. All of the choices of pushout such that $k_\bullet$, $l_\bullet$ are basis-preserving are equivalent up to a relabelling of basis elements in each component, hence (coherent) row and column permutation of the differential matrices. Computationally, this means that the choice of pushout is defined up to relabelling qubits, $Z$-checks and $X$-checks, and hence properties like code distance and being LDPC are well-defined for such pushouts. This will be important later. 

Lemma~\ref{lem:basis_preserve} also applies to coequalisers, following Remark~\ref{rem:push_coeq}.

\begin{lemma}\label{lem:abelian}
$\Chains$ is an Abelian category, and thus is finitely complete and cocomplete, meaning that it has all finite limits and colimits.
\end{lemma}

While this is well-known, we sketch a proof of this lemma in Appendix~\ref{app:lims}. There we also sketch some additional limits for completeness, but we only need colimits in this work. This lemma does not mean that every span has a \textit{basis-preserving} pushout, but whenever there is a basis-preserving span there is a basis-preserving pushout. The same applies to coequalisers. From now on every example of a span we use will be basis-preserving, so we assume that the pushouts and coequalisers are also basis-preserving and will no longer mention this property.

\subsection{Generic code surgery}

We now give a general set of definitions for surgery between arbitrary compatible CSS codes; the condition for compatibility is very weak here. Working at this level of generality means that we cannot prove very much about the output codes or relevant logical maps. As a consequence, we will then focus on particular surgeries which make use of `gluing' or `tearing' along logical $\overline{Z}$ or $\overline{X}$ operators in Section~\ref{sec:operator_surgery}.

\begin{definition}
Let $(C_\bullet,  C^\bullet)$, $(D_\bullet,  D^\bullet)$ and $(A_\bullet, A^\bullet)$ be CSS codes, such that there is a basis-preserving span of chain complexes
\[\begin{tikzcd}
A_\bullet \arrow[r, rightarrow, "g_\bullet"]\arrow[d, rightarrow, "f_\bullet"'] & D_\bullet\\
C_\bullet & &
\end{tikzcd}\]
The $Z$-type merged code of $(C_\bullet,  C^\bullet)$ and $(D_\bullet,  D^\bullet)$ along $f_\bullet, g_\bullet$ is the code $(Q_\bullet, Q^\bullet)$ such that $Q_\bullet$ is the pushout of the above diagram.
\end{definition}

Recall from Remark~\ref{rem:push_coeq} that we can view any pushout as a coequaliser. We thus have
\[\begin{tikzcd}
A_\bullet \arrow[r, "\iota_C\circ f_\bullet",shift left=1.5ex]\arrow[r, "\iota_D\circ g_\bullet"',shift right=1.5ex] & (C\oplus D)_\bullet \arrow[r, "\coeq_\bullet"] &Q_\bullet
\end{tikzcd}\]
and we call $\coeq_\bullet$ the $Z$-merge chain map. We can bundle this up into a $Z$-merge code map:
\begin{equation}\begin{tikzcd}\label{eq:Z_merge_code_map}
\ \arrow[d, "\CF_{\overline{Z}}"', Rightarrow] & ((C\oplus D)_\bullet, (C\oplus D)^\bullet)\arrow[d, "\coeq_\bullet"', shift right=3ex]\\
\ & (Q_\bullet, Q^\bullet)\arrow[u, "\coeq^\bullet"', shift right=3ex]
\end{tikzcd}\end{equation}
We then call $\coeq^\bullet: Q^\bullet\rightarrow (C\oplus D)^\bullet$ an $X$-split cochain map, and hence we have an $X$-split code map too:
\begin{equation}\begin{tikzcd}
\ \arrow[d, "\CF_{\overline{X}}"', Leftarrow] & ((C\oplus D)_\bullet, (C\oplus D)^\bullet)\arrow[d, "\coeq_\bullet"', shift right=3ex]\\
\ & (Q_\bullet, Q^\bullet)\arrow[u, "\coeq^\bullet"', shift right=3ex]
\end{tikzcd}\end{equation}

\begin{definition}
Let $(C_\bullet, C^\bullet)$, $(D_\bullet, D^\bullet)$ and $(A_\bullet, A^\bullet)$ be CSS codes, such that there is a span of cochain complexes
\[\begin{tikzcd}
A^\bullet \arrow[r, rightarrow, "g^\bullet"]\arrow[d, rightarrow, "f^\bullet"'] & D^\bullet\\
C^\bullet & &
\end{tikzcd}\]
the $X$-type merged code of $(C_\bullet, C^\bullet)$ and $(D_\bullet, D^\bullet)$ along $f^\bullet, g^\bullet$ is the code $(Q_\bullet, Q^\bullet)$ such that $Q^\bullet$ is the pushout of the above diagram.
\end{definition}
We have an $X$-merge cochain map and thus $X$-merge code map using the coequaliser picture, so
\[\begin{tikzcd}
\ \arrow[d, "\CE_{\overline{X}}"', Leftarrow] & (Q_\bullet, Q^\bullet)\arrow[d, "\coeq_\bullet"', shift right=3ex]\\
\ & ((C\oplus D)_\bullet, (C\oplus D)^\bullet)\arrow[u, "\coeq^\bullet"', shift right=3ex]
\end{tikzcd}\]
We also have a $Z$-split chain map and the $Z$-split code map $\CE_{\overline{Z}}$ by taking the opposite.
\[\begin{tikzcd}
\ \arrow[d, "\CE_{\overline{Z}}"', Rightarrow] & (Q_\bullet, Q^\bullet)\arrow[d, "\coeq_\bullet"', shift right=3ex]\\
\ & ((C\oplus D)_\bullet, (C\oplus D)^\bullet)\arrow[u, "\coeq^\bullet"', shift right=3ex]
\end{tikzcd}\]
This is rather abstract, so let's see a small concrete example.
\begin{example}\label{ex:small_pushout_codes}
Consider the following pushout of square lattices:
\[\begin{tikzpicture}
	\begin{pgfonlayer}{nodelayer}
		\node [style=boundary vertex] (0) at (-4, 1) {};
		\node [style=boundary vertex] (1) at (-2, 1) {};
		\node [style=none] (2) at (-1, 1) {};
		\node [style=none] (3) at (1, 1) {};
		\node [style=none] (4) at (-3, -0.5) {};
		\node [style=none] (5) at (-3, -2.5) {};
		\node [style=none] (6) at (-4, -6) {};
		\node [style=none] (7) at (-2, -6) {};
		\node [style=none] (8) at (2, 2) {};
		\node [style=none] (9) at (4, 2) {};
		\node [style=boundary vertex] (10) at (2, 0) {};
		\node [style=boundary vertex] (11) at (4, 0) {};
		\node [style=boundary vertex] (12) at (-2, -4) {};
		\node [style=boundary vertex] (13) at (-4, -4) {};
		\node [style=none] (14) at (3, -1) {};
		\node [style=none] (15) at (3, -3) {};
		\node [style=none] (16) at (-1, -5) {};
		\node [style=none] (17) at (1, -5) {};
		\node [style=none] (18) at (2, -4) {};
		\node [style=none] (19) at (4, -4) {};
		\node [style=boundary vertex] (20) at (2, -6) {};
		\node [style=boundary vertex] (21) at (4, -6) {};
		\node [style=none] (22) at (2, -8) {};
		\node [style=none] (23) at (4, -8) {};
		\node [style=boundary vertex] (24) at (4, -6) {};
		\node [style=boundary vertex] (25) at (2, -6) {};
		\node [style={white_dot}] (26) at (-3, -5) {};
		\node [style={white_dot}] (27) at (3, 1) {};
		\node [style={white_dot}] (28) at (3, -5) {};
		\node [style={white_dot}] (29) at (3, -7) {};
	\end{pgfonlayer}
	\begin{pgfonlayer}{edgelayer}
		\draw [style={blue_edge}] (0) to (1);
		\draw [style=arrow] (2.center) to (3.center);
		\draw [style=arrow] (4.center) to (5.center);
		\draw (13) to (6.center);
		\draw (12) to (7.center);
		\draw (8.center) to (10);
		\draw (9.center) to (11);
		\draw [style={blue_edge}] (13) to (12);
		\draw [style={blue_edge}] (10) to (11);
		\draw [style=arrow] (14.center) to (15.center);
		\draw [style=arrow] (16.center) to (17.center);
		\draw (18.center) to (20);
		\draw (19.center) to (21);
		\draw [style={blue_edge}] (20) to (21);
		\draw (25) to (22.center);
		\draw (24) to (23.center);
		\draw [style={blue_edge}] (25) to (24);
	\end{pgfonlayer}
\end{tikzpicture}\]
We have not properly formalised pushouts of square lattices in the main body for brevity, but we do so in Appendix~\ref{app:cells}. Informally, we are just `gluing along' the graph in the top left corner, where the edges to be glued are coloured in blue.

We can consider this pushout to be in $\Chains$ \footnote{Categorically, this is because there is a cocontinuous functor from the appropriate category of square lattices to $\Chains$.}, giving the pushout:
\[\begin{tikzcd}
A_\bullet \arrow[r, "g_\bullet"]\arrow[d, "f_\bullet"'] & D_\bullet \arrow[d,"q_\bullet"]\\
C_\bullet \arrow[r,"p_\bullet"'] & Q_\bullet\arrow[ul, phantom, "\usebox\pushout", very near start]
\end{tikzcd}\]
with
\[\begin{tikzcd}A_\bullet = \F_2 \arrow[r, "\del_{1}^{A_\bullet}"] & \F_2^2\end{tikzcd}\]
\[\begin{tikzcd}C_\bullet = \F_2 \arrow[r, "\del_{2}^{C_\bullet}"] & \F_2^3\arrow[r,"\del_{1}^{C_\bullet}"] & \F_2^2\end{tikzcd}\]
\[\begin{tikzcd}D_\bullet = \F_2 \arrow[r, "\del_{2}^{D_\bullet}"] & \F_2^3\arrow[r,"\del_{1}^{D_\bullet}"] & \F_2^2\end{tikzcd}\]
and
\[\del_{1}^{A_\bullet} = \begin{pmatrix}1\\1\end{pmatrix};\quad 
\del_{2}^{C_\bullet} = \del_{2}^{D_\bullet} = \begin{pmatrix}1\\1\\1 \end{pmatrix}; \quad
\del_{1}^{C_\bullet} = \begin{pmatrix}1 & 1 & 0\\0&1&1 \end{pmatrix};\quad 
\del_{1}^{D_\bullet} = \begin{pmatrix}1 & 0 & 1\\0&1&1 \end{pmatrix}.
\]
One can see from the cell complexes that we have
\[\begin{tikzcd}Q_\bullet = \F_2^2 \arrow[r, "\del_{2}^{Q_\bullet}"] &\F_2^5 \arrow[r,"\del_{1}^{Q_\bullet}"] & \F_2^2\end{tikzcd}\]
with 
\[\del_{2}^{Q_\bullet} = \begin{pmatrix}1&0\\1&1\\1&0\\0&1\\0&1 \end{pmatrix}; 
\quad \del_{1}^{Q_\bullet} = \begin{pmatrix}1&1&0&1&0\\0&1&1&0&1\end{pmatrix}\]
Rather than compute the pushout maps, let us instead give the coequaliser $\coeq_\bullet$:
\[\begin{tikzcd}(C\oplus D)_{2}\arrow[r, "\del^{(C\oplus D)_\bullet}_{2}"]\arrow[d, "\coeq_{2}"] & (C\oplus D)_{1}\arrow[r, "\del^{(C\oplus D)_\bullet}_{1}"]\arrow[d, "\coeq_{1}"] & (C\oplus D)_{0}\arrow[d,"\coeq_{0}"]\\
Q_{2}\arrow[r, "\del^{Q_\bullet}_2"] & Q_1\arrow[r, "\del^{Q_\bullet}_{1}"] & Q_{0}\end{tikzcd}
\]
We immediately see that $\coeq_2 = \id$. For the other two surjections we have
\[\coeq_1 = \begin{pmatrix}1&0&0&0&0&0\\0&1&0&0&0&1\\0&0&1&0&0&0\\0&0&0&1&0&0\\0&0&0&0&1&0\end{pmatrix};\quad \coeq_{0} = \begin{pmatrix}1&0&1&0\\0&1&0&1\end{pmatrix} \]
Finally we interpret all the chain complexes in this pushout as being the $Z$-type complexes of CSS codes $(A_\bullet, A^\bullet)$, $(C_\bullet,  C^\bullet)$ etc. Thus we have a $Z$-merge code map $\CF_{\overline{Z}}$, with an interpretation $M$ as a $\C$-linear map, using $\coeq_1$ and $\coeq_1^\intercal$. We refrain from writing out the full $32$-by-$64$ matrix, but as a ZX-diagram using gates from $\{{\rm CNOT}, \ket{+}, \bra{0}\}$ we have simply
\[\begin{tikzpicture}
	\begin{pgfonlayer}{nodelayer}
		\node [style=none] (13) at (-0.25, 1) {$=$};
		\node [style=none] (14) at (0.5, 2.5) {};
		\node [style=none] (15) at (0.5, 1) {};
		\node [style=none] (16) at (0.5, -5) {};
		\node [style=X dot] (18) at (1.5, 1) {};
		\node [style=Z dot] (20) at (1.5, -5) {};
		\node [style=none] (21) at (3.5, 2.5) {};
		\node [style=none] (22) at (3.5, 1) {};
		\node [style=Z dot] (23) at (2.5, -5) {};
		\node [style=none] (24) at (-5, 1) {};
		\node [style=none] (25) at (-1, 1) {};
		\node [style=gate] (26) at (-3, 1) {$M$};
		\node [style=none] (27) at (-4.25, 1.5) {$6$};
		\node [style=none] (28) at (-1.75, 1.5) {$5$};
		\node [style=none] (29) at (-4.5, 0.75) {};
		\node [style=none] (30) at (-4, 1.25) {};
		\node [style=none] (31) at (-2, 0.75) {};
		\node [style=none] (32) at (-1.5, 1.25) {};
		\node [style=none] (33) at (0.5, -0.5) {};
		\node [style=none] (34) at (3.5, -0.5) {};
		\node [style=none] (35) at (0.5, -2) {};
		\node [style=none] (36) at (3.5, -2) {};
		\node [style=none] (37) at (0.5, -3.5) {};
		\node [style=none] (38) at (3.5, -3.5) {};
		\node [style=none] (39) at (1.5, 1) {};
		\node [style=none] (40) at (1.5, -5) {};
		\node [style=Z dot] (41) at (1.5, 1) {};
		\node [style=X dot] (42) at (1.5, -5) {};
		\node [style=X dot] (43) at (2.5, -5) {};
	\end{pgfonlayer}
	\begin{pgfonlayer}{edgelayer}
		\draw (22.center) to (18);
		\draw (16.center) to (20);
		\draw (20) to (23);
		\draw (15.center) to (18);
		\draw (24.center) to (26);
		\draw (26) to (25.center);
		\draw (29.center) to (30.center);
		\draw (31.center) to (32.center);
		\draw (14.center) to (21.center);
		\draw (33.center) to (34.center);
		\draw (35.center) to (36.center);
		\draw (37.center) to (38.center);
		\draw (39.center) to (40.center);
	\end{pgfonlayer}
\end{tikzpicture}\]
We know from Lemma~\ref{lem:chain_map_rest} that this map must restrict to a map on logical qubits. However, easy calculations show that $H_1((C\oplus D)_\bullet) = 0$, while $H_1(Q_\bullet) = 1$. That is, in the code $((C\oplus D)_\bullet, (C\oplus D)^\bullet)$ there are no logical qubits -- there are still operators which show up as errors and some which don't, but all of those which don't are products of $Z$ or $X$ stabiliser generators. By Corollary~\ref{cor:zero_inputs} and Corollary~\ref{cor:restriction_maps} the logical map in $\FHilb$ is then just $\ket{+}$. This trivially preserves both $\overline{Z}$ and $\overline{X}$ operators, although its opposite code map $\CF_{\overline{X}}$ does not preserve $\overline{Z}$ operators.
\end{example}

This example was very simple, but the idea extends in quite a general way. To give an idea of how general this notion of CSS code surgery is, consider the balanced product codes from \cite{BE1, PK1, PK2}. 
The balanced product of codes is by definition a coequaliser in $\Chains$, and so we can convert it into a pushout using routine category theory. The coequaliser is
\[\begin{tikzcd}(C\tens A\tens D)_\bullet \arrow[r, "g_\bullet"', shift right=1.5ex]\arrow[r, "f_\bullet", shift left=1.5ex]& (C\tens D)_\bullet \arrow[r, "\coeq_\bullet"] &(C\tens_A D)_\bullet\end{tikzcd}\]
where $g_\bullet$ and $f_\bullet$ represent left and right actions of $A_\bullet$ respectively; in all the cases from \cite{BE1, PK1, PK2} these actions are basis-preserving. We have not explicitly defined the tensor product of chain complexes in the main body for brevity, but see Appendix~\ref{app:tensor}. Then to this coequaliser we can associate a pushout,
\[\begin{tikzcd}
((C\tens A\tens D)\oplus (C\tens D))_\bullet \arrow[r, "(g_\bullet\ |\ \id_\bullet)"]\arrow[d, "(f_\bullet\ |\ \id_\bullet)"'] & (C\tens D)_\bullet \arrow[d,"l_\bullet"]\\
(C\tens D)_\bullet \arrow[r,"k_\bullet"'] & (C\tens_A D)_\bullet\arrow[ul, phantom, "\usebox\pushout", very near start]
\end{tikzcd}\]
where one can check that the universal property is the same in both cases. Thus we can think of a balanced product as a merge of tensor product codes, with the apex being two adjacent tensor product codes. As the maps in the span are evidently not monic, the merge is of a distinctly different sort from Example~\ref{ex:small_pushout_codes}, and also the $\overline{Z}$- and $\overline{X}$-merges we will describe in Section~\ref{sec:operator_surgery}.

It would be convenient if we could guarantee some properties of pushouts in general; for example, if the pushout of LDPC codes was also LDPC, or if the homologies were always preserved. Unfortunately, the definition is general enough that neither of these are true. We discuss this in slightly greater detail in Appendix~\ref{app:pushouts_props}, but the gist is that we need to stipulate some additional conditions to guarantee bounds on these quantities.

\subsection{Surgery along a logical operator}\label{sec:operator_surgery}

The procedure of merging here is closely related to that of `welding' in \cite{Mich}. Our focus is not just on the resultant codes, but the maps on physical and logical data. On codes generated from square lattices, the merges here will correspond to a pushout along a `string' through the lattice.

\begin{definition}\label{def:op_subcomplex}
Let $\begin{tikzcd}C_\bullet = C_{2}\arrow[r, "\del_1"]& C_1\arrow[r, "\del_{1}"]& C_{0}\end{tikzcd}$ be a length 2 chain complex. Let $v\in C_1$ be a vector such that $v \in \ker(\del^{C_\bullet}_{1})\backslash \im(\del^{C_\bullet}_{2})$. We now construct the {\rm logical operator subcomplex} $V_\bullet$. This has:
\[\tilde{V}_1 = {\rm supp\ } v; \quad \del^{V_\bullet}_{1} = \del^{C_\bullet}_{1}\restriction_{{\rm supp\ } v}; \quad \tilde{V_{0}} = \bigcup_{u \in \im(\del^{V_\bullet}_{1})} {\rm supp\ } u \]
where ${\rm supp\ } v$ is the set of basis vectors in the support of $v$, and $\del_i\restriction_S$ is the restriction of a differential to a subset $S$ of its domain. All other components and differentials of $V_\bullet$ are zero.
\end{definition}
There is a monic $f_\bullet : V_\bullet \hookrightarrow C_\bullet$ given by the inclusion maps of $V_1 \subseteq C_1$ etc.
\begin{definition}\label{def:monic_span}
Let $V_\bullet$ be a logical operator subcomplex of two chain complexes \[\begin{tikzcd}C_\bullet = C_{2}\arrow[r, "\del_2"]& C_1\arrow[r, "\del_{1}"]& C_{0}\end{tikzcd}\] and \[\begin{tikzcd}D_\bullet = D_{2}\arrow[r, "\del_2"]& D_1\arrow[r, "\del_{1}"]& D_{0}\end{tikzcd}\] simultaneously, so there is a basis-preserving monic span
\[\begin{tikzcd}
V_\bullet \arrow[r, hookrightarrow, "g_\bullet"]\arrow[d, hookrightarrow, "f_\bullet"'] & D_\bullet\\
C_\bullet & &
\end{tikzcd}\]
This monic span has the pushout
\[\begin{tikzcd}
V_\bullet \arrow[r, hookrightarrow, "g_\bullet"]\arrow[d, hookrightarrow, "f_\bullet"'] & D_\bullet \arrow[d,"q_\bullet"]\\
C_\bullet \arrow[r,"p_\bullet"'] & Q_\bullet\arrow[ul, phantom, "\usebox\pushout", very near start]
\end{tikzcd}\]
with components
\[Q_2 = C_2 \oplus D_2;\quad Q_1 = C_1\oplus D_1/(\supp\ v \sim \supp\ w);\quad Q_{0} = C_{0} \oplus D_{0}/ (\tilde{f_0}\sim \tilde{g_0}),\]
where $w$ is the logical operator associated to $\im(g_1)\in D_1$.
\end{definition}

The construction here is inspired by \cite{Coh}.

\begin{definition}
Let $(C_\bullet,  C^\bullet)$ and $(D_\bullet,  D^\bullet)$ be CSS codes. Let $(V_\bullet, V^\bullet)$ be a CSS code such that $V_\bullet$ is a logical operator subcomplex of $C_\bullet$ and $D_\bullet$; this means that $(V_\bullet, V^\bullet)$ can be seen as merely a classical code, as $V_2=0$. Then the $Z$-type merged CSS code $(Q_\bullet, Q^\bullet)$ is called the $\overline{Z}$-{\rm merged code} of $(C_\bullet,  C^\bullet)$ and $(D_\bullet,  D^\bullet)$ along $(V_\bullet, V^\bullet)$.
\end{definition}

\begin{definition}{\rm (Separation)}\label{def:separation}
Let $V_\bullet$ be a logical operator subcomplex such that for the inclusion maps $f_\bullet$ and $g_\bullet$, $\im(f_1)$ and $\im(g_1)$ contain only one operator in $Z_1(C_\bullet)$, $Z_1(D_\bullet)$ respectively, with those operators being $v$, $w$. Then we say that these operators are separated, as they contain no other logicals in their support, and the pushout satisfies the separation property.
\end{definition}

The intuition here, following \cite{Coh}, is that it is convenient when the logical operators we glue along do not themselves contain any nontrivial logical operators belonging to a different logical qubit; if they do, the gluing procedure may yield a more complicated output code, as we could be merging along multiple logical operators simultaneously. In Appendix~\ref{app:octagon} we demonstrate that it is possible for this condition to not be satisfied, using a patch of octagonal surface code. Additionally, we do not want the gluing procedure to send any logical $\overline{Z}$ operators to stabilisers.

We would like to study not only the resultant code given some $\overline{Z}$-merge, but also the map on the logical space. We can freely switch between pushouts and coequalisers. Recall the $Z$-merge code map $\CF_{\overline{Z}}$ from Equation~\ref{eq:Z_merge_code_map}. We call this a $\overline{Z}$-merge code map when the merge is along a $\overline{Z}$-operator as above, and from now on we assume that all merges are separated.

\begin{lemma}\label{lem:separated_result}
Let $(Q_\bullet, Q^\bullet)$ be a separated $\overline{Z}$-merged code with parameters $\llbracket n_Q, k_Q, d_Q \rrbracket$, and let $\llbracket n_C, k_C, d_C\rrbracket$, $\llbracket n_D, k_D, d_D\rrbracket$ be the parameters of $(C_\bullet,  C^\bullet)$ and $(D_\bullet,  D^\bullet)$ respectively. Let $n_V = \dim V_0$. Then
\[n_Q = n_C + n_D - n_V;\quad k_Q \geq k_C + k_D -1\]
Further, let $\{[u]_i\}$ and $\{[v]_j\}$ be the bases for $H_1(C_\bullet)$ and $H_1(D_\bullet)$ respectively, and say w.l.o.g. that $u \in [u]_1$ and $v \in [v]_1$ are the vectors quotiented by the pushout. Then $H_1(Q_\bullet)$ has a basis $\{[w]_l\}$ for $l \leq k_Q$, where $[w]_1 = [u]_1 = [v]_1$, $[w]_l = [u]_l$ when $1 < l \leq k_C$ and $[w]_{l} = [v]_{l-k_C+1}$ for $k_C < l \leq k_C + k_D -1$.
\end{lemma}
\proof
$n_Q$ is immediate by the definition.
Given $u \in [u]_1$ and $v \in [v]_1$, any other representatives $y \in [u]_1$, $x \in [v]_1$ belong to the same equivalence class in $H_1(Q_\bullet)$, as $y \sim u \sim v \sim x$.

All other equivalence classes remain distinct, as they would be in $((C \oplus D)_\bullet, (C \oplus D)^\bullet)$.

However, it is possible to introduce new equivalence classes, without a preimage in $H_1((C \oplus D)_\bullet)$. Despite $\coeq_\bullet$ being surjective, the lift $H_1(\coeq_\bullet)$ is not always surjective, as the restriction of $\coeq_1$ to $\ker(\del^{(C \oplus D)_\bullet}_1)$ is not always surjective.
\endproof

This last case is subtle, and rarely occurs with small codes or topological codes. We present an explicit example in Appendix~\ref{app:merge_bigger}. Should it be useful, we can swap to subsystem codes after the merge, and not store any data in the new logical qubits, relegating them to gauge qubits, as done in \cite{Coh}.

\begin{lemma}\label{lem:Z_map_matrix}
Let the $\overline{Z}$-merge code map
\[\begin{tikzcd}
\ \arrow[d, "\CF_{\overline{Z}}"', Rightarrow] & ((C\oplus D)_\bullet, (C\oplus D)^\bullet)\arrow[d, "\coeq_\bullet"', shift right=3ex]\\
\ & (Q_\bullet, Q^\bullet)\arrow[u, "\coeq^\bullet"', shift right=3ex]
\end{tikzcd}\]
of a separated $\overline{Z}$-merged code have its interpretation $M$ as a $\C$-linear map. Then $M$ acts as
\[\begin{tikzpicture}
	\begin{pgfonlayer}{nodelayer}
		\node [style=Z dot] (0) at (0, 0) {};
		\node [style=none] (1) at (1, 0) {};
		\node [style=none] (2) at (-1, 0.5) {};
		\node [style=Z dot] (3) at (0, 0) {};
		\node [style=none] (4) at (-1, -0.5) {};
	\end{pgfonlayer}
	\begin{pgfonlayer}{edgelayer}
		\draw [bend left] (2.center) to (0);
		\draw (0) to (1.center);
		\draw [bend right] (4.center) to (3);
	\end{pgfonlayer}
\end{tikzpicture} = \begin{pmatrix}
1 & 0 & 0 & 0\\
0 & 0 & 0 & 1
\end{pmatrix}\]
on each pair of qubits in $((C\oplus D)_\bullet, (C\oplus D)^\bullet)$ which are equivalent in $(Q_\bullet, Q^\bullet)$ and $M$ acts as identity on all other qubits.
\end{lemma}
\proof
$M$ must have the following maps on Paulis on each pair of qubits being merged:
\[Z\tens I \mapsto Z;\quad I\tens Z\mapsto Z;\quad X\tens X \mapsto X\]
which uniquely defines the matrix above. In other words we have $\ket{00}\mapsto \ket{0}$, $\ket{11}\mapsto \ket{1}$, $\ket{01} \mapsto \ket{0}, \ket{10}\mapsto \ket{0}$ etc, which has the convenient presentation as the ZX diagram on the left above.
\endproof

\begin{lemma}\label{lem:map_logical_Zs}
Let $(Q_\bullet, Q^\bullet)$ be a separated $\overline{Z}$-merged code of $(C_\bullet,  C^\bullet)$ and $(D_\bullet,  D^\bullet)$ along $(V_\bullet, V^\bullet)$. Call $f=H_1(\coeq_\bullet)$. Then
\[f([u]_i + [v]_j) = [w]_l\]
where $[w]_l$ was defined in Lemma~\ref{lem:separated_result}.
\end{lemma}
This is obvious by considering the surjection in question and using Lemma~\ref{lem:separated_result}. It essentially says that on the pair of logical operators in $((C\oplus D)_\bullet, (C\oplus D)^\bullet)$ which are being quotiented together, $\CF_{\overline{Z}}$ acts as:
\[\overline{Z}\tens \overline{I} \mapsto \overline{Z};\quad \overline{I}\tens \overline{Z}\mapsto \overline{Z};\quad \overline{X}\tens \overline{X} \mapsto \overline{X}\]
where the map on $X$s is inferred from the dual. In the case where new logical qubits are introduced, as described in Lemma~\ref{lem:separated_result}, it can be easily checked that these are initialised in the logical $\ket{+}$ state, as they are not in the image of the $H_1(\coeq_\bullet)$.

\begin{lemma}\label{lem:merge_distance} Let the merged code have no new logical qubits, i.e. $k_Q = k_C + k_D -1$. Then,
\[d^X_Q \geq \min(d^X_C, d^X_D)\]
\end{lemma}
\proof
By considering the code map $\CF_{\overline{Z}}$, we see that any $\overline{X}$ logical operator $u$ in $(Q_\bullet, Q^\bullet)$ has a preimage $w$ which is also an $\overline{X}$ logical operator in $((C\oplus D)_\bullet, (C\oplus D)^\bullet)$, s.t. $|w| \leq |u|$. This is because $\coeq^\bullet$ can be restricted to $f^\intercal = H^1(\coeq^\bullet)$, and any logical in $((C\oplus D)_\bullet, (C\oplus D)^\bullet)$ has sublogicals in $(C_\bullet, C^\bullet)$ and $(D_\bullet, D^\bullet)$.
\endproof
The proof fails when the merged code has new logical qubits, as there can be a $\overline{X}$ logical operator $u$ whose image under $\coeq^\bullet$ is in $[0]$, which gives no bound on the weight of $u$.

\begin{remark}\label{rem:distance_fail}
Note that we do not in general have a lower bound on $d^Z_Q$ in terms of $d^Z_C$ and $d^Z_D$. We can see this from the discussion in Section~\ref{sec:code_maps}. Given the code map $\CF_{\overline{Z}}$, the chain map $f_1: (C\oplus D)_1\rightarrow Q_1$ restricts to $H_1(f)$, but this does not preclude there being other vectors in $(C\oplus D)_1\backslash \ker \del^{(C\oplus D)_\bullet}_{1}$ which are mapped into one of the equivalence classes in $H_1(Q_\bullet)$. In computational terms, while we cannot have detectable $X$ operators in the initial codes which are mapped to logicals by the code map $\CF_{\overline{Z}}$, this is unfortunately possible with detectable $Z$ operators. We illustrate this with an example in Appendix~\ref{app:not_distance_preserving}.
\end{remark}

We now show that, if we consider two codes to be merged as instances of LDPC families, their combined $\overline{Z}$-merged code code is also LDPC. Recall Definition~\ref{def:weights}.
\begin{lemma}{\rm (LDPC)}\label{lem:LDPC_conservation}
Say our input codes $(C_\bullet,  C^\bullet)$, $(D_\bullet,  D^\bullet)$ have maximal weights of generators labelled $w^Z_C$, $w^X_C$ and $w^Z_D$, $w^X_D$ respectively. Let $(Q_\bullet, Q^\bullet)$ be a separated $\overline{Z}$-merged code of $(C_\bullet,  C^\bullet)$ and $(D_\bullet,  D^\bullet)$ along $(V_\bullet, V^\bullet)$. Then
\[w^Z_Q = \max(w^Z_C, w^Z_D);\quad w^X_Q < w^X_C + w^X_D.\]
Similarly, letting the input codes have maximal number of shared generators on a single qubit $q^Z_C$, $q^X_C$ and $q^Z_D$, $q^X_D$ we have
\[q^Z_Q \leq q^Z_C + q^Z_D;\quad q^X_Q = \max(q^X_C,q^X_D)\]
\end{lemma}
\proof
None of the $Z$-type generators are quotiented by a $\overline{Z}$-merge map, so $w^Z_Q = w^Z_{(C\oplus D)} = \max(w^Z_C, w^Z_D)$. For the $X$-type generators, in the worst case the two generators which are made to be equivalent by the merge are the highest weight ones. For these generators to appear in $V_{0}$ they must have at least two qubits in each of their support which is in $V_1$, and thus these qubits are merged together, so $w^X_Q < w^X_C + w^X_D$.

Next, using again the fact that none of the $Z$-type generators are quotiented, a single qubit could in the worst case be the result of merging two qubits in $(C_\bullet,  C^\bullet)$ and $(D_\bullet,  D^\bullet)$ which each have the maximal number of shared $Z$-type generators, so $q^Z_Q \leq q^Z_C + q^Z_D$. For the $X$ case, if a qubit is in $V_1$ then all $X$-type generators it is in the support of must appear in $V_{0}$. Therefore, when any two qubits are merged all of their $X$-type generators are also merged. Thus $q^X_Q = q^X_{(C\oplus D)} = \max(q^X_C,q^X_D)$.
\endproof

Note that as $w^Z$, $w^X$ and $q^Z$, $q^X$ are at worst additive in those of the input codes, the $\overline{Z}$-merge of two  LDPC codes is still LDPC, assuming the pushout is still well-defined using matching $\overline{Z}$ operators for each member of the code families. Next, we dualise everything, and talk about $\overline{X}$-merges.

\begin{definition}
Let $(C_\bullet,  C^\bullet)$ and $(D_\bullet,  D^\bullet)$ be CSS codes. Let $(V_\bullet, V^\bullet)$ be a CSS code such that $V^\bullet$ is a logical operator subcomplex of $C^\bullet$ and $D^\bullet$, and $Q^\bullet$ is the merged complex along $V^\bullet$. Then the CSS code $(Q_\bullet, Q^\bullet)$ is called the $\overline{X}$-{\rm merged code} of $(C_\bullet,  C^\bullet)$ and $(D_\bullet,  D^\bullet)$ along $(V_\bullet, V^\bullet)$.
\end{definition}
In this case we glue along an $\overline{X}$ logical operator instead. The notions of separation, Lemma~\ref{lem:separated_result} and Lemma~\ref{lem:LDPC_conservation} carry over by transposing appropriately.

An $\overline{X}$-merge map $\CE_{\overline{X}}$ can be defined similarly, and a similar result as Lemma~\ref{lem:Z_map_matrix} applies to separated $\overline{X}$-merged codes.
\begin{lemma}\label{lem:X_map_matrix}
Let the $\overline{X}$-merge code map of a separated $\overline{X}$-merged code have its interpretation $L$ as a $\C$-linear map. Then $L$ acts as
\[\begin{tikzpicture}
	\begin{pgfonlayer}{nodelayer}
		\node [style=Z dot] (0) at (0, 0) {};
		\node [style=none] (1) at (1, 0) {};
		\node [style=none] (2) at (-1, 0.5) {};
		\node [style=Z dot] (3) at (0, 0) {};
		\node [style=none] (4) at (-1, -0.5) {};
		\node [style=X dot] (5) at (0, 0) {};
	\end{pgfonlayer}
	\begin{pgfonlayer}{edgelayer}
		\draw [bend left] (2.center) to (0);
		\draw (0) to (1.center);
		\draw [bend right] (4.center) to (3);
	\end{pgfonlayer}
\end{tikzpicture} = \frac{1}{\sqrt{2}}\begin{pmatrix}
1 & 0 & 0 & 1\\
0 & 1 & 1 & 0
\end{pmatrix}\]
on each pair of qubits in $((C\oplus D)_\bullet, (C\oplus D)^\bullet)$ which are equivalent in $(Q_\bullet, Q^\bullet)$, i.e. $\ket{++}\mapsto \ket{+}$, $\ket{--}\mapsto \ket{-}$, and $L$ acts as identity on all other qubits.
\end{lemma}
\proof
This time, $L$ must have the maps
\[X\tens I\mapsto X; \quad I\tens X\mapsto X;\quad Z\tens Z\mapsto Z\]
\endproof

Similarly, the maps on logical operators are
\[\overline{X}\tens \overline{I} \mapsto \overline{X};\quad \overline{I}\tens \overline{X}\mapsto \overline{X};\quad \overline{Z}\tens \overline{Z} \mapsto \overline{Z}\]
and, when new logical qubits are generated, they are initialised in the $\ket{0}$ state.

Having discussed $\overline{Z}$- and $\overline{X}$-merged codes, we briefly mention splits. These are just the opposite code maps to $\CF_{\overline{Z}}$ and $\CE_{\overline{X}}$. In both cases, all the mappings are determined entirely by Lemma~\ref{lem:separated_result} by taking transposes or adjoints when appropriate.

\begin{remark}
In practice, when the CSS codes in question hold multiple logical qubits it may be preferable to merge/split along multiple disjoint $\overline{Z}$ or $\overline{X}$ operators at the same time. Such a protocol is entirely viable within our framework, and requires only minor tweaks to the above results. The same is true should one wish to merge/split along operators within the same code.
\end{remark}

We now look at a short series of examples.

\subsection{Examples of surgery}\label{sec:surgery_examples}

\subsubsection{Lattice surgery}\label{sec:lattice_surgery}
Lattice surgery is the prototypical instance of CSS code surgery. It starts with patches of surface code and then employs separated splits and merges to perform non-unitary logical operations \cite{HFDM}. The presentation we give of lattice surgery is idiosyncratic, in the sense that we perform the merges on physical edges/qubits, whereas the standard method is to introduce additional edges between patches to join them together. We remedy this in Section~\ref{sec:practical}.

Consider the pushout of cell complexes below:
\[\begin{tikzpicture}
	\begin{pgfonlayer}{nodelayer}
		\node [style=none] (0) at (2, 0) {};
		\node [style=none] (1) at (2, -6) {};
		\node [style=none] (2) at (4, -6) {};
		\node [style=none] (3) at (4, 0) {};
		\node [style=boundary vertex] (4) at (2, -2) {};
		\node [style=boundary vertex] (5) at (4, -2) {};
		\node [style=boundary vertex] (6) at (2, -4) {};
		\node [style=boundary vertex] (7) at (4, -4) {};
		\node [style=boundary vertex] (8) at (2, -2) {};
		\node [style=boundary vertex] (9) at (2, -4) {};
		\node [style=boundary vertex] (10) at (4, -4) {};
		\node [style=boundary vertex] (11) at (4, -2) {};
		\node [style=none] (12) at (3, 4) {};
		\node [style=none] (13) at (3, 10) {};
		\node [style=boundary vertex] (14) at (3, 8) {};
		\node [style=boundary vertex] (15) at (3, 6) {};
		\node [style=boundary vertex] (16) at (3, 6) {};
		\node [style=boundary vertex] (17) at (3, 8) {};
		\node [style=none] (42) at (10, 10) {};
		\node [style=none] (43) at (10, 4) {};
		\node [style=none] (44) at (12, 4) {};
		\node [style=none] (45) at (12, 10) {};
		\node [style=boundary vertex] (46) at (10, 8) {};
		\node [style=boundary vertex] (47) at (12, 8) {};
		\node [style=boundary vertex] (48) at (10, 6) {};
		\node [style=boundary vertex] (49) at (12, 6) {};
		\node [style=boundary vertex] (50) at (10, 8) {};
		\node [style=boundary vertex] (51) at (10, 6) {};
		\node [style=boundary vertex] (52) at (12, 6) {};
		\node [style=boundary vertex] (53) at (12, 8) {};
		\node [style=none] (54) at (9, 0) {};
		\node [style=none] (55) at (9, -6) {};
		\node [style=none] (56) at (11, -6) {};
		\node [style=none] (57) at (11, 0) {};
		\node [style=boundary vertex] (58) at (9, -2) {};
		\node [style=boundary vertex] (59) at (11, -2) {};
		\node [style=boundary vertex] (60) at (13, -2) {};
		\node [style=boundary vertex] (61) at (9, -4) {};
		\node [style=boundary vertex] (62) at (11, -4) {};
		\node [style=boundary vertex] (63) at (13, -4) {};
		\node [style=boundary vertex] (64) at (9, -2) {};
		\node [style=boundary vertex] (65) at (9, -4) {};
		\node [style=boundary vertex] (66) at (11, -4) {};
		\node [style=boundary vertex] (67) at (11, -2) {};
		\node [style=none] (68) at (13, 0) {};
		\node [style=none] (69) at (13, -6) {};
		\node [style=boundary vertex] (70) at (13, -2) {};
		\node [style=boundary vertex] (71) at (13, -4) {};
		\node [style=none] (72) at (3, 3) {};
		\node [style=none] (73) at (3, 1) {};
		\node [style=none] (74) at (6, 7) {};
		\node [style=none] (75) at (8, 7) {};
		\node [style=none] (76) at (11, 3) {};
		\node [style=none] (77) at (11, 1) {};
		\node [style=none] (80) at (6, -3) {};
		\node [style=none] (81) at (8, -3) {};
		\node [style={white_dot}] (86) at (11, 7) {};
		\node [style={white_dot}] (87) at (11, 5) {};
		\node [style={white_dot}] (88) at (11, 9) {};
		\node [style={white_dot}] (89) at (3, -1) {};
		\node [style={white_dot}] (90) at (3, -3) {};
		\node [style={white_dot}] (91) at (3, -5) {};
		\node [style={white_dot}] (92) at (10, -5) {};
		\node [style={white_dot}] (93) at (10, -3) {};
		\node [style={white_dot}] (94) at (10, -1) {};
		\node [style={white_dot}] (95) at (12, -1) {};
		\node [style={white_dot}] (96) at (12, -3) {};
		\node [style={white_dot}] (97) at (12, -5) {};
	\end{pgfonlayer}
	\begin{pgfonlayer}{edgelayer}
		\draw (0.center) to (8);
		\draw (8) to (9);
		\draw (9) to (1.center);
		\draw (6) to (7);
		\draw (4) to (5);
		\draw (45.center) to (53);
		\draw (53) to (52);
		\draw (52) to (44.center);
		\draw (48) to (49);
		\draw (46) to (47);
		\draw (54.center) to (64);
		\draw (64) to (65);
		\draw (65) to (55.center);
		\draw (61) to (62);
		\draw (62) to (63);
		\draw (58) to (59);
		\draw (59) to (60);
		\draw (68.center) to (70);
		\draw (70) to (71);
		\draw (71) to (69.center);
		\draw [style=arrow] (72.center) to (73.center);
		\draw [style=arrow] (76.center) to (77.center);
		\draw [style=arrow] (74.center) to (75.center);
		\draw [style=arrow] (80.center) to (81.center);
		\draw [style={blue_edge}] (13.center) to (17);
		\draw [style={blue_edge}] (17) to (16);
		\draw [style={blue_edge}] (16) to (12.center);
		\draw [style={blue_edge}] (43.center) to (51);
		\draw [style={blue_edge}] (51) to (50);
		\draw [style={blue_edge}] (50) to (42.center);
		\draw [style={blue_edge}] (3.center) to (11);
		\draw [style={blue_edge}] (11) to (10);
		\draw [style={blue_edge}] (10) to (2.center);
		\draw [style={blue_edge}] (57.center) to (67);
		\draw [style={blue_edge}] (67) to (66);
		\draw [style={blue_edge}] (66) to (56.center);
	\end{pgfonlayer}
\end{tikzpicture}\]
As before, we informally consider this to be `gluing along' the graph in the top left, but for completeness it is formalised in Appendix~\ref{app:cells}. By considering the pushout to be in $\Chains$, we have:
\[\begin{tikzcd}
V_\bullet \arrow[r, "g_\bullet",hookrightarrow]\arrow[d, "f_\bullet"',hookrightarrow] & D_\bullet \arrow[d,"q_\bullet"]\\
C_\bullet \arrow[r,"p_\bullet"'] & Q_\bullet\arrow[ul, phantom, "\usebox\pushout", very near start]
\end{tikzcd}\]
Letting $\coeq_\bullet : (C \oplus D)_\bullet \rightarrow Q_\bullet$ be the relevant coequaliser map, we see that $\CF_{\overline{Z}} = (\coeq_\bullet, \coeq^\bullet)$ constitutes a separated $\overline{Z}$-merge map. In particular, observe that $\CF_{\overline{Z}}$ sends the logical operators: 
\begin{align*}
&\overline{Z} \tens \overline{I} \mapsto \overline{Z}\\
&\overline{I} \tens \overline{Z} \mapsto \overline{Z}\\
&\overline{X} \tens \overline{X} \mapsto \overline{X}
\end{align*}
as predicted by Lemma~\ref{lem:map_logical_Zs}.

The first two give $H_1(\coeq_\bullet) = \begin{pmatrix}1 & 1\end{pmatrix}$ and the last $H^1(\coeq^\bullet) = \begin{pmatrix}1\\1 \end{pmatrix}$. $\CF_{\overline{Z}}$ is evidently $\overline{Z}$-preserving but not $\overline{X}$-preserving, as $\overline{X} \tens \overline{I}$ is taken to an operation which is detected by the $Z$ stabilisers. Observe that we end up with a greater cosystolic distance of $(Q_\bullet, Q^\bullet)$ than we started with in $((C \oplus D)_\bullet,(C \oplus D)^\bullet)$.

If we instead consider the pair $(\coeq_\bullet, \coeq^\bullet)$ as an $\overline{X}$-preserving code map $\CF_{\overline{X}}$, then it is a separated $\overline{X}$-split map. In terms of cell complexes we would have \footnote{Pedantically, this is a morphism in the opposite category of cell complexes $\OACC^{\rm op}$.}
\[\begin{tikzpicture}
	\begin{pgfonlayer}{nodelayer}
		\node [style=none] (0) at (-6.5, 3) {};
		\node [style=none] (1) at (-6.5, -3) {};
		\node [style=none] (2) at (-4.5, -3) {};
		\node [style=none] (3) at (-4.5, 3) {};
		\node [style=boundary vertex] (4) at (-6.5, 1) {};
		\node [style=boundary vertex] (5) at (-4.5, 1) {};
		\node [style=boundary vertex] (6) at (-2.5, 1) {};
		\node [style=boundary vertex] (7) at (-6.5, -1) {};
		\node [style=boundary vertex] (8) at (-4.5, -1) {};
		\node [style=boundary vertex] (9) at (-2.5, -1) {};
		\node [style=boundary vertex] (10) at (-6.5, 1) {};
		\node [style=boundary vertex] (11) at (-6.5, -1) {};
		\node [style=boundary vertex] (12) at (-4.5, -1) {};
		\node [style=boundary vertex] (13) at (-4.5, 1) {};
		\node [style=none] (14) at (-2.5, 3) {};
		\node [style=none] (15) at (-2.5, -3) {};
		\node [style=boundary vertex] (16) at (-2.5, 1) {};
		\node [style=boundary vertex] (17) at (-2.5, -1) {};
		\node [style=none] (18) at (0, 0) {$\rightarrow$};
		\node [style=none] (19) at (2.5, 3) {};
		\node [style=none] (20) at (2.5, -3) {};
		\node [style=none] (21) at (4.5, -3) {};
		\node [style=none] (22) at (4.5, 3) {};
		\node [style=boundary vertex] (23) at (2.5, 1) {};
		\node [style=boundary vertex] (24) at (4.5, 1) {};
		\node [style=boundary vertex] (25) at (2.5, -1) {};
		\node [style=boundary vertex] (26) at (4.5, -1) {};
		\node [style=boundary vertex] (27) at (2.5, 1) {};
		\node [style=boundary vertex] (28) at (2.5, -1) {};
		\node [style=boundary vertex] (29) at (4.5, -1) {};
		\node [style=boundary vertex] (30) at (4.5, 1) {};
		\node [style=none] (31) at (8.5, 3) {};
		\node [style=none] (32) at (8.5, -3) {};
		\node [style=none] (33) at (6.5, -3) {};
		\node [style=none] (34) at (6.5, 3) {};
		\node [style=boundary vertex] (35) at (8.5, 1) {};
		\node [style=boundary vertex] (36) at (6.5, 1) {};
		\node [style=boundary vertex] (37) at (8.5, -1) {};
		\node [style=boundary vertex] (38) at (6.5, -1) {};
		\node [style=boundary vertex] (39) at (8.5, 1) {};
		\node [style=boundary vertex] (40) at (8.5, -1) {};
		\node [style=boundary vertex] (41) at (6.5, -1) {};
		\node [style=boundary vertex] (42) at (6.5, 1) {};
		\node [style={white_dot}] (43) at (-5.5, 2) {};
		\node [style={white_dot}] (44) at (-3.5, 2) {};
		\node [style={white_dot}] (45) at (-3.5, 0) {};
		\node [style={white_dot}] (46) at (-5.5, 0) {};
		\node [style={white_dot}] (47) at (-5.5, -2) {};
		\node [style={white_dot}] (48) at (-3.5, -2) {};
		\node [style={white_dot}] (49) at (3.5, 0) {};
		\node [style={white_dot}] (50) at (3.5, 2) {};
		\node [style={white_dot}] (51) at (3.5, -2) {};
		\node [style={white_dot}] (52) at (7.5, -2) {};
		\node [style={white_dot}] (53) at (7.5, 0) {};
		\node [style={white_dot}] (54) at (7.5, 2) {};
	\end{pgfonlayer}
	\begin{pgfonlayer}{edgelayer}
		\draw (0.center) to (10);
		\draw (10) to (11);
		\draw (11) to (1.center);
		\draw (7) to (8);
		\draw (8) to (9);
		\draw (4) to (5);
		\draw (5) to (6);
		\draw (14.center) to (16);
		\draw (16) to (17);
		\draw (17) to (15.center);
		\draw [style={blue_edge}] (3.center) to (13);
		\draw [style={blue_edge}] (13) to (12);
		\draw [style={blue_edge}] (12) to (2.center);
		\draw (19.center) to (27);
		\draw (27) to (28);
		\draw (28) to (20.center);
		\draw (25) to (26);
		\draw (23) to (24);
		\draw [style={blue_edge}] (22.center) to (30);
		\draw [style={blue_edge}] (30) to (29);
		\draw [style={blue_edge}] (29) to (21.center);
		\draw (31.center) to (39);
		\draw (39) to (40);
		\draw (40) to (32.center);
		\draw (37) to (38);
		\draw (35) to (36);
		\draw [style={blue_edge}] (34.center) to (42);
		\draw [style={blue_edge}] (42) to (41);
		\draw [style={blue_edge}] (41) to (33.center);
	\end{pgfonlayer}
\end{tikzpicture}\]

We similarly have a separated $\overline{X}$-merge map and separated $\overline{Z}$-split map with the obvious forms by dualising appropriately.

\begin{remark}
While it is convenient to choose logical operators along patch boundaries to glue along, so that the complexes can all be embedded on the 2D plane, this is not necessary. One could intersect two patches along any matching operator.
\end{remark}

Recall the toric code from Example~\ref{ex:homological}. We can merge two copies of the code along a logical $\overline{Z}$ operator, which corresponds to an essential cycle of each torus. The resultant code will then look like two tori intersecting, depending somewhat on the choices of essential cycle:
\[\begin{tikzpicture}
	\begin{pgfonlayer}{nodelayer}
		\node [style=none] (0) at (-0.75, 0.5) {};
		\node [style=none] (1) at (0.25, 0.5) {};
		\node [style=none] (2) at (-2.25, 0.5) {};
		\node [style=none] (3) at (1.75, 0.5) {};
		\node [style=none] (4) at (1.75, 0.5) {};
		\node [style=none] (5) at (2.75, 0.5) {};
		\node [style=none] (6) at (0.25, 0.5) {};
		\node [style=none] (7) at (4.25, 0.5) {};
		\node [style=none] (8) at (-2.5, 8) {};
		\node [style=none] (9) at (-1.5, 8) {};
		\node [style=none] (10) at (-4, 8) {};
		\node [style=none] (11) at (0, 8) {};
		\node [style=none] (12) at (-1.5, 8) {};
		\node [style=none] (13) at (0, 8) {};
		\node [style=none] (14) at (0, 8) {};
		\node [style=none] (15) at (-1.5, 8) {};
		\node [style=none] (16) at (4.5, 8) {};
		\node [style=none] (17) at (3.5, 8) {};
		\node [style=none] (18) at (6, 8) {};
		\node [style=none] (19) at (2, 8) {};
		\node [style=none] (20) at (3.5, 8) {};
		\node [style=none] (21) at (2, 8) {};
		\node [style=none] (22) at (2, 8) {};
		\node [style=none] (23) at (3.5, 8) {};
		\node [style=none] (24) at (1, 5.5) {};
		\node [style=none] (25) at (1, 3.5) {};
	\end{pgfonlayer}
	\begin{pgfonlayer}{edgelayer}
		\draw [bend left=90, looseness=1.75] (0.center) to (1.center);
		\draw [bend right=90, looseness=1.75] (0.center) to (1.center);
		\draw [bend left=90, looseness=1.75] (2.center) to (3.center);
		\draw [bend right=90, looseness=1.75] (2.center) to (3.center);
		\draw [bend left=90, looseness=1.75] (4.center) to (5.center);
		\draw [bend right=90, looseness=1.75] (4.center) to (5.center);
		\draw [bend left=90, looseness=1.75] (6.center) to (7.center);
		\draw [bend right=90, looseness=1.75] (6.center) to (7.center);
		\draw [style={blue_edge}, bend left=90, looseness=1.25] (6.center) to (4.center);
		\draw [style={blue_edge}, bend right=90, looseness=1.25] (6.center) to (4.center);
		\draw [bend left=90, looseness=1.75] (8.center) to (9.center);
		\draw [bend right=90, looseness=1.75] (8.center) to (9.center);
		\draw [bend left=90, looseness=1.75] (10.center) to (11.center);
		\draw [bend right=90, looseness=1.75] (10.center) to (11.center);
		\draw [style={blue_edge}, bend left=90, looseness=1.25] (15.center) to (14.center);
		\draw [style={blue_edge}, bend right=90, looseness=1.25] (15.center) to (14.center);
		\draw [bend left=90, looseness=1.75] (16.center) to (17.center);
		\draw [bend right=90, looseness=1.75] (16.center) to (17.center);
		\draw [bend left=90, looseness=1.75] (18.center) to (19.center);
		\draw [bend right=90, looseness=1.75] (18.center) to (19.center);
		\draw [style={blue_edge}, bend left=90, looseness=1.25] (23.center) to (22.center);
		\draw [style={blue_edge}, bend right=90, looseness=1.25] (23.center) to (22.center);
		\draw [style=arrow] (24.center) to (25.center);
	\end{pgfonlayer}
\end{tikzpicture}\]
The $\overline{Z}$-merge map on logical qubits will be the same as for patches.

\subsubsection{Shor code surgery}\label{sec:shor_surgery}

Of course, the pushout we take does not have to come from square lattices. Let $C_\bullet$ and $D_\bullet$ be two copies of Shor codes from Example~\ref{ex:shor}.\footnote{The Shor code can be constructed as a cellulation of the projective plane, so it is actually not wholly dissimilar from the lattice codes \cite{FM}.} We can perform merges between them. We give two examples. First, for a $\overline{Z}$-merge, we take the logical $\overline{Z}$ operator $\overline{Z} = \bigotimes_i^8 Z_i$ and apply Definition~\ref{def:op_subcomplex} to get the logical operator subcomplex:
\[\begin{tikzcd}V_\bullet = V_1\arrow[r, "P_X"] &V_{0}\end{tikzcd}\]
with $V_1 = \F_2^9$, $V_{0} = \F_2^2$, and all other components zero. This is just $C_\bullet$ from Example~\ref{ex:shor} truncated to be length 1, as this logical $\overline{Z}$ operator has support on all physical qubits. The monic chain map $f_\bullet$ given by inclusion into the Shor code is just
\[\begin{tikzcd}0\arrow[r, "0"]\arrow[d, "0"] & V_1\arrow[r, "P_X"]\arrow[d, "\id"] & V_{0}\arrow[d,"\id"]\\
C_2\arrow[r, "P_Z^\intercal"] & C_{1}\arrow[r, "P_X"] & C_{0}\end{tikzcd}
\]
and the same for $g_\bullet$. The pushout of
\[\begin{tikzcd}
V_\bullet \arrow[r, hookrightarrow, "g_\bullet"]\arrow[d, hookrightarrow, "f_\bullet"'] & D_\bullet\\
C_\bullet & &
\end{tikzcd}\]
will then be
\[\begin{tikzcd}
Q_\bullet = \F_2^{12} \arrow[r, "\del^{Q_\bullet}_2"] &\F_2^9 \arrow[r, "\del^{Q_\bullet}_{1}"] & \F_2^2
\end{tikzcd}\]
where $\del^{Q_\bullet}_{1} = P_X$ and $\del^{Q_\bullet}_{2} = \begin{pmatrix}P_Z^\intercal | P_Z^\intercal \end{pmatrix}$. We have ended up with virtually the same code as the Shor code, except that we have a duplicate for every $Z$-type generator, i.e. every measurement of $Z$ stabilisers is performed twice and the result noted separately. While this example is very simple, it highlights that the result of a merge can have somewhat subtle features, such as duplicating measurements, which the two input codes do not. This merge did not use separated logical operators.

For our second case, we use a different (but equivalent) logical operator, $\overline{Z} = Z_1\tens Z_4\tens Z_7$. We still glue two copies of the Shor code, but now we have $V_1 = \F_2^3$, $V_{0} = \F_2^2$ and $\del^{V_\bullet}_{1} = \begin{pmatrix} 1&1&0\\1&0&1\end{pmatrix}$. That is, our logical operator subcomplex is just the repetition code from Example~\ref{ex:rep_code}. We then have
\[\begin{tikzcd} 0\arrow[r, "0"]\arrow[d, "0"] & V_1\arrow[r, "\del^{V_\bullet}_{1}"]\arrow[d, "f_1"] & V_{0}\arrow[d,"\id"]\\
C_2\arrow[r, "P_Z^\intercal"] & C_{1}\arrow[r, "P_X"] & C_{0}\end{tikzcd}
\]
where 
\[f_1 = \begin{pmatrix} 1&0&0\\0&0&0\\0&0&0\\0&1&0\\0&0&0\\0&0&0\\0&0&1\\0&0&0\\0&0&0\end{pmatrix}\]
and the same for $g_1$, forming again a monic span of chain complexes. The resultant $\overline{Z}$-merged code is then
\[\begin{tikzcd}
Q_\bullet = \F_2^{12} \arrow[r, "\del^{Q_\bullet}_2"] &\F_2^{15} \arrow[r, "\del^{Q_\bullet}_{1}"] & \F_2^2
\end{tikzcd}\]
and the large matrices $\del^{Q_\bullet}_2$ and $\del^{Q_\bullet}_{1}$ are easily obtained by quotienting out rows and columns from $\del^{C_\bullet}_2\oplus \del^{D_\bullet}_2$ and $\del^{C_\bullet}_{1}\oplus \del^{D_\bullet}_{1}$.

\begin{remark}
The protocols for performing $\overline{X}$ and $\overline{Z}$ measurements by generalised lattice surgery in \cite{Coh} can be seen as using separated $\overline{Z}$- and $\overline{X}$-merged codes, with the caveat that they don't perform the merge maps; instead they initialise fresh qubits in the ancillary hypergraph code and measure all stabiliser generators. The present work has overlap with their protocols, but we do not subsume them; for example their $\overline{X}\tens \overline{Z}$ and $\overline{Y}$ measurement methods are outside of our formalism, as they lead to non-CSS codes.
\end{remark}

\section{Error-corrected logical operations}\label{sec:practical}

We now describe how our abstract formalism leads to a general set of error-corrected logical operations for CSS codes. We consider this to be a good application of the homological algebraic formalism, as we suspect these logical operations would be challenging to derive without the machinery of $\Chains$. \footnote{An alternative approach could be to use Tanner graphs.} So far in our description of code maps there are two main assumptions baked in: that one can perform linear maps between CSS codes (a) deterministically and (b) while maintaining error-correction, both of which are desired for performing quantum computation.

For assumption (a), we can only implement code maps which are interpreted as an isometry deterministically. If they are not, instead we must perform measurements on physical qubits. Recall from Proposition~\ref{prop:CNOT_circuit} that every code map has an interpretation constructed from CNOTs and some additional states and effects taken from $\{\ket{+}, \bra{0}\}$ for a $\overline{Z}$-preserving code map or $\{\bra{+}, \ket{0}\}$ for an $\overline{X}$-preserving code map. This means that in order to implement the code map non-deterministically, one need only apply CNOTs and measure some qubits in the $Z$-basis (for a $\overline{Z}$-preserving code map) or the $X$-basis ($\overline{X}$-preserving code map). Of course, should we acquire the undesired measurement result, we induce errors in our code map. There is no protocol for correcting these errors in all generality. For assumption (b), there is no protocol for performing arbitrary CNOT circuits on physical qubits in a code fault-tolerantly. However, when performing CSS code surgery which is a separated $\overline{Z}$- or $\overline{X}$-merge, we have a protocol which addresses both (a) and (b).

\subsection{Procedure summary}
Our procedure for performing an error-corrected $\overline{Z}\tens\overline{Z}$ measurement is as follows:
\begin{enumerate}
\item Find a matching $\overline{Z}$ logical operator which belongs to both initial codes, in the sense of Definition~\ref{def:op_subcomplex}.
\item Verify that this logical operator satisfies the separation property of Definition~\ref{def:separation} in both codes.
\item Verify that the merge is bounded below, in the sense of Definition~\ref{def:distance_preserving} below.
\item Perform the merge as described in Proposition~\ref{prop:fault_tolerant_sandwich}.
\end{enumerate}
We do not know how difficult it will be in general to perform the verification in steps (2) and (3) for codes (or families of codes) of interest.

\subsection{Full description of procedure}

We will first describe gauge fixing. Luckily this does not become an additional condition, as we will show it coincides precisely with separation. For reasons of brevity we do not describe the connection between lattice surgery and gauge fixing, but refer the interested reader to \cite{VLCABT}. Briefly, we will consider the whole system to be a subsystem code, and fix the gauges of the $\overline{Z}$ operators we are gluing along.

\begin{definition}\label{def:gauge_fixable}
Let $C_\bullet$ be a chain complex and $u$ be a representative of the equivalence class $[u]\in H_1(C_\bullet)$, which is a basis vector of $H_1(C_\bullet)$. Let $x$ be a vector in $C_1$ such that $|x| = 1$ and $x\cdot u = 1$. We say that $x$ is a qubit in the support of $u$. Recall from Lemma~\ref{lem:duality_basis} that $u$ has a unique paired basis vector $[v]\in H^1( C^\bullet)$ such that $[u]\cdot [v] = 1$. It is possible to {\rm safely correct} a qubit $x$ when there is a vector $v \in [v]$ such that $x\cdot v =1$ and $y \cdot v = 0$ for all other qubits $y$ in the support of $u$. We say that $u$ is {\rm gauge-fixable} when it is possible to safely correct all qubits in the support of $u$.
\end{definition}

\begin{example}\label{ex:shor_fixing}
Consider the Shor code from Example~\ref{ex:shor} and Section~\ref{sec:shor_surgery}. The $\overline{Z}$ operator 
\[v = \begin{pmatrix}1&1&1&1&1&1&1&1&1\end{pmatrix}^\intercal\]
has qubits in its support for which it is not possible to safely correct, as there are only 4 representatives of the nonzero equivalence class $[w]\in H^1( C^\bullet)$ but 9 qubits for which being able to safely correct is necessary. However, it is possible to safely correct all qubits in the support of the $\overline{Z}$ operator 
\[u = \begin{pmatrix}1&0&0&1&0&0&1&0&0 \end{pmatrix}^\intercal,\]
where $u\in [v]$, with the fixing operators:
\[\begin{pmatrix}1&1&1&0&0&0&0&0&0 \end{pmatrix}^\intercal;\quad \begin{pmatrix}0&0&0&1&1&1&0&0&0 \end{pmatrix}^\intercal;\quad \begin{pmatrix}0&0&0&0&0&0&1&1&1 \end{pmatrix}^\intercal\]
\end{example}

The same definition of gauge-fixability applies if we exchange $X$ and $Z$ appropriately.

\begin{lemma}\label{lem:gauge_fix_separation}
Every separated logical operator is also gauge-fixable, and vice versa.
\end{lemma}
\proof
See Appendix~\ref{app:separation_gauge_fixability}.
\endproof

Next we will require the tensor product of chain complexes, for which see Appendix~\ref{app:tensor}.

\begin{definition}\label{def:tensor_sandwich}
Let $V_\bullet = \begin{tikzcd}V_1\arrow[r] & V_{0}\end{tikzcd}$ and $P_\bullet = \begin{tikzcd}P_1 \arrow[r, "\begin{pmatrix}1\\1\end{pmatrix}"] & P_0\end{tikzcd}$ be length 1 chain complexes. Then we can make the tensor product chain complex $W_\bullet = (P \tens V)_\bullet$. Explicitly,
\[W_\bullet = \begin{tikzcd}W_2 \arrow[r] &W_1 \arrow[r] & W_{0}\end{tikzcd}\]
with 
\[W_2 = P_1\tens V_1 = V_1;\quad W_1 = (P_0\tens V_1)\oplus (P_{1}\tens V_{0}) = (\F_2^2\tens V_1)\oplus V_{0};\quad W_{0} = P_0\tens V_{0} = \F_2^2 \tens V_{0}\]
Also, $\del_{2}^{W_\bullet} = \begin{pmatrix}\id_{V_1}\\\id_{V_1}\\ \del_{1}^{V_\bullet}\end{pmatrix}$ and $\del_{1}^{W_\bullet} = \begin{pmatrix} \id_{\F_2^2}\tens \del_{1}^{V_\bullet} &\del^{P_\bullet}_1 \tens \id_{V_{0}} \end{pmatrix} = \begin{pmatrix}\del_{1}^{V_\bullet} & 0 & \id_{V_{0}}\\ 0 & \del_{1}^{V_\bullet} & \id_{V_{0}}\end{pmatrix}$.
\end{definition}

In the case where $V_\bullet$ is a string along a patch of surface code, say of the form:
\[\begin{tikzpicture}
	\begin{pgfonlayer}{nodelayer}
		\node [style=boundary vertex] (6) at (-4.25, 1) {};
		\node [style=boundary vertex] (7) at (-4.25, -1) {};
		\node [style=none] (8) at (-4.25, -3) {};
		\node [style=none] (9) at (-4.25, 3) {};
		\node [style=none] (10) at (-5.25, 0) {$V_\bullet \rightsquigarrow $};
	\end{pgfonlayer}
	\begin{pgfonlayer}{edgelayer}
		\draw (6) to (7);
		\draw (7) to (8.center);
		\draw (9.center) to (6);
	\end{pgfonlayer}
\end{tikzpicture}\]
then $W_\bullet$ will be of the form
\[\begin{tikzpicture}
	\begin{pgfonlayer}{nodelayer}
		\node [style=none] (0) at (-1, 3) {};
		\node [style=none] (1) at (1, 3) {};
		\node [style=none] (2) at (-1, 1) {};
		\node [style=none] (3) at (1, 1) {};
		\node [style=none] (4) at (-1, -1) {};
		\node [style=none] (5) at (1, -1) {};
		\node [style=none] (10) at (-2.5, 0) {$W_\bullet \rightsquigarrow$};
		\node [style=boundary vertex] (11) at (-1, 1) {};
		\node [style=boundary vertex] (12) at (1, 1) {};
		\node [style=boundary vertex] (13) at (1, -1) {};
		\node [style=boundary vertex] (14) at (-1, -1) {};
		\node [style={white_dot}] (17) at (0, 2) {};
		\node [style={white_dot}] (18) at (0, 0) {};
		\node [style={white_dot}] (19) at (0, -2) {};
		\node [style=none] (20) at (-1, -3) {};
		\node [style=none] (21) at (1, -3) {};
	\end{pgfonlayer}
	\begin{pgfonlayer}{edgelayer}
		\draw (0.center) to (2.center);
		\draw (2.center) to (3.center);
		\draw (3.center) to (1.center);
		\draw (2.center) to (4.center);
		\draw (5.center) to (4.center);
		\draw (5.center) to (3.center);
		\draw (14) to (20.center);
		\draw (21.center) to (13);
	\end{pgfonlayer}
\end{tikzpicture}\]
as a square lattice, see Definition~\ref{def:box_product}. We can see this as the `intermediate section' used to perform lattice surgery.

\begin{lemma}\label{lem:w_weights}
Let $V_\bullet$ be a $\overline{Z}$ logical operator subcomplex of a chain complex $C_\bullet$, and let $V_\bullet$ satisfy the separation property from Definition~\ref{def:separation}. Then
\[w_W^X = w_C^X + 1;\quad w^Z_W = q^X_C + 2; \quad q^X_W = \max(q^X_C, 2); \quad q^Z_W = w^X_C\]
and $\dim H_1(W_\bullet) = \dim H_1(V_\bullet)=1$.
\end{lemma}
\proof
Observe that $\del_{1}^{V_\bullet}$ has maximum row weight $w^X_C$ and column weight $q^X_C$. Then inspect the matrices $\del_{2}^{W_\bullet}$ and $\del_{1}^{W_\bullet}$ from Definition~\ref{def:weights}. For $\dim H_1(W_\bullet)$, we use the K{\"u}nneth formula, for which see Lemma~\ref{lem:hom_factorA}, which in this case says $H_1((P \tens V)_\bullet) = (H_{0}(P_\bullet)\tens H_1(V_\bullet))\oplus (H_1(P_\bullet) \tens H_{0}(V_\bullet))$. We then have
\[\dim H_{0}(P_\bullet) = 1; \quad \dim H_1(P_\bullet) = 0; \quad \dim H_0(V_\bullet) = 0; \quad \dim H_{1}(V_\bullet) = 1\]
where the last comes from the fact that $V_{0} = \im(\del_{1}^{V_\bullet})$, using Definition~\ref{def:op_subcomplex}. $\dim H_1(V_\bullet) = 1$ as $B_1(V_\bullet) =0$ and $Z_1(V_\bullet) = 1$.
\endproof

\begin{definition}\label{def:two_pushouts}
Let $V_\bullet$ be a simultaneous $\overline{Z}$ logical operator subcomplex of both $C_\bullet$ and $D_\bullet$, satisfying the separation property. Then define the `sandwiched code' $(T_\bullet, T^\bullet)$, with $T_\bullet$ as the pushout of a pushout:
\[\begin{tikzcd}&V_\bullet \arrow[r, hookrightarrow] \arrow[d, hookrightarrow]& C_\bullet \arrow[d]\\
V_\bullet \arrow[d, hookrightarrow]\arrow[r, hookrightarrow] & W_\bullet \arrow[r] & R_\bullet \arrow[d]\arrow[ul, phantom, "\usebox\pushout", very near start]\\
D_\bullet \arrow[rr] & & T_\bullet\arrow[ul, phantom, "\usebox\pushout", very near start]\end{tikzcd}\]
where the middle term is $W_\bullet = (P \tens V)_\bullet$ from Definition~\ref{def:tensor_sandwich} above, and the two inclusion maps $V_\bullet \hookrightarrow W_\bullet$ map $V_1$ into each of the copies of $V_1$ in $W_1$, and the same for $V_{0}$.
All maps in the pushouts are basis-preserving, and one can check that they are all monic.
\end{definition}

Colloquially, we are gluing first one side of the code $W_\bullet$ to $C_\bullet$, and then the other side to $D_\bullet$. \footnote{We could equally do it the other way, in which case the two pushouts would be flipped, but this does not change $T_\bullet$.}

\begin{lemma}\label{lem:sandwich_parameters}
The `sandwiched code' $(T_\bullet, T^\bullet)$ has
\[n_T = n_C + n_D + r;\quad k_T \geq k_C + k_D - 1 \]
and
\[w^X_{T} \leq w^X_{C\oplus D} +1; \quad w^Z_{T} \leq 
\max(w^Z_{C\oplus D}, q^X_{C\oplus D}+2);\quad q^Z_T \leq q^Z_{C\oplus D} + w^X_{C\oplus D}; \quad q^X_T = \max(q^X_{C\oplus D}, 2).\]
If $k_T = k_C + k_D - 1$ then $d^X_T \geq \min(d^X_C, d^X_D)$.
\end{lemma}
\proof
For $n_T$, just apply Lemma~\ref{lem:separated_result} twice. For $k_T$, use Lemma~\ref{lem:w_weights} and apply Lemma~\ref{lem:separated_result} twice.

For $d^X_T$, we first show that $(R_\bullet, R^\bullet)$ has $d^X_R \geq d^X_C$. Every $\overline{X}$ operator in $(W_\bullet, W^\bullet)$ must anticommute with the $\overline{Z}$ operator used to construct $V_\bullet$, and thus must have support on those qubits. In addition, it must have a matched $\overline{X}$ operator in $(C_\bullet,  C^\bullet)$, which also has support on those qubits. As the only other $\overline{X}$ operators in $(R_\bullet, R^\bullet)$ are those in $(C_\bullet,  C^\bullet)$ which are unaffected by the merge, having no support on the qubits being merged, $d^X_R \geq d^X_C$. Then, if $k_T = k_C + k_D - 1$, $d^X_T \geq \min(d^X_C, d^X_D)$ using Lemma~\ref{lem:merge_distance}.

For $w^X_{T}$, the pushouts will glue each $X$ type stabiliser generator in $W_\bullet$ into those in $C_\bullet$ and $D_\bullet$ in such a way that they will have exactly one extra qubit in the support, by the product construction of $W_\bullet$; we can see this from $\del_{1}^{W_\bullet}$ in Definition~\ref{def:tensor_sandwich}, as there is exactly a single 1 which is not part of the $\del_{1}^{V_\bullet}$ in any given row of the matrix.

For $w^Z_{T}, q^Z_T$ and $q^X_T$ we just use Lemma~\ref{lem:w_weights} and apply Lemma~\ref{lem:LDPC_conservation} twice.
\endproof

The intuition here is that rather than gluing two codes $(C_\bullet,  C^\bullet)$ and $(D_\bullet,  D^\bullet)$ together directly along a logical operator, we have made a low distance hypergraph code $(W_\bullet, W^\bullet)$ and used that to sandwich the codes. A consequence of the above lemma is that this `sandwiching' procedure maps LDPC codes to LDPC codes.
Importantly, under suitable conditions the two pushouts let us perform a code map on logical qubits in an error-corrected manner.

\begin{definition}\label{def:distance_preserving}
Let $(T_\bullet, T^\bullet)$ have no logical $\overline{Z}$ operators with weight lower than $d_{C\oplus D}$. Then we say that the merged code has \emph{distance bounded below}.
\end{definition}

\begin{remark}
Note that the only $Z$ operators which can lower the distance are those with support on the logical $\overline{Z}$ which is used to construct $V_\bullet$, as all others will be unchanged by the quotient. The condition for a merge to have distance bounded below is quite a tricky one, as we do not know of a way to check this easily. Because of Lemma~\ref{lem:sandwich_parameters}, this problem is isolated to $\overline{Z}$ operators, as the distance is guaranteed to be bounded below for $\overline{X}$ operators.
\end{remark}

\begin{proposition}\label{prop:fault_tolerant_sandwich}
Let $(C_\bullet,  C^\bullet)$ and $(D_\bullet,  D^\bullet)$ be CSS codes which share a separated $\overline{Z}$ operator on $m$ physical qubits and $r$ $X$-type stabiliser generators each; let the relevant logical qubits be $i$ and $j$, and let $V_\bullet$ be the logical operator subcomplex of $C_\bullet$ and $D_\bullet$ such that the codes admit a separated $\overline{Z}$-merge. Further, let $d$ be the code distance of $((C\oplus D)_\bullet, (C\oplus D)^\bullet)$, and let the merged code $(T_\bullet, T^\bullet)$ have distance bounded below. Then there is an error-corrected procedure with distance $d$ for implementing a $\overline{Z}\tens \overline{Z}$ measurement on the pair $i$, $j$ of logical qubits, which gives the code $(T_\bullet, T^\bullet)$. This procedure requires $r$ auxiliary clean qubits and an additional $m$ $Z$-type stabiliser generators.
\end{proposition}
\proof
We aim to go from the code $((C\oplus D)_\bullet, (C\oplus D)^\bullet)$ to $(T_\bullet, T^\bullet)$. The code map we apply to physical qubits is as follows. We call the physical qubits in the support of the logical operators to be glued together the \textit{participating} qubits. We initialise a fresh qubit in the $\ket{+}$ state for each pairing of $X$-measurements on the two logical operators of qubits $i$ and $j$, that is for each qubit in $(W_\bullet,W^\bullet)$ which is not glued to a qubit in $(C_\bullet,  C^\bullet)$ or $(D_\bullet,  D^\bullet)$. 

We now modify the stabilisers to get to $(T_\bullet, T^\bullet)$. To start, change the $X$ stabiliser generators with support on the participating qubits to have one additional fresh qubit each, so that each pairing of $X$-measurements shares one fresh qubit. We add a new $Z$ stabiliser generator with weight $a+2$ for each participating qubit in one of the logical operators to be glued, where $a$ is the number of $X$ type generators of which that physical qubit is in the support. One can see this using Definition~\ref{def:tensor_sandwich}, as on the middle code $(W_\bullet, W^\bullet)$ we have
\[P_Z = (\del_2^{W_\bullet})^\intercal = \begin{pmatrix}\id_{\F_2^m} & \id_{\F_2^m} & (\del^{V_\bullet}_{1})^\intercal\end{pmatrix}\]
We then measure $d$ rounds of all stabilisers. All of the qubits in the domain of the last block of $P_Z$ above are those which were initialised to $\ket{+}$. The only other qubits which contribute to the new $Z$ stabiliser generators are those on either side of the sandwiched code, i.e. those along the $\overline{Z}$ logical operators of qubits $i$ and $j$. Each of the physical qubits in the support of these logical operators is measured exactly once by the new $Z$ stabiliser generators, and they are measured in pairs, one from each side; therefore performing these measurements and recording the total product is equivalent to measuring $\overline{Z}\tens \overline{Z}$. We will now check this, and verify that it maintains error-correction.

Let the outcome of a new $Z$-type measurement be $c_{\lambda} \in \{1,-1\}$, and the overall outcome $c_L = \prod_{\lambda \leq m} c_{\lambda}$. Whenever $c_{\lambda} = -1$ we apply the gauge fixing operator $X_{\lambda} = \bigotimes_{(i \in v\ |\ v_i = 1)} X_i$ for the specified $v \in C^0$ (or one could choose a gauge fixing operator using $D^0$ instead). We let $X_{c_L} = \prod_{(\lambda\ |\ c_{\lambda} = -1)}X_{\lambda}$. On participating physical qubits, the merge is then
\[X_{c_L} \prod_{\lambda}\frac{I + c_{\lambda}Z}{2} = \prod_{\lambda}\frac{I + Z}{2} X_{c_L}\]
where we abuse notation somewhat to let $I$ and $Z$ here refer to tensor products thereof. As each $X_{\lambda}$ belongs to the same equivalence class of logical $\overline{X}$ operators in $H_1(C_\bullet)$, if $c_L = 1$ then $X_{c_L}$ acts as identity on the logical space; if $c_L = -1$ then $X_{c_L}$ acts as $\overline{X}$ on logical qubit $i$ in the code before merging. One can then see that these two branches are precisely the branches of the logical $\overline{Z}\tens \overline{Z}$ measurement. As the measurements were performed using $d$ rounds of stabilisers, and the gauge fixing operators each have support on at least $d$ qubits, the overall procedure is error-protected with code distance $d$.

We also check that the procedure is insensitive to errors in the initialisation of fresh qubits. If a qubit is initialised instead to $\ket{-}$, or equivalently suffers a $Z$ error, then the new $Z$ stabiliser measurements are insensitive to this change, and it will just show up at the $X$ measurements on either side of the fresh qubit. If it suffers some other error, say sending it to $\ket{1}$, then each new stabiliser measurement with that qubit in its support may have its result flipped. By construction of $V_\bullet$, each fresh qubit is in the support of an even number of new $Z$ stabiliser measurements, and so initialising the fresh qubits incorrectly will not change $c_L$.
\endproof

As ZX diagrams, the branches are:
\[\begin{tikzpicture}
	\begin{pgfonlayer}{nodelayer}
		\node [style=Z dot] (0) at (0, 0) {};
		\node [style=none] (1) at (1, 0) {};
		\node [style=none] (2) at (-2, 1) {};
		\node [style=Z dot] (3) at (0, 0) {};
		\node [style=none] (4) at (-2, -1) {};
	\end{pgfonlayer}
	\begin{pgfonlayer}{edgelayer}
		\draw [bend left, looseness=0.75] (2.center) to (0);
		\draw (0) to (1.center);
		\draw [bend right, looseness=0.75] (4.center) to (3);
	\end{pgfonlayer}
\end{tikzpicture}\ ;\quad \begin{tikzpicture}
	\begin{pgfonlayer}{nodelayer}
		\node [style=Z dot] (0) at (0, 0) {};
		\node [style=none] (1) at (1, 0) {};
		\node [style=none] (2) at (-2, 1) {};
		\node [style=Z dot] (3) at (0, 0) {};
		\node [style=none] (4) at (-2, -1) {};
		\node [style=X phase dot] (5) at (-1, 0.75) {$\pi$};
	\end{pgfonlayer}
	\begin{pgfonlayer}{edgelayer}
		\draw [bend left, looseness=0.75] (2.center) to (0);
		\draw (0) to (1.center);
		\draw [bend right, looseness=0.75] (4.center) to (3);
	\end{pgfonlayer}
\end{tikzpicture}\]
on logical qubits $i$ and $j$, and all other logical qubits in the code are acted on as identity. We can freely choose which logical qubit may have the red $\pi$ spider, as it will differ only up to a red $\pi$ -- i.e. a logical $\overline{X}$ -- on the output logical qubit. In practice, depending on the code there will typically be cheaper ways of fixing the gauges than using an $\overline{X}$ logical operator for each $-1$ outcome, as there could be an $\overline{X}$ logical operator which has support on multiple of the qubits belonging to new stabilisers. Moreover, one can just update the Pauli frame rather than apply any actual $\overline{X}$ logical operators. The ability to do so is necessary, however, so that the $-1$ outcome is well-defined.

The protocol obviates the problem of performing the code map on \emph{physical} qubits deterministically, as the only non-isometric transformations we perform are measurements of stabiliser generators. However, the code map on \emph{logical} qubits is still not isometric, hence we have a logical measurement.

For the prototypical example of lattice surgery we then have:
\[\begin{tikzpicture}
	\begin{pgfonlayer}{nodelayer}
		\node [style=none] (0) at (2, 3) {};
		\node [style=none] (1) at (2, -3) {};
		\node [style=none] (2) at (4, -3) {};
		\node [style=none] (3) at (4, 3) {};
		\node [style=boundary vertex] (4) at (2, 1) {};
		\node [style=boundary vertex] (5) at (4, 1) {};
		\node [style=boundary vertex] (7) at (2, -1) {};
		\node [style=boundary vertex] (8) at (4, -1) {};
		\node [style=boundary vertex] (10) at (2, 1) {};
		\node [style=boundary vertex] (11) at (2, -1) {};
		\node [style=boundary vertex] (12) at (4, -1) {};
		\node [style=boundary vertex] (13) at (4, 1) {};
		\node [style=none] (18) at (0, 0) {$\rightarrow$};
		\node [style=none] (19) at (-8, 3) {};
		\node [style=none] (20) at (-8, -3) {};
		\node [style=none] (21) at (-6, -3) {};
		\node [style=none] (22) at (-6, 3) {};
		\node [style=boundary vertex] (23) at (-8, 1) {};
		\node [style=boundary vertex] (24) at (-6, 1) {};
		\node [style=boundary vertex] (25) at (-8, -1) {};
		\node [style=boundary vertex] (26) at (-6, -1) {};
		\node [style=boundary vertex] (27) at (-8, 1) {};
		\node [style=boundary vertex] (28) at (-8, -1) {};
		\node [style=boundary vertex] (29) at (-6, -1) {};
		\node [style=boundary vertex] (30) at (-6, 1) {};
		\node [style=none] (31) at (-2, 3) {};
		\node [style=none] (32) at (-2, -3) {};
		\node [style=none] (33) at (-4, -3) {};
		\node [style=none] (34) at (-4, 3) {};
		\node [style=boundary vertex] (35) at (-2, 1) {};
		\node [style=boundary vertex] (36) at (-4, 1) {};
		\node [style=boundary vertex] (37) at (-2, -1) {};
		\node [style=boundary vertex] (38) at (-4, -1) {};
		\node [style=boundary vertex] (39) at (-2, 1) {};
		\node [style=boundary vertex] (40) at (-2, -1) {};
		\node [style=boundary vertex] (41) at (-4, -1) {};
		\node [style=boundary vertex] (42) at (-4, 1) {};
		\node [style=none] (43) at (8, 3) {};
		\node [style=none] (44) at (8, -3) {};
		\node [style=none] (45) at (6, -3) {};
		\node [style=none] (46) at (6, 3) {};
		\node [style=boundary vertex] (47) at (8, 1) {};
		\node [style=boundary vertex] (48) at (6, 1) {};
		\node [style=boundary vertex] (49) at (8, -1) {};
		\node [style=boundary vertex] (50) at (6, -1) {};
		\node [style=boundary vertex] (51) at (8, 1) {};
		\node [style=boundary vertex] (52) at (8, -1) {};
		\node [style=boundary vertex] (53) at (6, -1) {};
		\node [style=boundary vertex] (54) at (6, 1) {};
		\node [style={white_dot}] (55) at (-7, 2) {};
		\node [style={white_dot}] (56) at (-7, 0) {};
		\node [style={white_dot}] (57) at (-7, -2) {};
		\node [style={white_dot}] (58) at (-3, -2) {};
		\node [style={white_dot}] (59) at (-3, 0) {};
		\node [style={white_dot}] (60) at (-3, 2) {};
		\node [style={white_dot}] (61) at (3, 2) {};
		\node [style={white_dot}] (62) at (5, 2) {};
		\node [style={white_dot}] (63) at (7, 2) {};
		\node [style={white_dot}] (64) at (7, 0) {};
		\node [style={white_dot}] (65) at (5, 0) {};
		\node [style={white_dot}] (66) at (3, 0) {};
		\node [style={white_dot}] (67) at (3, -2) {};
		\node [style={white_dot}] (68) at (5, -2) {};
		\node [style={white_dot}] (69) at (7, -2) {};
	\end{pgfonlayer}
	\begin{pgfonlayer}{edgelayer}
		\draw (0.center) to (10);
		\draw (10) to (11);
		\draw (11) to (1.center);
		\draw (7) to (8);
		\draw (4) to (5);
		\draw [style={blue_edge}] (3.center) to (13);
		\draw [style={blue_edge}] (13) to (12);
		\draw [style={blue_edge}] (12) to (2.center);
		\draw (19.center) to (27);
		\draw (27) to (28);
		\draw (28) to (20.center);
		\draw (25) to (26);
		\draw (23) to (24);
		\draw [style={blue_edge}] (22.center) to (30);
		\draw [style={blue_edge}] (30) to (29);
		\draw [style={blue_edge}] (29) to (21.center);
		\draw (31.center) to (39);
		\draw (39) to (40);
		\draw (40) to (32.center);
		\draw (37) to (38);
		\draw (35) to (36);
		\draw [style={blue_edge}] (34.center) to (42);
		\draw [style={blue_edge}] (42) to (41);
		\draw [style={blue_edge}] (41) to (33.center);
		\draw (43.center) to (51);
		\draw (51) to (52);
		\draw (52) to (44.center);
		\draw (49) to (50);
		\draw (47) to (48);
		\draw [style={blue_edge}] (46.center) to (54);
		\draw [style={blue_edge}] (54) to (53);
		\draw [style={blue_edge}] (53) to (45.center);
		\draw (13) to (54);
		\draw (12) to (53);
	\end{pgfonlayer}
\end{tikzpicture}\]
We also look at a less obvious example, that of error-corrected surgery of the Shor code, in Appendix~\ref{app:shor_merge}.

By dualising appropriately one can perform an $\overline{X}$-merge by sandwiching in a similar manner. We can also do the `inverse` of the merge operation:
\begin{corollary}
Let $(T_\bullet, T^\bullet)$ be a CSS code formed by sandwiching codes $(C_\bullet,  C^\bullet)$ and $(D_\bullet,  D^\bullet)$ together along a $\overline{Z}$ operator. Then there is an error-corrected procedure to implement a code map on logical qubits $\CE_{\overline{X}}$ from $(T_\bullet, T^\bullet)$ to $((C\oplus D)_\bullet, (C\oplus D)^\bullet)$.
\end{corollary}
\proof
As the initial code is already a sandwiched code we can just take the opposite of sandwiching. We delete the qubits belonging to the intermediate code $(W_\bullet, W^\bullet)$ but not $(C_\bullet,  C^\bullet)$ or $(D_\bullet,  D^\bullet)$ by measuring them out in the $X$-basis. The code map $\CE_{\overline{X}}$ on participating logical qubits is
\[\begin{tikzpicture}
	\begin{pgfonlayer}{nodelayer}
		\node [style=Z dot] (0) at (0, 0) {};
		\node [style=none] (1) at (-1, 0) {};
		\node [style=none] (2) at (2, 1) {};
		\node [style=Z dot] (3) at (0, 0) {};
		\node [style=none] (4) at (2, -1) {};
	\end{pgfonlayer}
	\begin{pgfonlayer}{edgelayer}
		\draw [bend right, looseness=0.75] (2.center) to (0);
		\draw (0) to (1.center);
		\draw [bend left, looseness=0.75] (4.center) to (3);
	\end{pgfonlayer}
\end{tikzpicture}\]
by following precisely the same logic as for traditional lattice surgery \cite{HFDM}.
\endproof
Again, by dualising appropriately we get the last split operation.

Given a procedure for making $\overline{Z}\tens\overline{Z}$ and $\overline{X}\tens\overline{X}$ logical measurements and the isometries from splits, one can easily construct a logical CNOT between suitable CSS codes following, say, \cite{BH} and observing that the same ZX diagrammatic arguments apply. Augmented with some Clifford single-qubit gates and non-stabiliser logical states one can then perform universal computation. As opposed to some other methods of performing entangling gates with CSS codes, e.g. transversal 2-qubit gates, the schemes above require only the $m$ qubits from the respective $\overline{Z}$ or $\overline{X}$ operators to participate, and we expect $m \ll n$ for practical codes. Unlike that of \cite{Coh}, our method does not require a large ancillary hypergraph product code, which can have significantly worse encoding rate and code distance scaling than the LDPC codes holding data. Our method does not require the code to be `self-ZX-dual' in the sense of \cite{Burt1}, and unlike \cite{HJY} our method does not require the code to be defined on any kind of manifold, and is purely algebraic in description; moreover, these works study single qubit gates and measurements rather than entangling gates.

\section{Conclusions and further work}
We believe our constructions are flexible and conceptually quite simple. The immediate next step is to benchmark our CSS code surgery against other methods of performing entangling logical gates and characterise which CSS codes admit separated logical operators such that the merges have distance bounded below.

The pushouts we gave along logical operators are the most obvious cases. By taking pushouts of more interesting spans other maps on logical data can be obtained, although by Proposition~\ref{prop:CNOT_circuit} and Corollary~\ref{cor:restriction_maps} all code maps as we defined them are limited and do not allow for universal quantum computation; we also do not know whether other pushouts would allow the maps on logical data to be performed fault-tolerantly. 

In this paper we assumed that the two codes being `glued' are different codes, but the same principles apply if we have only one code we would like to perform internal surgery on. In this case, the correct universal construction to use should be a coequaliser. There may be other uses of colimits in $\Chains$. The methods of constructing families of quantum LDPC codes in \cite{PK1, PK2, BE1} use a balanced product, which is also a coequaliser, and the instances used there could be generalised. The quotient is with respect to actions of a finite group $G$ or ring $R$. This could be changed to some other differential graded algebra, for example the algebras studied in quantum Riemannian geometry \cite{BeMaj}. Therein, for a first-order differential calculus $A_\bullet$ one has, say, a group algebra $\F_2 G$ as the $A_0$ component, but then the $A_1$ component is an $\F_2 G$-bimodule. The initial codes $C_\bullet$ and $D_\bullet$ are then right- and left-modules over $A_\bullet$, which in the language of noncommutative geometry are sections of a vector bundle. One must ensure that the actions of $A_\bullet$ on the initial codes are basis-preserving in the sense of Definition~\ref{def:basis_preserving} for the resultant balanced product chain complexes to have a reasonable interpretation as quantum codes. This is always true for free modules, but can also hold for non-free modules. A balanced product code of this form cannot have asymptotically better $\llbracket n,k,d\rrbracket$ parameters than those of \cite{PK1}, up to constant factors, as their construction already saturates the relevant bounds. However, it would be interesting to see if one can obtain better parameters for concrete instances. Calculating the homologies of such codes is likely to require tools such as spectral sequences. This area is largely unexplored, as basis-dependent notions like code-distance are not typically of interest to mathematical physicists studying noncommutative geometry, and the process of selecting an instance of a code which has a consistent definition of code distance from an iso-class is categorically `evil'.

It should be possible to extend the definitions of $\overline{X}$- and $\overline{Z}$-merges straightforwardly to include metachecks \cite{Cam}, say by specifying that the logical operator subcomplex $V_\bullet$ now runs from $V_1$ to $V_{-1}$, so it has $X$-checks and then metachecks on $X$-checks, but we have not proved how this affects metachecks in the merged code.

There are several ways in which our constructions could be generalised to other codes. The obvious generalisation is to qudit CSS codes. For qudits of prime dimension $q$, everything should generalise fairly straightforwardly using a different finite field $\F_q$ but in this case the cell complexes will require additional data in the form of an orientation on edges, as is familiar for qudit surface codes. When $q$ is not prime, one formalism for CSS codes with dimension $q$ is chain complexes in $\Z_q$-$\mathtt{FFMod}$, the category of free finite modules over the ring $\Z_q$. As $\Z_q$ is not generally a domain this complicates the homological algebra.

Second, if we wish to upgrade to more general stabiliser codes we can no longer use chain complexes. The differential composition $P_X P_Z^\intercal$ is a special case of the symplectic product $\omega(M,N) = M\omega N^\intercal$ for $\omega = \begin{pmatrix} 0_n &I_n\\ -I_n&0_n \end{pmatrix}$ \cite{Haah}, but by generalising to such a product we lose the separation of $Z$ and $X$ stabilisers to form a pair of differentials. It is unclear what the appropriate notion of a quotient along an $\overline{X}$ or $\overline{Z}$ operator is for such codes.

For quantum codes which are not stabiliser but are based on cell complexes, such as the Kitaev model \cite{Kit}, there are no stabiliser generators, but the codes are still `CSS-like', in the sense that vertices correspond to actions of the group algebra $\C G$ and faces actions of the function algebra $\C(G)$, with each measurement outcome corresponding to an irreducible representation of the quantum double $D(G) = \C(G)\lcross\ \C G$. More generally we can replace $\C G$ and $\C(G)$ with $H$ and $H^*$ for any semisimple Hopf algebra $H$ \cite{Meu,CowMa1} while retaining the relevant features of the model. Just as there are no stabiliser generators, there are no longer $Z$ and $X$-operators, but there are ribbon operators. As special cases there are ribbon operators which correspond to actions of only $\C G$ or $\C(G)$. The first author recently generalised lattice surgery to Kitaev models \cite{CowMa2}, albeit with some caveats. In the same way that CSS codes generalise stabiliser codes based on cell complexes, there should be a general class of commuting projector models using the quantum double, which are not necessarily defined on a tessellated manifold. We speculate that the notion of `gluing' along, say, a $\C G$ operator could work for such commuting projector models. These would give merges and splits which correspond to the multiplications and comultiplications of the relevant Hopf algebras.

\section{Acknowledgements}
AC thanks Aleks Kissinger for helpful discussions about Lemma~\ref{lem:duality_basis}, and both Aleks Kissinger and John van de Wetering for helpful discussions about Proposition~\ref{prop:CNOT_circuit}. AC also thanks the Wolfson Harrison UK Research Council Quantum Foundation Scholarship for making this work possible.

We are grateful to Christophe Vuillot for spotting a crucial error in an earlier version of this paper, and for providing us with the illustrative example in Appendix~\ref{app:not_distance_preserving}. We are also grateful to anonymous reviewers, whose thoughtful and rigorous feedback helped improve the paper substantially.

\appendix

\section{Limits and colimits in $\Chains$}\label{app:lims}
\begin{lemma}\label{lem:kerns}
$\Chains$ has all kernels and cokernels.
\end{lemma}
\proof
Recall that $\MatF$ has all kernels and cokernels, i.e. subspaces and quotient spaces. Then given a chain map $f : C_\bullet \rightarrow D_\bullet$ we define ${\rm ker}(f)$ with
\[\begin{tikzcd}
\cdots \arrow[r, dotted] & K_{n+1} \arrow[r, dotted, "\del^{K_\bullet}_n"] \arrow[d, "{\rm ker}(f_{n+1})"'] & K_n \arrow[d, "{\rm ker}(f_n)"] \arrow[r, dotted]& \cdots\\
\cdots \arrow[r]& C_{n+1} \arrow[r, "\del^{C_\bullet}_n"] \arrow[d, "f_{n+1}"'] & C_n \arrow[d, "f_n"] \arrow[r]& \cdots\\
\cdots \arrow[r]& D_{n+1} \arrow[r, "\del^{D_\bullet}_n"] & D_n \arrow[r]& \cdots
\end{tikzcd}\]
where $\del^{K_\bullet}_n$ always exists and is uniquely defined, because
\[f_n\circ \del^{C_\bullet}_n\circ {\rm ker}(f_{n+1}) = \del^{D_\bullet}_n\circ f_{n+1}\circ {\rm ker}(f_{n+1}) = 0\]
and so by the universal property of ${\rm ker}(f_{n})$ there is a unique matrix $\del^{K_\bullet}_n: K_{n+1}\rightarrow K_n$.  These satisfy $\del^{K_\bullet}_n\circ\del^{K_\bullet}_{n+1} = 0$ as
\[{\rm ker}(f_n)\circ\del^{K_\bullet}_n\circ\del^{K_\bullet}_{n+1}=\del^{C_\bullet}_n\circ\del^{C_\bullet}_{n+1}\circ {\rm ker}(f_{n+2})=0\]
and then kernels are monic. $K_n = \{v \in C_n\ |\ f_n(v)=0\}$ by the definition of kernels in $\MatF$. Given the correct choice of basis, $\del^{K_\bullet}_n$ is thus just $\del^{C_\bullet}_n\circ {\rm ker}(f_{n+1})$ as a matrix but without the all-zero rows which map into $C_n/K_n$.

That ${\rm ker}(f)$ is a genuine kernel in $\Chains$ is straightforward to check but we do not give further details.

The reversed argument applies for cokernels, giving quotient complexes $D_\bullet/{\rm im}(f)$ with components $D_n/{\rm im}(f_n)$ etc.
\endproof

\begin{remark}
As $\Chains$ is additive, equalisers and coequalisers can be seen as special cases of kernels and cokernels by defining ${\rm eq}(f,g) = {\rm ker}(f-g)$ and ${\rm coeq}(f,g) = {\rm coker}(f-g)$, for $f,g : C_\bullet \rightarrow D_\bullet$. For the chain complex part $E_\bullet$ of an equaliser we have components $E_n = \{c\ |\ f(c)=g(c)\}\subseteq C_n$. For the chain complex part $F_\bullet$ of a coequaliser, we have components $F_n = D_n/f(c)\sim g(c)$, for $c\in C_n$.
\end{remark}

We now sketch a proof of Lemma~\ref{lem:abelian}.
\proof
Recall that an Abelian category is an additive category such that:
\begin{enumerate}
\item Every morphism has a kernel and cokernel.
\item Every monomorphism is the kernel of its cokernel.
\item Every epimorphism is the cokernel of its kernel.
\end{enumerate}
The first is just Lemma~\ref{lem:kerns}, and the other two follow using the fact that they hold degree-wise in $\MatF$. 
\endproof

We will now spell out pullbacks. While they can be defined using equalisers and products we construct them explicitly, as it is easy to do so.

\begin{definition}\label{def:pullback}
The pullback of chain maps $f: X_\bullet \rightarrow Z_\bullet$ and $g:Y_\bullet \rightarrow Z_\bullet$ gives the chain complex $W_\bullet$, where each component is the pullback $W_n$ of $f_n$ and $g_n$. The differentials $\del^{W_\bullet}_n$ are given by the unique map from each component's pullback. Specifically, if we have the pullback
\[\begin{tikzcd}
W_\bullet \arrow[r, "w"]\arrow[d, "v"']\arrow[dr, phantom, "\usebox\pullback", very near start] & Y_\bullet \arrow[d,"g"]\\
X_\bullet \arrow[r,"f"'] & Z_\bullet
\end{tikzcd}\]
then for degrees $n, n+1$ we have
\[\begin{tikzcd}
W_n\arrow[rrr,"w_n"]\arrow[ddd,"v_n"'] & & & Y_n\arrow[ddd, "g_n"]\\
& W_{n+1}\arrow[ul, dotted, "\del^{W_\bullet}_n"]\arrow[r, "w_{n+1}"]\arrow[d, "v_{n+1}"']\arrow[dr, phantom, "\usebox\pullback", very near start] & Y_{n+1}\arrow[ur, "\del^{Y_\bullet}_n"]\arrow[d, "g_{n+1}"] &\\
& X_{n+1}\arrow[dl,"\del^{X_\bullet}_n"]\arrow[r,"f_{n+1}"'] & Z_{n+1}\arrow[dr, "\del^{Z_\bullet}_n"] &\\
X_n\arrow[rrr, "f_n"'] & & & Z_n
\end{tikzcd}\]
where 
\[W_n = \{(x,y)\ |\ f_n(x) = g_n(y)\} \subseteq X_n\oplus Y_n; \quad v_n(x,y)=x \in X_n;\quad w_n(x,y)=y\in Y_n.\] 
As $f_n\circ \del^{X_\bullet}_n\circ v_{n+1} = g_n\circ \del^{Y_\bullet}_n\circ w_{n+1}$ and the outer square is a pullback, there is a unique matrix $\del^{W_\bullet}_n$. One can check by diagram chasing that the differentials $\del^{W_\bullet}_n\circ \del^{W_\bullet}_{n+1} = 0$, and then that this is indeed a pullback in $\Chains$.
\end{definition}

\section{Tensor structure of $\Chains$}\label{app:tensor}
\begin{definition}\label{def:tensor}\cite[Sec.~2.7]{Weib}
Let $C_\bullet,D_\bullet$ be chain complexes in $\Chains$. Define $(C\tens D)_{\bullet}$ with components
\[(C\tens D)_n = \bigoplus_{i+j=n}C_i \tens D_j \]
where the latter tensor product is the normal tensor product in $\MatF$. Differentials between components are given by
\[\del^{(C\tens D)_\bullet}_{n} = \line(1,0){10} {}_{i+j=n}\begin{pmatrix}\id_{C_i} \tens \del^{D_\bullet}_j | \del^{C_\bullet}_i \tens \id_{D_j}\end{pmatrix}\]
where the horizontal line $\line(1,0){10}$ indicates that all these matrices are stacked vertically, which we illustrate in an example below. One can check that $\del^{(C\tens D)_\bullet}_{n}\circ\del^{(C\tens D)_\bullet}_{n+1}=0 \pmod 2$, as desired.

Also define the object $\mathbf{1}_\bullet \in \Chains$ as
\[\begin{tikzcd}
\mathbf{1}_\bullet = \cdots\arrow[r]& 0 \arrow[r]& \mathbf{1}_0 \arrow[r]& 0 \arrow[r]& \cdots
\end{tikzcd}\]
where $\mathbf{1}_0 = \F_2$, and all other $\mathbf{1}_i$ are $0$.
\end{definition}

One can check that $(C\tens D)_{\bullet}$ is a $\F_2$-linear monoidal product $\tens$ in $\Chains$, which follows from associativity and distributivity of $\oplus$ and $\tens$ in $\MatF$. For the unit, observe that
\[(C\tens \mathbf{1})_n = C_n \tens 1 = C_n;\quad \del^{(C\tens \mathbf{1})_\bullet}_n = \begin{pmatrix}\id_{C_n} \tens \del^{\mathbf{1}_\bullet}_0 | \del^{C_\bullet}_n \tens \id_{\mathbf{1}_0}\end{pmatrix} = \del^{C_\bullet}_n.\]

\begin{example}\label{ex:tensor_complex}
Consider two chain complexes of length 1:
\[\begin{tikzcd}C_\bullet = \cdots\arrow[r]& 0\arrow[r]& C_1 \arrow[r, "\del^{C_\bullet}_1"]& C_0\arrow[r]& 0 \arrow[r]& \cdots\end{tikzcd}\]
\[\begin{tikzcd}D_\bullet = \cdots\arrow[r]& 0\arrow[r]& D_1 \arrow[r, "\del^{D_\bullet}_1"]& D_0\arrow[r]& 0 \arrow[r]& \cdots\end{tikzcd}\]
In this case we have
\[(C\tens D)_0 = C_0\tens D_0;\quad (C\tens D)_1 = (C_1\tens D_0)\oplus(C_0\tens D_1); \quad (C\tens D)_2 = C_1\tens D_1\]
for nonzero components, and
\[\del^{(C\tens D)_\bullet}_1 = (\id_{C_0} \tens \del^{D_\bullet}_1|\del^{C_\bullet}_1\tens \id_{D_0});\quad
\del^{(C\tens D)_\bullet}_2 = \begin{pmatrix}\underline{\del^{C_\bullet}_1\tens \id_{D_1}}\\ \id_{C_1} \tens \del^{D_\bullet}_1\end{pmatrix}\]
for nonzero differentials. Then 
\[\del^{(C\tens D)_\bullet}_1\circ \del^{(C\tens D)_\bullet}_2 = \del^{C_\bullet}_1\tens \del^{D_\bullet}_1 + \del^{C_\bullet}_1\tens \del^{D_\bullet}_1 = 0 \pmod 2\]
as the matrix partitions factor upon multiplication.
\end{example}

This example illustrates an interesting property of $\tens$ in $\Chains$: both $C_\bullet, D_\bullet$ have only one nonzero differential, but $(C\tens D)_\bullet$ has two. It is easy to see that given two complexes of lengths $s,t$ the tensor product will have length $s+t$.

\begin{lemma}\label{lem:hom_factorA}\cite{Weib}
\[H_n((C\tens D)_\bullet) \cong \bigoplus_{i+j=n} H_i(C_\bullet)\tens H_j(D_\bullet)\]
\end{lemma}
That is, the homology subspaces factor through the tensor product conveniently. This is also called the K{\"u}nneth formula. The manner in which the homology factors through does not make $H_n(-)$ a monoidal functor with respect to the tensor product.

The tensor product is used to build codes from other CSS codes \cite{AC}.
\section{Graphs and cell complexes}\label{app:cells}

In this appendix we give some categorical background on abstract cell complexes. This is not necessary to define CSS code surgery, but codes obtained from cell complexes are an important motivating example, as they include surface codes, toric codes \cite{Kit}, hyperbolic codes \cite{BVCKT} and the expander lifted product codes from \cite{PK1}. In general, if a CSS code comes from tessellating a manifold, it is likely to use cell complexes. Cell complexes are important in the study of topological spaces, and many of the constructions of CSS codes, such as balanced/lifted products, can also be phrased in the language of topology, but we stick to cell complexes for brevity. As a warm-up, we describe certain categories of graphs, and then move on to a specific kind of cell complex.

Let $\Gamma$ be a finite simple undirected graph. Recall that as a simple graph, $\Gamma$ has at most one edge between any two vertices and no self-loops on vertices. $\Gamma$ can be defined as a pair of sets, $V(\Gamma)$ and $E(\Gamma)$, with $E(\Gamma) \subseteq 2^{V(\Gamma)}$, the powerset of vertices, where each $e \in E(\Gamma)$ has 2 elements i.e. it can be expressed as $e = \{v_1, v_2\}$. An example of a graph is $\CC_n$, the cycle graph with $n$ vertices and edges. We will also use $\CP_n$, the path graph with $n$ edges and $n+1$ vertices.

\begin{definition}
Let $\Grph$ be the category of finite simple undirected graphs. A morphism $\Gamma \rightarrow \Delta$ in $\Grph$ is a function $f : V(\Gamma)\rightarrow V(\Delta)$ such that $\{v_1, v_2\} \in E(\Gamma) \implies \{f(v_1), f(v_2)\} \in E(\Delta)$, i.e. the function respects the incidence of edges.
\end{definition}

$\Grph$ has several different products and other categorical features. We are particularly interested in colimits. $\Grph$ has a coproduct $\Gamma + \Delta$ being the disjoint union, with $V(\Gamma + \Delta) = V(\Gamma) \sqcup V(\Delta)$ and $E(\Gamma + \Delta) = E(\Gamma) \sqcup E(\Delta)$. It also has an initial object $I$ given by the empty graph. However, $\Grph$ is not cocomplete, as it does not have all pushouts. 
\begin{example}\label{ex:counter_push}
As a counterexample \cite{mathover}, given the diagram
\[\begin{tikzpicture}
	\begin{pgfonlayer}{nodelayer}
		\node [style=boundary vertex] (0) at (-2, 1) {};
		\node [style=boundary vertex] (1) at (0, 1) {};
		\node [style=none] (2) at (-1, 0) {};
		\node [style=none] (3) at (-1, -2) {};
		\node [style=none] (4) at (1, 1) {};
		\node [style=none] (5) at (3, 1) {};
		\node [style=boundary vertex] (6) at (-2, -3) {};
		\node [style=boundary vertex] (7) at (0, -3) {};
		\node [style=boundary vertex] (8) at (4, 1) {};
	\end{pgfonlayer}
	\begin{pgfonlayer}{edgelayer}
		\draw [style=arrow] (2.center) to (3.center);
		\draw [style=arrow] (4.center) to (5.center);
		\draw (6) to (7);
	\end{pgfonlayer}
\end{tikzpicture}\]
no cocone exists, as the graphs are not allowed self-loops. Therefore, no pushout exists.
\end{example} 
One can easily see that there are diagrams for which pushouts do exist, though.

More than just graphs, we would like to allow for \textit{open} graphs, i.e. graphs which may have edges which connect to only one vertex, but are not self-loops. For example,
\[\begin{tikzpicture}
	\begin{pgfonlayer}{nodelayer}
		\node [style=boundary vertex] (0) at (-1.25, 0) {};
		\node [style=boundary vertex] (1) at (0.75, 0) {};
		\node [style=none] (2) at (2.75, 0) {};
		\node [style=none] (3) at (-3.25, 0) {};
	\end{pgfonlayer}
	\begin{pgfonlayer}{edgelayer}
		\draw (0) to (1);
		\draw (1) to (2.center);
		\draw (3.center) to (0);
	\end{pgfonlayer}
\end{tikzpicture}\]
We call $\CG_3$ the 3rd \textit{open path graph}, where the $n$th open path graph $\CG_n$ has $n$ edges in a line with $n-1$ vertices between them. We now give a particular formalisation of open graphs.

\begin{definition}\label{def:open_graph}
Let $\Gamma$ be a finite simple undirected graph with two disjoint vertex sets $V(\Gamma)$ and $B(\Gamma)$, where $E(\Gamma) \subseteq 2^{V(\Gamma)\cup B(\Gamma)}$. We then say that $\Gamma$ is an open graph. We call $V(\Gamma)$ the internal vertices and $B(\Gamma)$ the boundary vertices.
\end{definition}

So in the picture of $\CG_3$ above there are vertices at either end of the open wires, but they are considered `invisible', i.e. they belong to $B(\Gamma)$. 

\begin{definition}
Let $\OGrph$ be the category of open graphs. A morphism $\Gamma \rightarrow \Delta$ in $\OGrph$ is a function $f: V(\Gamma)\cup B(\Gamma) \rightarrow V(\Delta)\cup B(\Delta)$ such that $\{v_1, v_2\} \in E(\Gamma) \implies \{f(v_1), f(v_2)\} \in E(\Delta)$ and $f(x) \in V(\Delta) \iff x \in V(\Gamma)$.
\end{definition}

This restriction disallows internal vertices from being `created' or `deleted' by a graph morphism by converting them to boundary vertices. $\OGrph$ has very similar properties to $\Grph$. Its initial object is the empty open graph. $\OGrph$ has a coproduct, where $V(\Gamma + \Delta) = V(\Gamma)\sqcup V(\Delta)$ and $B(\Gamma + \Delta) = B(\Gamma)\sqcup B(\Delta)$. Like $\Grph$, $\OGrph$ is not cocomplete, as Example~\ref{ex:counter_push} also works in the setting of open graphs. It is obvious that $\Grph$ is a subcategory of $\OGrph$.

We now move on to cell complexes, in particular abstract cubical complexes. These are abstract cell complexes which are `square', unlike their `triangular' relatives simplicial complexes.

\begin{definition}
The abstract $d$-cube is the set $\{0,1\}^d$, with the $0$-cube $\{0,1\}^0 := \{0\}$. A face of the abstract $d$-cube is a product $A_1\times\cdots \times A_d$, where each $A_i$ is a nonempty subset of $\{0,1\}$.
\end{definition}

\begin{definition}\cite{Far}
Let $S$ be a finite set and let $\Omega$ be a collection of nonempty subsets of $S$ such that:
\begin{itemize}
\item $\Omega$ covers $S$.
\item For $X, Y \in \Omega$, $X \cap Y \in \Omega$ or $X \cap Y = \emptyset$.
\item For each $X \in \Omega$, there is a bijection from $X$ to the abstract $d$-cube for some choice of $d$, such that any $Y \subset X$ is in $\Omega$ iff it is mapped to a face of the $d$-cube.
\end{itemize}
Then $\Omega$ is an abstract cubical complex.
\end{definition}

Abstract cubical complexes are combinatorial versions of cubical complexes, meaning they are stripped of their associated geometry. The elements in $\Omega$ are still called \textit{faces}. We can consider $\Omega$ to be a graded poset, with subset inclusion as the partial order, and the grading $\dim (X) =  \log_2|X|$. We also call this grading the dimension $d$ of $X$, and we call $X$ a $d$-face. The set of $d$-faces in $\Omega$ is called $\Omega_d$. There is a relation $\Omega_d \rightarrow \Omega_{d-1}$ taking a $d$-face to its $(d-1)$-face subsets.

We call the vertex set $V(\Omega) = S = \Omega_0$, and also define the dimension of a cubical complex
\[\dim (\Omega) = \max_{X \in \Omega} \dim (X)\]
The $d$-skeleton of $\Omega$ is the maximal subcomplex $\Upsilon \subseteq \Omega$ such that $\dim (\Upsilon) = d$. The $1$-skeleton of an abstract cubical complex is a finite simple undirected graph. The $2$-skeleton of an abstract cubical complex is `like' a square lattice, in that it has $2$-faces which each have 4 $0$-faces as subsets and 4 $1$-faces.

\begin{definition}
Let $\ACC$ be the category of abstract cubical complexes. A morphism $f : \Omega \rightarrow \Upsilon$ in $\ACC$ is a function $f: V(\Omega) \rightarrow V(\Upsilon)$, such that $\{x,\cdots,y\}\in \Omega_d \implies \{f(x),\cdots,f(y)\}\in \Upsilon_d$, i.e. incidence is preserved at each dimension.
\end{definition}

Similar to $\Grph$, $\ACC$ has coproduct given by $(\Omega + \Upsilon)_i = \Omega_i \sqcup \Upsilon_i$ and an initial object $I = \emptyset$, and does not generally have pushouts, where we can reuse the same counterexample as $\Grph$. Another categorical property we highlight here is that $\ACC$ has a monoidal product called the \textit{box product}.

\begin{definition}\label{def:box_product}
Let $\Upsilon\ \Box\ \Omega$ be the box product of abstract cubical complexes. Then 
\[(\Upsilon\ \Box\ \Omega)_n = \sum_{i+j=n} \Upsilon_i \times \Omega_j.\]
\end{definition}
We now check that $\Upsilon\ \Box\ \Omega$ is indeed an abstract cubical complex. 
\proof
First, it has a vertex set $V(\Upsilon\ \Box\ \Omega) = V(\Upsilon)\times V(\Omega)$, and thus trivially covers $\Upsilon_0\times \Omega_0$. Second, let $X \times Y \in \Upsilon_i\times \Omega_j$ and $T\times U \in \Upsilon_k\times\Omega_l$. This has
$(X\times Y)\cap (T\times U) = (X\cap T)\times (Y\cap U)$ which is either in $\Upsilon_m \times \Omega_n$ for some $m \leq i, m\leq k$ and $n\leq j, n\leq l$, and thus $(X\cap T)\times (Y\cap U) \in \Upsilon\ \Box\ \Omega$, or $(X\cap T)\times (Y\cap U) = \emptyset$. Third, if $X$ and $Y$ each have a bijection to an $i$-cube and $j$-cube respectively, then $X\times Y$ has a bijection to an $(i+j)$-cube. Any $W \subset X\times Y$ can be expressed as $T\times U$, for $T \subset X$ and $U\subset Y$. Then $W$ is in $\Omega\ \Box\ \Upsilon$ iff $T$ is mapped to a face of the $i$-cube and $U$ to a face of the $j$-cube, thus $W$ to a face of the $(i+j)$-cube.
\endproof

Let us compile this into a more digestible form for the case when $\Upsilon$ and $\Omega$ are both graphs. Given vertices $(u,u')$ and $(v,v')$ in $V(\Upsilon\ \Box\ \Omega)$, the 1-face $\{(u,u'),(v,v')\}\in(\Upsilon\ \Box\ \Omega)_1$ iff $(u = v\ \&\ (u',v')\in \Omega)$ or $((u,v)\in \Upsilon\ \&\ u'=v')$. Then $(\Upsilon\ \Box\ \Omega)_2 \cong E(\Upsilon)\times E(\Omega)$. The 1-skeleton of $\Upsilon\ \Box\ \Omega$ is just the normal box product of graphs \cite{HS}.

\begin{example}\label{ex:box_cycles}
Let $\CC_m$ and $\CC_n$ be cycle graphs with $m$ and $n$ vertices respectively, considered as abstract cubical complexes. Then $\CT = \CC_m\ \Box\ \CC_n$ admits an embedding as a square lattice on the torus, and has $\dim(\CC_m\ \Box\ \CC_n) = 2$. Setting $m = n =3$ we have
\[\begin{tikzpicture}
	\begin{pgfonlayer}{nodelayer}
		\node [style=boundary vertex] (5) at (-3, 3) {};
		\node [style=boundary vertex] (6) at (-1, 3) {};
		\node [style=boundary vertex] (7) at (1, 3) {};
		\node [style=boundary vertex] (8) at (-3, 1) {};
		\node [style=boundary vertex] (9) at (-1, 1) {};
		\node [style=boundary vertex] (10) at (1, 1) {};
		\node [style=boundary vertex] (11) at (-3, -1) {};
		\node [style=boundary vertex] (12) at (-1, -1) {};
		\node [style=boundary vertex] (13) at (1, -1) {};
		\node [style=none] (14) at (3, 3) {};
		\node [style=none] (15) at (3, 1) {};
		\node [style=none] (16) at (3, -1) {};
		\node [style=none] (17) at (1, -3) {};
		\node [style=none] (18) at (-1, -3) {};
		\node [style=none] (19) at (-3, -3) {};
		\node [style={grey_dot}] (20) at (-3, -3) {};
		\node [style={grey_dot}] (21) at (-1, -3) {};
		\node [style={grey_dot}] (22) at (1, -3) {};
		\node [style={grey_dot}] (23) at (3, -1) {};
		\node [style={grey_dot}] (24) at (3, 1) {};
		\node [style={grey_dot}] (25) at (3, 3) {};
		\node [style={white_dot}] (36) at (-2, 2) {};
		\node [style={white_dot}] (37) at (0, 2) {};
		\node [style={white_dot}] (38) at (2, 2) {};
		\node [style={white_dot}] (39) at (2, 0) {};
		\node [style={white_dot}] (40) at (0, 0) {};
		\node [style={white_dot}] (41) at (-2, 0) {};
		\node [style={white_dot}] (42) at (-2, -2) {};
		\node [style={white_dot}] (43) at (0, -2) {};
		\node [style={white_dot}] (44) at (2, -2) {};
	\end{pgfonlayer}
	\begin{pgfonlayer}{edgelayer}
		\draw (5) to (8);
		\draw (8) to (9);
		\draw (9) to (10);
		\draw (10) to (7);
		\draw (7) to (6);
		\draw (6) to (9);
		\draw (5) to (6);
		\draw (8) to (11);
		\draw (11) to (12);
		\draw (12) to (9);
		\draw (12) to (13);
		\draw (13) to (10);
		\draw (7) to (14.center);
		\draw (10) to (15.center);
		\draw (13) to (16.center);
		\draw (13) to (17.center);
		\draw (12) to (18.center);
		\draw (11) to (19.center);
	\end{pgfonlayer}
\end{tikzpicture}\]
where the grey dots indicate periodic boundary conditions and the white circles specify 2-faces. This example comes up in the form of the toric code in Section~\ref{sec:codes}.
\end{example}

Obviously, $\Grph$ is a subcategory of $\ACC$.

We are also interested in \textit{open} abstract cubical complexes.
\begin{definition}
Let $\Upsilon$ be an open abstract cubical complex. $\Upsilon$ is an abstract cubical complex where $\Upsilon_0$ is divided into two disjoint vertex sets $V(\Upsilon)$ and $B(\Upsilon)$.
\end{definition}

The 1-skeleton of an open abstract cubical complex is an open graph.

\begin{definition}
Let $\OACC$ be the category of open abstract cubical complexes. A morphism $f: \Omega \rightarrow \Upsilon$ in $\OACC$ is a function $f: V(\Omega)\cup B(\Omega) \rightarrow V(\Upsilon)\cup B(\Upsilon)$ such that $f(x) \in V(\Upsilon) \iff x \in V(\Omega)$ and $\{x,\cdots,y\}\in \Omega_d \implies \{f(x),\cdots,f(y)\}\in \Upsilon_d$.
\end{definition}

As in our previous examples, $\OACC$ has the obvious coproduct and initial object, and does not have pushouts in general.

\begin{example}\label{ex:box_paths}
Let $\Upsilon$ be a `patch', a square lattice with two rough and two smooth boundaries:
\[\begin{tikzpicture}
	\begin{pgfonlayer}{nodelayer}
		\node [style=none] (15) at (-2, 3) {};
		\node [style=none] (16) at (-2, -3) {};
		\node [style=none] (17) at (0, -3) {};
		\node [style=none] (18) at (0, 3) {};
		\node [style=boundary vertex] (21) at (-2, 1) {};
		\node [style=boundary vertex] (22) at (0, 1) {};
		\node [style=boundary vertex] (23) at (2, 1) {};
		\node [style=boundary vertex] (24) at (-2, -1) {};
		\node [style=boundary vertex] (25) at (0, -1) {};
		\node [style=boundary vertex] (26) at (2, -1) {};
		\node [style=boundary vertex] (27) at (-2, 1) {};
		\node [style=boundary vertex] (28) at (-2, -1) {};
		\node [style=boundary vertex] (29) at (0, -1) {};
		\node [style=boundary vertex] (30) at (0, 1) {};
		\node [style=none] (31) at (2, 3) {};
		\node [style=none] (32) at (2, -3) {};
		\node [style=boundary vertex] (33) at (2, 1) {};
		\node [style=boundary vertex] (34) at (2, -1) {};
		\node [style={white_dot}] (35) at (-1, 2) {};
		\node [style={white_dot}] (36) at (1, 2) {};
		\node [style={white_dot}] (37) at (1, 0) {};
		\node [style={white_dot}] (38) at (-1, 0) {};
		\node [style={white_dot}] (39) at (-1, -2) {};
		\node [style={white_dot}] (40) at (1, -2) {};
	\end{pgfonlayer}
	\begin{pgfonlayer}{edgelayer}
		\draw (15.center) to (27);
		\draw (27) to (28);
		\draw (28) to (16.center);
		\draw (18.center) to (30);
		\draw (30) to (29);
		\draw (29) to (17.center);
		\draw (24) to (25);
		\draw (25) to (26);
		\draw (21) to (22);
		\draw (22) to (23);
		\draw (31.center) to (33);
		\draw (33) to (34);
		\draw (34) to (32.center);
	\end{pgfonlayer}
\end{tikzpicture}\]
This patch has 6 2-faces, 13 1-faces and 6 0-faces.
\end{example}

\begin{example}\label{ex:small_pushout}
We can perform the pushout of two smaller open abstract cubical complexes to acquire a patch:
\[\begin{tikzpicture}
	\begin{pgfonlayer}{nodelayer}
		\node [style=boundary vertex] (0) at (-4, 1) {};
		\node [style=boundary vertex] (1) at (-2, 1) {};
		\node [style=none] (2) at (-1, 1) {};
		\node [style=none] (3) at (1, 1) {};
		\node [style=none] (4) at (-3, -0.5) {};
		\node [style=none] (5) at (-3, -2.5) {};
		\node [style=none] (6) at (-4, -6) {};
		\node [style=none] (7) at (-2, -6) {};
		\node [style=none] (8) at (2, 2) {};
		\node [style=none] (9) at (4, 2) {};
		\node [style=boundary vertex] (10) at (2, 0) {};
		\node [style=boundary vertex] (11) at (4, 0) {};
		\node [style=boundary vertex] (12) at (-2, -4) {};
		\node [style=boundary vertex] (13) at (-4, -4) {};
		\node [style=none] (14) at (3, -1) {};
		\node [style=none] (15) at (3, -3) {};
		\node [style=none] (16) at (-1, -5) {};
		\node [style=none] (17) at (1, -5) {};
		\node [style=none] (18) at (2, -4) {};
		\node [style=none] (19) at (4, -4) {};
		\node [style=boundary vertex] (20) at (2, -6) {};
		\node [style=boundary vertex] (21) at (4, -6) {};
		\node [style=none] (22) at (2, -8) {};
		\node [style=none] (23) at (4, -8) {};
		\node [style=boundary vertex] (24) at (4, -6) {};
		\node [style=boundary vertex] (25) at (2, -6) {};
		\node [style={white_dot}] (26) at (-3, -5) {};
		\node [style={white_dot}] (27) at (3, 1) {};
		\node [style={white_dot}] (28) at (3, -5) {};
		\node [style={white_dot}] (29) at (3, -7) {};
	\end{pgfonlayer}
	\begin{pgfonlayer}{edgelayer}
		\draw [style={blue_edge}] (0) to (1);
		\draw [style=arrow] (2.center) to (3.center);
		\draw [style=arrow] (4.center) to (5.center);
		\draw (13) to (6.center);
		\draw (12) to (7.center);
		\draw (8.center) to (10);
		\draw (9.center) to (11);
		\draw [style={blue_edge}] (13) to (12);
		\draw [style={blue_edge}] (10) to (11);
		\draw [style=arrow] (14.center) to (15.center);
		\draw [style=arrow] (16.center) to (17.center);
		\draw (18.center) to (20);
		\draw (19.center) to (21);
		\draw [style={blue_edge}] (20) to (21);
		\draw (25) to (22.center);
		\draw (24) to (23.center);
		\draw [style={blue_edge}] (25) to (24);
	\end{pgfonlayer}
\end{tikzpicture}\]
where the apex is $\CP_1$, the blue edge indicates where the apex is mapped to, and the bottom right open abstract cubical complex is the object of the pushout.
\end{example}

\begin{example}\label{ex:pushout_patches}
Let $\CG_3$ be the open path graph, and let $\Omega$ be a patch. Then we have a pushout
\[\begin{tikzpicture}
	\begin{pgfonlayer}{nodelayer}
		\node [style=none] (0) at (2, 0) {};
		\node [style=none] (1) at (2, -6) {};
		\node [style=none] (2) at (4, -6) {};
		\node [style=none] (3) at (4, 0) {};
		\node [style=boundary vertex] (4) at (2, -2) {};
		\node [style=boundary vertex] (5) at (4, -2) {};
		\node [style=boundary vertex] (6) at (2, -4) {};
		\node [style=boundary vertex] (7) at (4, -4) {};
		\node [style=boundary vertex] (8) at (2, -2) {};
		\node [style=boundary vertex] (9) at (2, -4) {};
		\node [style=boundary vertex] (10) at (4, -4) {};
		\node [style=boundary vertex] (11) at (4, -2) {};
		\node [style=none] (12) at (3, 4) {};
		\node [style=none] (13) at (3, 10) {};
		\node [style=boundary vertex] (14) at (3, 8) {};
		\node [style=boundary vertex] (15) at (3, 6) {};
		\node [style=boundary vertex] (16) at (3, 6) {};
		\node [style=boundary vertex] (17) at (3, 8) {};
		\node [style=none] (42) at (10, 10) {};
		\node [style=none] (43) at (10, 4) {};
		\node [style=none] (44) at (12, 4) {};
		\node [style=none] (45) at (12, 10) {};
		\node [style=boundary vertex] (46) at (10, 8) {};
		\node [style=boundary vertex] (47) at (12, 8) {};
		\node [style=boundary vertex] (48) at (10, 6) {};
		\node [style=boundary vertex] (49) at (12, 6) {};
		\node [style=boundary vertex] (50) at (10, 8) {};
		\node [style=boundary vertex] (51) at (10, 6) {};
		\node [style=boundary vertex] (52) at (12, 6) {};
		\node [style=boundary vertex] (53) at (12, 8) {};
		\node [style=none] (54) at (9, 0) {};
		\node [style=none] (55) at (9, -6) {};
		\node [style=none] (56) at (11, -6) {};
		\node [style=none] (57) at (11, 0) {};
		\node [style=boundary vertex] (58) at (9, -2) {};
		\node [style=boundary vertex] (59) at (11, -2) {};
		\node [style=boundary vertex] (60) at (13, -2) {};
		\node [style=boundary vertex] (61) at (9, -4) {};
		\node [style=boundary vertex] (62) at (11, -4) {};
		\node [style=boundary vertex] (63) at (13, -4) {};
		\node [style=boundary vertex] (64) at (9, -2) {};
		\node [style=boundary vertex] (65) at (9, -4) {};
		\node [style=boundary vertex] (66) at (11, -4) {};
		\node [style=boundary vertex] (67) at (11, -2) {};
		\node [style=none] (68) at (13, 0) {};
		\node [style=none] (69) at (13, -6) {};
		\node [style=boundary vertex] (70) at (13, -2) {};
		\node [style=boundary vertex] (71) at (13, -4) {};
		\node [style=none] (72) at (3, 3) {};
		\node [style=none] (73) at (3, 1) {};
		\node [style=none] (74) at (6, 7) {};
		\node [style=none] (75) at (8, 7) {};
		\node [style=none] (76) at (11, 3) {};
		\node [style=none] (77) at (11, 1) {};
		\node [style=none] (80) at (6, -3) {};
		\node [style=none] (81) at (8, -3) {};
		\node [style={white_dot}] (86) at (11, 7) {};
		\node [style={white_dot}] (87) at (11, 5) {};
		\node [style={white_dot}] (88) at (11, 9) {};
		\node [style={white_dot}] (89) at (3, -1) {};
		\node [style={white_dot}] (90) at (3, -3) {};
		\node [style={white_dot}] (91) at (3, -5) {};
		\node [style={white_dot}] (92) at (10, -5) {};
		\node [style={white_dot}] (93) at (10, -3) {};
		\node [style={white_dot}] (94) at (10, -1) {};
		\node [style={white_dot}] (95) at (12, -1) {};
		\node [style={white_dot}] (96) at (12, -3) {};
		\node [style={white_dot}] (97) at (12, -5) {};
	\end{pgfonlayer}
	\begin{pgfonlayer}{edgelayer}
		\draw (0.center) to (8);
		\draw (8) to (9);
		\draw (9) to (1.center);
		\draw (6) to (7);
		\draw (4) to (5);
		\draw (45.center) to (53);
		\draw (53) to (52);
		\draw (52) to (44.center);
		\draw (48) to (49);
		\draw (46) to (47);
		\draw (54.center) to (64);
		\draw (64) to (65);
		\draw (65) to (55.center);
		\draw (61) to (62);
		\draw (62) to (63);
		\draw (58) to (59);
		\draw (59) to (60);
		\draw (68.center) to (70);
		\draw (70) to (71);
		\draw (71) to (69.center);
		\draw [style=arrow] (72.center) to (73.center);
		\draw [style=arrow] (76.center) to (77.center);
		\draw [style=arrow] (74.center) to (75.center);
		\draw [style=arrow] (80.center) to (81.center);
		\draw [style={blue_edge}] (13.center) to (17);
		\draw [style={blue_edge}] (17) to (16);
		\draw [style={blue_edge}] (16) to (12.center);
		\draw [style={blue_edge}] (43.center) to (51);
		\draw [style={blue_edge}] (51) to (50);
		\draw [style={blue_edge}] (50) to (42.center);
		\draw [style={blue_edge}] (3.center) to (11);
		\draw [style={blue_edge}] (11) to (10);
		\draw [style={blue_edge}] (10) to (2.center);
		\draw [style={blue_edge}] (57.center) to (67);
		\draw [style={blue_edge}] (67) to (66);
		\draw [style={blue_edge}] (66) to (56.center);
	\end{pgfonlayer}
\end{tikzpicture}\]
\end{example}

This example comes up in the context of lattice surgery on surface codes.
Evidently, both $\OGrph$ and $\ACC$ are subcategories of $\OACC$, and one can define a box product for $\OACC$ in the same way as we did for $\ACC$ in Definition~\ref{def:box_product}.

One can define quantum codes using abstract cell complexes more generally, but abstract cubical complexes are the specific type which we make use of in examples in Section~\ref{sec:codes} and onwards. We now relate the above cell complexes to chain complexes by way of functors.

\begin{definition}\label{def:abstract_complex}
Given an abstract cubical complex $\Omega$ we can define the incidence chain complex $C_\bullet$ in $\Chains$, where each nonzero component has a basis $\tilde{C}_{n}=\Omega_n$,  and each nonzero differential $\del^{C_\bullet}_{n+1}$ takes an $n+1$-face to its $n$-dimensional subsets. The differential is thus a matrix with a 1 where an $n$-face is contained within an $(n+1)$-face, and 0 elsewhere. It is an elementary fact that every $(d-2)$-face in a $d$-face is the intersection of exactly 2 $(d-1)$-faces, thus $\del^{C_\bullet}_{n-1}\circ\del^{C_\bullet}_{n} = 0 \mod 2$. Clearly, the incidence chain complex of a dimension 1 abstract cubical complex is just the incidence matrix of a simple undirected graph.
\end{definition}

We can do essentially the same thing given an open abstract cubical complex $\Upsilon$. In this case, each nonzero component has a basis $\tilde{C}_{n} = \{X \in \Omega_n\ |\ X \not\subseteq B(\Omega)\}$, that is we ignore all faces which are made up only of boundary vertices, and differentials are the same matrices as above, with a 1 where an $n$-face which is not a subset of $B(\Omega)$ (and therefore would be `invisible') is contained in an $(n+1)$-face. It is easy to see that we still have $\del^{C_\bullet}_{n}\circ\del^{C_\bullet}_{n+1} = 0 \mod 2$, as making vertices `invisible' corresponds to deleting rows in $\del^{C_\bullet}_{1}$, edges rows in $\del^{C_\bullet}_2$ etc. The incidence chain complex of a dimension 1 open abstract cubical complex is the incidence matrix of an open graph.

\begin{definition}\label{def:faith_functor}
Let $C_\bullet$ and $D_\bullet$ be the incidence chain complexes of two abstract cubical complexes $\Omega$ and $\Upsilon$ with a morphism $f: \Omega \rightarrow \Upsilon$, and set $\tilde{C}_{0}, \tilde{D}_{0}$ as $V(\Omega), V(\Upsilon)$ respectively. This induces a chain map $g_\bullet : C_\bullet \rightarrow D_\bullet$, with the matrix $g_{1}$ given by $f$, and all matrices on higher components generated inductively. Degrees $i < 1$ are assumed to be zero.

As a consequence, we can define a functor $\varphi:\ACC\rightarrow \Chains$, sending each abstract cell complex to its free chain complex as described in Definition~\ref{def:abstract_complex}. One can check that $\varphi(f) \in \Hom(\varphi(\Omega), \varphi(\Upsilon))$ for any morphism $f : \Omega \rightarrow \Upsilon$ between abstract cubical complexes. $\varphi$ is faithful but not full, as there exist morphisms, such as the zero morphism, which are not in the image of $\varphi$.
\end{definition}

\begin{definition}\label{def:faith_functor2}
There is also a functor $\vartheta: \OACC \rightarrow \Chains$. On objects, this again follows Definition~\ref{def:abstract_complex}. On morphisms this is the same as $\varphi$ except it must obviously ignore maps between boundary vertices everywhere. Thus $\vartheta$ is not faithful.
\end{definition}

\begin{example}
Let $\Omega$ and $\Upsilon$ be two abstract cubical complexes. Then $\varphi(\Omega + \Upsilon) = \varphi(\Omega)\oplus \varphi(\Upsilon)$, which is easy to check. Similarly, $\varphi(\emptyset) = \mathbf{0}_\bullet$. The same is true of $\vartheta$, except that $\vartheta(\Xi) = \mathbf{0}_\bullet$ for any $\Xi$ with $V(\Xi) = \emptyset$.
\end{example}

\begin{lemma}\label{lem:colimit_preserved}
The functors $\varphi$ and $\vartheta$ are cocontinuous i.e. they preserve colimits.
\end{lemma}
\proof
We give a proof sketch here. We know already that $\varphi$ preserves coproducts so it is sufficient to check that it preserves pushouts. Let
\[\begin{tikzcd}
\Xi \arrow[r, "g"]\arrow[d, "f"'] & \Upsilon \arrow[d,"l"]\\
\Omega \arrow[r,"k"'] & \chi\arrow[ul, phantom, "\usebox\pushout", very near start]
\end{tikzcd}\]
be a pushout in $\ACC$. Then $\chi_0 = \Omega_0\sqcup \Upsilon_0/f\sim g$, and we have elements in $\chi_n$ of the form $([x],\cdots,[y])$, which can be seen as pushouts at each dimension. Also, $(x,\cdots, y) \in \Omega_n \implies ([x],\cdots,[y])\in \chi_n$, and the same for $\Upsilon_n$.
Then $\tilde{\varphi(\chi)}_{0} = \chi_0$. We then have basis elements of the form $([x],\cdots,[y]) \in \tilde{\varphi(\chi)}_{n}$, and differentials have their obvious form. If we take the diagram in $\Chains$:
\[\begin{tikzcd}
\varphi(\Xi) \arrow[r, "\varphi(g)"]\arrow[d, "\varphi(f)"'] & \varphi(\Upsilon)\\
\varphi(\Omega)
\end{tikzcd}\]
Then we have $Q_\bullet$ as the pushout. Basis elements in $Q_n$ are then also of the form $([x],\cdots,[y])$ for $[x], [y] \in \chi_n$. The differentials also match up correctly, and so $Q_\bullet = \varphi(\chi)$.

The same checks apply if we take $\vartheta: \OACC \rightarrow \Chains$ instead. Observe that in this case $f$ and $g$ may have images only in $B(\Omega)$ and $B(\Upsilon)$, in which case $\Xi$ must have empty $V(\Xi)$. Then the pushout in $\Chains$ will just be a direct sum, i.e. the pushout with $\vartheta(\Xi) = \mathbf{0}_\bullet$ as the apex.

Recall that $\ACC$ and $\OACC$ do not themselves have all pushouts, and therefore all colimits, but $\varphi$ and $\vartheta$ preserve those which they do have.
\endproof

\begin{definition}\label{def:shift_indices}
For any chain complex $C_\bullet$ we have also the $p$th translation $C[p]_\bullet$, where all indices are shifted down by $p$, i.e. $C[p]_{n} = C_{n+p}$ and $\del^{C[p]_\bullet}_{n}=\del^{C_\bullet}_{n+p}$. This extends to an invertible endofunctor $p : \Chains\rightarrow\Chains$ in the obvious way.
\end{definition}

\begin{lemma}\label{lem:box_product}
Let $\Upsilon$ and $\Omega$ be two open abstract cubical complexes. Recalling the functor $\vartheta :\OACC\rightarrow \Chains$ from Definition~\ref{def:faith_functor}, we have $\vartheta(\Upsilon\ \Box\ \Omega) = \vartheta(\Upsilon)\otimes \vartheta(\Omega)$, so $\vartheta$ is a monoidal functor, but for moral purposes it is.
\end{lemma}

\section{Pushouts and properties of codes}\label{app:pushouts_props}
Here we describe a few problems with using general pushouts to construct new quantum codes, even when the spans are basis-preserving.
First, in a certain sense the pushout of LDPC codes is not necessarily LDPC. To illustrate this, consider the following pushout of graphs:
\[\begin{tikzpicture}
	\begin{pgfonlayer}{nodelayer}
		\node [style=none] (2) at (-1, -1) {};
		\node [style=none] (3) at (-1, -3) {};
		\node [style=none] (4) at (1, 1) {};
		\node [style=none] (5) at (3, 1) {};
		\node [style=boundary vertex] (6) at (4, 2) {};
		\node [style=boundary vertex] (7) at (6, 2) {};
		\node [style=boundary vertex] (10) at (4, 1) {};
		\node [style=boundary vertex] (11) at (6, 1) {};
		\node [style=boundary vertex] (12) at (4, 0) {};
		\node [style=boundary vertex] (13) at (6, 0) {};
		\node [style={blue_dot}] (14) at (-1, 0) {};
		\node [style={blue_dot}] (15) at (-1, 1) {};
		\node [style={blue_dot}] (16) at (-1, 2) {};
		\node [style={blue_dot}] (17) at (4, 0) {};
		\node [style={blue_dot}] (18) at (4, 1) {};
		\node [style={blue_dot}] (19) at (4, 2) {};
		\node [style=boundary vertex] (20) at (-2, -4) {};
		\node [style=boundary vertex] (21) at (0, -4) {};
		\node [style=boundary vertex] (22) at (-2, -5) {};
		\node [style=boundary vertex] (23) at (0, -5) {};
		\node [style=boundary vertex] (24) at (-2, -6) {};
		\node [style=boundary vertex] (25) at (0, -6) {};
		\node [style={blue_dot}] (28) at (-2, -4) {};
		\node [style=none] (29) at (5, -1) {};
		\node [style=none] (30) at (5, -3) {};
		\node [style=none] (31) at (1, -5) {};
		\node [style=none] (32) at (3, -5) {};
		\node [style=boundary vertex] (33) at (4, -4) {};
		\node [style=boundary vertex] (34) at (6, -4) {};
		\node [style=boundary vertex] (35) at (4, -4) {};
		\node [style=boundary vertex] (36) at (6, -5) {};
		\node [style=boundary vertex] (37) at (4, -4) {};
		\node [style=boundary vertex] (38) at (6, -6) {};
		\node [style={blue_dot}] (39) at (4, -4) {};
		\node [style={blue_dot}] (40) at (4, -4) {};
		\node [style={blue_dot}] (41) at (4, -4) {};
		\node [style=boundary vertex] (42) at (4, -4) {};
		\node [style=boundary vertex] (43) at (6, -7) {};
		\node [style={blue_dot}] (44) at (4, -4) {};
		\node [style=boundary vertex] (45) at (4, -8) {};
		\node [style=boundary vertex] (46) at (6, -8) {};
		\node [style=boundary vertex] (47) at (4, -9) {};
		\node [style=boundary vertex] (48) at (6, -9) {};
	\end{pgfonlayer}
	\begin{pgfonlayer}{edgelayer}
		\draw [style=arrow] (2.center) to (3.center);
		\draw [style=arrow] (4.center) to (5.center);
		\draw (6) to (7);
		\draw (10) to (11);
		\draw (12) to (13);
		\draw (20) to (21);
		\draw (22) to (23);
		\draw (24) to (25);
		\draw [style=arrow] (29.center) to (30.center);
		\draw [style=arrow] (31.center) to (32.center);
		\draw (33) to (34);
		\draw (35) to (36);
		\draw (37) to (38);
		\draw (42) to (43);
		\draw (45) to (46);
		\draw (47) to (48);
	\end{pgfonlayer}
\end{tikzpicture}\]
where the light dots indicate the graph morphisms.
As $\vartheta$ is cocontinuous this pushout exists also in $\Chains$. There, it represents a merge of two binary classical codes, although we can consider a binary linear code to just be a CSS code without any $Z$ measurements. As a consequence, we have two initial codes with $P_X$ having maximal weights 1 each, and the merged code has maximal weight 4. Evidently, one can scale this with the size of the input graphs: here, the input graphs each have 3 edges, but if there are graphs with $m$ edges each (and weight 1) and the apex with $m$ vertices (and weight 0) then the pushout graph will have maximal weight $m+1$. As a consequence the family of pushout graphs as $m$ scales is not bounded above by a constant, and so the corresponding family of codes is not LDPC.

\begin{conjec}
Let
\[\begin{tikzcd}
A_\bullet \arrow[r, hookrightarrow, "g_\bullet"]\arrow[d, hookrightarrow, "f_\bullet"'] & D_\bullet\\
C_\bullet & &
\end{tikzcd}\]
be a basis-preserving monic span in $\Chains$, and let $Q_\bullet$ be the pushout chain complex of this monic span. Further, let the monic span be a representative of a family of monic spans which are parameterised by some $n\in \N$, and let $A_\bullet$, $C_\bullet$ and $D_\bullet$ be the $Z$-type complexes of quantum LDPC codes. Then $(Q_\bullet, Q^\bullet)$ is also LDPC.
\end{conjec}
Formulating this conjecture properly requires specifying what it means for a monic span to be parameterised.

Lastly, taking pushouts evidently preserves neither homologies nor code distances, as easy examples with lattice surgery demonstrate. Moreover, we do not know of a way of giving bounds on these quantities for general pushouts, although again we suspect it should be easier for monic spans.

\section{Octagonal surface code patch}\label{app:octagon}

Consider the following patch of surface code:
\[\begin{tikzpicture}
	\begin{pgfonlayer}{nodelayer}
		\node [style=none] (0) at (-3, 2) {};
		\node [style=none] (1) at (-3, 0) {};
		\node [style=none] (2) at (-1, -2) {};
		\node [style=none] (3) at (1, -2) {};
		\node [style=none] (4) at (3, 0) {};
		\node [style=none] (5) at (3, 2) {};
		\node [style=none] (6) at (1, 4) {};
		\node [style=none] (7) at (-1, 4) {};
		\node [style=none] (8) at (-0.75, 4) {};
		\node [style=none] (9) at (-0.5, 4) {};
		\node [style=none] (10) at (-0.25, 4) {};
		\node [style=none] (11) at (0, 4) {};
		\node [style=none] (12) at (0.25, 4) {};
		\node [style=none] (13) at (0.5, 4) {};
		\node [style=none] (14) at (0.75, 4) {};
		\node [style=none] (15) at (1, 4.5) {};
		\node [style=none] (16) at (-1, 4.5) {};
		\node [style=none] (17) at (-0.75, 4.5) {};
		\node [style=none] (18) at (-0.5, 4.5) {};
		\node [style=none] (19) at (-0.25, 4.5) {};
		\node [style=none] (20) at (0, 4.5) {};
		\node [style=none] (21) at (0.25, 4.5) {};
		\node [style=none] (22) at (0.5, 4.5) {};
		\node [style=none] (23) at (0.75, 4.5) {};
		\node [style=none] (24) at (-3.5, 0) {};
		\node [style=none] (25) at (-3.5, 2) {};
		\node [style=none] (26) at (-3.5, 1.75) {};
		\node [style=none] (27) at (-3.5, 1.5) {};
		\node [style=none] (28) at (-3.5, 1.25) {};
		\node [style=none] (29) at (-3.5, 1) {};
		\node [style=none] (30) at (-3.5, 0.75) {};
		\node [style=none] (31) at (-3.5, 0.5) {};
		\node [style=none] (32) at (-3.5, 0.25) {};
		\node [style=none] (33) at (-3, 0) {};
		\node [style=none] (34) at (-3, 2) {};
		\node [style=none] (35) at (-3, 1.75) {};
		\node [style=none] (36) at (-3, 1.5) {};
		\node [style=none] (37) at (-3, 1.25) {};
		\node [style=none] (38) at (-3, 1) {};
		\node [style=none] (39) at (-3, 0.75) {};
		\node [style=none] (40) at (-3, 0.5) {};
		\node [style=none] (41) at (-3, 0.25) {};
		\node [style=none] (42) at (3, 0) {};
		\node [style=none] (43) at (3, 2) {};
		\node [style=none] (44) at (3, 1.75) {};
		\node [style=none] (45) at (3, 1.5) {};
		\node [style=none] (46) at (3, 1.25) {};
		\node [style=none] (47) at (3, 1) {};
		\node [style=none] (48) at (3, 0.75) {};
		\node [style=none] (49) at (3, 0.5) {};
		\node [style=none] (50) at (3, 0.25) {};
		\node [style=none] (51) at (3.5, 0) {};
		\node [style=none] (52) at (3.5, 2) {};
		\node [style=none] (53) at (3.5, 1.75) {};
		\node [style=none] (54) at (3.5, 1.5) {};
		\node [style=none] (55) at (3.5, 1.25) {};
		\node [style=none] (56) at (3.5, 1) {};
		\node [style=none] (57) at (3.5, 0.75) {};
		\node [style=none] (58) at (3.5, 0.5) {};
		\node [style=none] (59) at (3.5, 0.25) {};
		\node [style=none] (60) at (1, -2) {};
		\node [style=none] (61) at (-1, -2) {};
		\node [style=none] (62) at (-0.75, -2) {};
		\node [style=none] (63) at (-0.5, -2) {};
		\node [style=none] (64) at (-0.25, -2) {};
		\node [style=none] (65) at (0, -2) {};
		\node [style=none] (66) at (0.25, -2) {};
		\node [style=none] (67) at (0.5, -2) {};
		\node [style=none] (68) at (0.75, -2) {};
		\node [style=none] (69) at (1, -2.5) {};
		\node [style=none] (70) at (-1, -2.5) {};
		\node [style=none] (71) at (-0.75, -2.5) {};
		\node [style=none] (72) at (-0.5, -2.5) {};
		\node [style=none] (73) at (-0.25, -2.5) {};
		\node [style=none] (74) at (0, -2.5) {};
		\node [style=none] (75) at (0.25, -2.5) {};
		\node [style=none] (76) at (0.5, -2.5) {};
		\node [style=none] (77) at (0.75, -2.5) {};
	\end{pgfonlayer}
	\begin{pgfonlayer}{edgelayer}
		\draw (7.center) to (0.center);
		\draw (0.center) to (1.center);
		\draw (1.center) to (2.center);
		\draw (2.center) to (3.center);
		\draw (3.center) to (4.center);
		\draw (4.center) to (5.center);
		\draw (5.center) to (6.center);
		\draw (6.center) to (7.center);
		\draw (16.center) to (7.center);
		\draw (8.center) to (17.center);
		\draw (18.center) to (9.center);
		\draw (19.center) to (10.center);
		\draw (20.center) to (11.center);
		\draw (21.center) to (12.center);
		\draw (22.center) to (13.center);
		\draw (23.center) to (14.center);
		\draw (15.center) to (6.center);
		\draw (34.center) to (25.center);
		\draw (26.center) to (35.center);
		\draw (36.center) to (27.center);
		\draw (37.center) to (28.center);
		\draw (38.center) to (29.center);
		\draw (39.center) to (30.center);
		\draw (40.center) to (31.center);
		\draw (41.center) to (32.center);
		\draw (33.center) to (24.center);
		\draw (42.center) to (43.center);
		\draw (52.center) to (43.center);
		\draw (44.center) to (53.center);
		\draw (54.center) to (45.center);
		\draw (55.center) to (46.center);
		\draw (56.center) to (47.center);
		\draw (57.center) to (48.center);
		\draw (58.center) to (49.center);
		\draw (59.center) to (50.center);
		\draw (51.center) to (42.center);
		\draw (60.center) to (61.center);
		\draw (70.center) to (61.center);
		\draw (62.center) to (71.center);
		\draw (72.center) to (63.center);
		\draw (73.center) to (64.center);
		\draw (74.center) to (65.center);
		\draw (75.center) to (66.center);
		\draw (76.center) to (67.center);
		\draw (77.center) to (68.center);
		\draw (69.center) to (60.center);
	\end{pgfonlayer}
\end{tikzpicture}\]
where the bristled edges are rough boundaries, and the diagonal edges are smooth boundaries. We have abstracted away from the actual cell complex as the tessellation is not important. $Z$-type logical operators take the form of strings extending from one rough boundary to another, e.g.
\[\begin{tikzpicture}
	\begin{pgfonlayer}{nodelayer}
		\node [style=none] (0) at (-2, 1) {};
		\node [style=none] (1) at (-2, -1) {};
		\node [style=none] (2) at (0, -3) {};
		\node [style=none] (3) at (2, -3) {};
		\node [style=none] (4) at (4, -1) {};
		\node [style=none] (5) at (4, 1) {};
		\node [style=none] (6) at (2, 3) {};
		\node [style=none] (7) at (0, 3) {};
		\node [style=none] (8) at (0.25, 3) {};
		\node [style=none] (9) at (0.5, 3) {};
		\node [style=none] (10) at (0.75, 3) {};
		\node [style=none] (11) at (1, 3) {};
		\node [style=none] (12) at (1.25, 3) {};
		\node [style=none] (13) at (1.5, 3) {};
		\node [style=none] (14) at (1.75, 3) {};
		\node [style=none] (15) at (2, 3.5) {};
		\node [style=none] (16) at (0, 3.5) {};
		\node [style=none] (17) at (0.25, 3.5) {};
		\node [style=none] (18) at (0.5, 3.5) {};
		\node [style=none] (19) at (0.75, 3.5) {};
		\node [style=none] (20) at (1, 3.5) {};
		\node [style=none] (21) at (1.25, 3.5) {};
		\node [style=none] (22) at (1.5, 3.5) {};
		\node [style=none] (23) at (1.75, 3.5) {};
		\node [style=none] (24) at (-2.5, -1) {};
		\node [style=none] (25) at (-2.5, 1) {};
		\node [style=none] (26) at (-2.5, 0.75) {};
		\node [style=none] (27) at (-2.5, 0.5) {};
		\node [style=none] (28) at (-2.5, 0.25) {};
		\node [style=none] (29) at (-2.5, 0) {};
		\node [style=none] (30) at (-2.5, -0.25) {};
		\node [style=none] (31) at (-2.5, -0.5) {};
		\node [style=none] (32) at (-2.5, -0.75) {};
		\node [style=none] (33) at (-2, -1) {};
		\node [style=none] (34) at (-2, 1) {};
		\node [style=none] (35) at (-2, 0.75) {};
		\node [style=none] (36) at (-2, 0.5) {};
		\node [style=none] (37) at (-2, 0.25) {};
		\node [style=none] (38) at (-2, 0) {};
		\node [style=none] (39) at (-2, -0.25) {};
		\node [style=none] (40) at (-2, -0.5) {};
		\node [style=none] (41) at (-2, -0.75) {};
		\node [style=none] (42) at (4, -1) {};
		\node [style=none] (43) at (4, 1) {};
		\node [style=none] (44) at (4, 0.75) {};
		\node [style=none] (45) at (4, 0.5) {};
		\node [style=none] (46) at (4, 0.25) {};
		\node [style=none] (47) at (4, 0) {};
		\node [style=none] (48) at (4, -0.25) {};
		\node [style=none] (49) at (4, -0.5) {};
		\node [style=none] (50) at (4, -0.75) {};
		\node [style=none] (51) at (4.5, -1) {};
		\node [style=none] (52) at (4.5, 1) {};
		\node [style=none] (53) at (4.5, 0.75) {};
		\node [style=none] (54) at (4.5, 0.5) {};
		\node [style=none] (55) at (4.5, 0.25) {};
		\node [style=none] (56) at (4.5, 0) {};
		\node [style=none] (57) at (4.5, -0.25) {};
		\node [style=none] (58) at (4.5, -0.5) {};
		\node [style=none] (59) at (4.5, -0.75) {};
		\node [style=none] (60) at (2, -3) {};
		\node [style=none] (61) at (0, -3) {};
		\node [style=none] (62) at (0.25, -3) {};
		\node [style=none] (63) at (0.5, -3) {};
		\node [style=none] (64) at (0.75, -3) {};
		\node [style=none] (65) at (1, -3) {};
		\node [style=none] (66) at (1.25, -3) {};
		\node [style=none] (67) at (1.5, -3) {};
		\node [style=none] (68) at (1.75, -3) {};
		\node [style=none] (69) at (2, -3.5) {};
		\node [style=none] (70) at (0, -3.5) {};
		\node [style=none] (71) at (0.25, -3.5) {};
		\node [style=none] (72) at (0.5, -3.5) {};
		\node [style=none] (73) at (0.75, -3.5) {};
		\node [style=none] (74) at (1, -3.5) {};
		\node [style=none] (75) at (1.25, -3.5) {};
		\node [style=none] (76) at (1.5, -3.5) {};
		\node [style=none] (77) at (1.75, -3.5) {};
		\node [style=none] (78) at (-1, 0.25) {};
		\node [style=none] (79) at (0.25, 1) {};
		\node [style=none] (80) at (1.25, 0.5) {};
		\node [style=none] (81) at (1.5, -0.5) {};
		\node [style=none] (83) at (0.75, -2) {};
	\end{pgfonlayer}
	\begin{pgfonlayer}{edgelayer}
		\draw (7.center) to (0.center);
		\draw (0.center) to (1.center);
		\draw (1.center) to (2.center);
		\draw (2.center) to (3.center);
		\draw (3.center) to (4.center);
		\draw (4.center) to (5.center);
		\draw (5.center) to (6.center);
		\draw (6.center) to (7.center);
		\draw (16.center) to (7.center);
		\draw (8.center) to (17.center);
		\draw (18.center) to (9.center);
		\draw (19.center) to (10.center);
		\draw (20.center) to (11.center);
		\draw (21.center) to (12.center);
		\draw (22.center) to (13.center);
		\draw (23.center) to (14.center);
		\draw (15.center) to (6.center);
		\draw (34.center) to (25.center);
		\draw (26.center) to (35.center);
		\draw (36.center) to (27.center);
		\draw (37.center) to (28.center);
		\draw (38.center) to (29.center);
		\draw (39.center) to (30.center);
		\draw (40.center) to (31.center);
		\draw (41.center) to (32.center);
		\draw (33.center) to (24.center);
		\draw (42.center) to (43.center);
		\draw (52.center) to (43.center);
		\draw (44.center) to (53.center);
		\draw (54.center) to (45.center);
		\draw (55.center) to (46.center);
		\draw (56.center) to (47.center);
		\draw (57.center) to (48.center);
		\draw (58.center) to (49.center);
		\draw (59.center) to (50.center);
		\draw (51.center) to (42.center);
		\draw (60.center) to (61.center);
		\draw (70.center) to (61.center);
		\draw (62.center) to (71.center);
		\draw (72.center) to (63.center);
		\draw (73.center) to (64.center);
		\draw (74.center) to (65.center);
		\draw (75.center) to (66.center);
		\draw (76.center) to (67.center);
		\draw (77.center) to (68.center);
		\draw (69.center) to (60.center);
		\draw [style={blue_edge}] (38.center) to (78.center);
		\draw [style={blue_edge}, in=210, out=0] (78.center) to (79.center);
		\draw [style={blue_edge}, bend left=60, looseness=1.50] (79.center) to (80.center);
		\draw [style={blue_edge}] (80.center) to (81.center);
		\draw [style={blue_edge}] (83.center) to (65.center);
		\draw [style={blue_edge}, in=-75, out=105] (83.center) to (81.center);
	\end{pgfonlayer}
\end{tikzpicture}\]
Two strings belong to the same equivalence class iff they are isotopic on the surface, allowing for the endpoints to slide up and down a rough boundary. There are exactly 3 nontrivial such classes out of which all other strings can be composed. As a consequence, this patch of surface code has logical space $V$ with $\dim V = 2^3=8$. \footnote{More generally, a patch with $2m$ edges, alternating rough and smooth, has $\dim V = 2^{m-1}$, i.e. the number of edges in a minimal spanning tree on the complete graph with $m$ vertices.}
We can choose a basis for this logical space, which has logical $\overline{Z}$ operators with representatives:
\[\input{octagonal_patch3.tikz}\]
where the middle operator can be smoothly deformed to a vertical line from top to bottom if desired. Recall that on the surface code an $\overline{X}$ operator anticommutes with a $\overline{Z}$ operator iff the strings cross an odd number of times. Thus, given the basis above, the duality pairing of Lemma~\ref{lem:duality_basis} forces a similar basis of $\overline{X}$ operators, with representatives:
\[\input{octagonal_patch4.tikz}\]
We see that $\overline{Z}_1$ is contained entirely within $\overline{Z}_2$ on physical qubits. Thus it is possible to construct a $\overline{Z}$ merge which is not separated, in the parlance of Definition~\ref{def:separation}. If we choose a different representative, by deforming $\overline{Z}_2$ to be a vertical line, then we can also perform a separated $\overline{Z}$ merge.

\section{A $\overline{Z}$-merge map which is not distance preserving}\label{app:not_distance_preserving}
Here we provide an illustrative example to show that it is possible to create $\overline{Z}$ operators in a $\overline{Z}$-merged code which are of lower weight than any logical operator in the initial code. Consider the following surface code patches:
\[\input{z_logical_counterexample.tikz}\]
where, as in the previous appendix, bristled edges represent rough boundaries and non-bristled edges represent smooth boundaries. There is a hole in each patch, with bristled edges around it. As a consequence, each patch has 2 logical qubits. We can assign $\overline{Z}$ logical operators $u_2$ and $v_2$, representatives of each equivalence class $[u_2]$ and $[v_2]$ from which all other classes can be composed, like so:
\[\begin{tikzpicture}
	\begin{pgfonlayer}{nodelayer}
		\node [style=none] (0) at (-1, 4) {};
		\node [style=none] (1) at (4, 4) {};
		\node [style=none] (2) at (-1, -5) {};
		\node [style=none] (3) at (4, -5) {};
		\node [style=none] (4) at (-0.5, 0) {};
		\node [style=none] (5) at (-0.5, -1) {};
		\node [style=none] (6) at (0.5, -1) {};
		\node [style=none] (7) at (0.5, 0) {};
		\node [style=none] (8) at (1, 4) {};
		\node [style=none] (9) at (-1, 4) {};
		\node [style=none] (10) at (-0.75, 4) {};
		\node [style=none] (11) at (-0.5, 4) {};
		\node [style=none] (12) at (-0.25, 4) {};
		\node [style=none] (13) at (0, 4) {};
		\node [style=none] (14) at (0.25, 4) {};
		\node [style=none] (15) at (0.5, 4) {};
		\node [style=none] (16) at (0.75, 4) {};
		\node [style=none] (17) at (1, 4.25) {};
		\node [style=none] (18) at (-1, 4.25) {};
		\node [style=none] (19) at (-0.75, 4.25) {};
		\node [style=none] (20) at (-0.5, 4.25) {};
		\node [style=none] (21) at (-0.25, 4.25) {};
		\node [style=none] (22) at (0, 4.25) {};
		\node [style=none] (23) at (0.25, 4.25) {};
		\node [style=none] (24) at (0.5, 4.25) {};
		\node [style=none] (25) at (0.75, 4.25) {};
		\node [style=none] (26) at (3.25, 4) {};
		\node [style=none] (27) at (1.25, 4) {};
		\node [style=none] (28) at (1.5, 4) {};
		\node [style=none] (29) at (1.75, 4) {};
		\node [style=none] (30) at (2, 4) {};
		\node [style=none] (31) at (2.25, 4) {};
		\node [style=none] (32) at (2.5, 4) {};
		\node [style=none] (33) at (2.75, 4) {};
		\node [style=none] (34) at (3, 4) {};
		\node [style=none] (35) at (3.25, 4.25) {};
		\node [style=none] (36) at (1.25, 4.25) {};
		\node [style=none] (37) at (1.5, 4.25) {};
		\node [style=none] (38) at (1.75, 4.25) {};
		\node [style=none] (39) at (2, 4.25) {};
		\node [style=none] (40) at (2.25, 4.25) {};
		\node [style=none] (41) at (2.5, 4.25) {};
		\node [style=none] (42) at (2.75, 4.25) {};
		\node [style=none] (43) at (3, 4.25) {};
		\node [style=none] (44) at (4, 4) {};
		\node [style=none] (45) at (3.5, 4) {};
		\node [style=none] (46) at (3.75, 4) {};
		\node [style=none] (47) at (4, 4) {};
		\node [style=none] (48) at (3.5, 4.25) {};
		\node [style=none] (49) at (3.75, 4.25) {};
		\node [style=none] (50) at (4, 4.25) {};
		\node [style=none] (51) at (-0.5, -1) {};
		\node [style=none] (52) at (-0.5, 0) {};
		\node [style=none] (53) at (-0.5, -0.25) {};
		\node [style=none] (54) at (-0.5, -0.5) {};
		\node [style=none] (55) at (-0.5, -0.75) {};
		\node [style=none] (56) at (-0.5, -1) {};
		\node [style=none] (57) at (-0.25, 0) {};
		\node [style=none] (58) at (-0.25, -0.25) {};
		\node [style=none] (59) at (-0.25, -0.5) {};
		\node [style=none] (60) at (-0.25, -0.75) {};
		\node [style=none] (61) at (-0.25, -1) {};
		\node [style=none] (62) at (-1, -5) {};
		\node [style=none] (63) at (4, -5) {};
		\node [style=none] (64) at (1, -5) {};
		\node [style=none] (65) at (-1, -5) {};
		\node [style=none] (66) at (-0.75, -5) {};
		\node [style=none] (67) at (-0.5, -5) {};
		\node [style=none] (68) at (-0.25, -5) {};
		\node [style=none] (69) at (0, -5) {};
		\node [style=none] (70) at (0.25, -5) {};
		\node [style=none] (71) at (0.5, -5) {};
		\node [style=none] (72) at (0.75, -5) {};
		\node [style=none] (73) at (1, -5.25) {};
		\node [style=none] (74) at (-1, -5.25) {};
		\node [style=none] (75) at (-0.75, -5.25) {};
		\node [style=none] (76) at (-0.5, -5.25) {};
		\node [style=none] (77) at (-0.25, -5.25) {};
		\node [style=none] (78) at (0, -5.25) {};
		\node [style=none] (79) at (0.25, -5.25) {};
		\node [style=none] (80) at (0.5, -5.25) {};
		\node [style=none] (81) at (0.75, -5.25) {};
		\node [style=none] (82) at (3.25, -5) {};
		\node [style=none] (83) at (1.25, -5) {};
		\node [style=none] (84) at (1.5, -5) {};
		\node [style=none] (85) at (1.75, -5) {};
		\node [style=none] (86) at (2, -5) {};
		\node [style=none] (87) at (2.25, -5) {};
		\node [style=none] (88) at (2.5, -5) {};
		\node [style=none] (89) at (2.75, -5) {};
		\node [style=none] (90) at (3, -5) {};
		\node [style=none] (91) at (3.25, -5.25) {};
		\node [style=none] (92) at (1.25, -5.25) {};
		\node [style=none] (93) at (1.5, -5.25) {};
		\node [style=none] (94) at (1.75, -5.25) {};
		\node [style=none] (95) at (2, -5.25) {};
		\node [style=none] (96) at (2.25, -5.25) {};
		\node [style=none] (97) at (2.5, -5.25) {};
		\node [style=none] (98) at (2.75, -5.25) {};
		\node [style=none] (99) at (3, -5.25) {};
		\node [style=none] (100) at (4, -5) {};
		\node [style=none] (101) at (3.5, -5) {};
		\node [style=none] (102) at (3.75, -5) {};
		\node [style=none] (103) at (4, -5) {};
		\node [style=none] (104) at (3.5, -5.25) {};
		\node [style=none] (105) at (3.75, -5.25) {};
		\node [style=none] (106) at (4, -5.25) {};
		\node [style=none] (107) at (0.5, -1) {};
		\node [style=none] (108) at (-0.5, -1) {};
		\node [style=none] (109) at (-0.25, -1) {};
		\node [style=none] (110) at (0, -1) {};
		\node [style=none] (111) at (0.25, -1) {};
		\node [style=none] (112) at (0.5, -1) {};
		\node [style=none] (113) at (-0.5, -0.75) {};
		\node [style=none] (114) at (-0.25, -0.75) {};
		\node [style=none] (115) at (0, -0.75) {};
		\node [style=none] (116) at (0.25, -0.75) {};
		\node [style=none] (117) at (0.5, -0.75) {};
		\node [style=none] (118) at (0.5, 0) {};
		\node [style=none] (119) at (-0.5, 0) {};
		\node [style=none] (120) at (-0.25, 0) {};
		\node [style=none] (121) at (0, 0) {};
		\node [style=none] (122) at (0.25, 0) {};
		\node [style=none] (123) at (0.5, 0) {};
		\node [style=none] (124) at (-0.5, -0.25) {};
		\node [style=none] (125) at (-0.25, -0.25) {};
		\node [style=none] (126) at (0, -0.25) {};
		\node [style=none] (127) at (0.25, -0.25) {};
		\node [style=none] (128) at (0.5, -0.25) {};
		\node [style=none] (129) at (0.5, 0) {};
		\node [style=none] (130) at (0.5, -1) {};
		\node [style=none] (131) at (0.5, -0.75) {};
		\node [style=none] (132) at (0.5, -0.5) {};
		\node [style=none] (133) at (0.5, -0.25) {};
		\node [style=none] (134) at (0.5, 0) {};
		\node [style=none] (135) at (0.25, -1) {};
		\node [style=none] (136) at (0.25, -0.75) {};
		\node [style=none] (137) at (0.25, -0.5) {};
		\node [style=none] (138) at (0.25, -0.25) {};
		\node [style=none] (139) at (0.25, 0) {};
		\node [style=none] (140) at (0.5, -2) {$u_2$};
		\node [style=none] (141) at (2.75, 2) {$v_2$};
	\end{pgfonlayer}
	\begin{pgfonlayer}{edgelayer}
		\draw (0.center) to (2.center);
		\draw (2.center) to (3.center);
		\draw (3.center) to (1.center);
		\draw (1.center) to (0.center);
		\draw (4.center) to (7.center);
		\draw (7.center) to (6.center);
		\draw (6.center) to (5.center);
		\draw (5.center) to (4.center);
		\draw (8.center) to (9.center);
		\draw (18.center) to (9.center);
		\draw (10.center) to (19.center);
		\draw (20.center) to (11.center);
		\draw (21.center) to (12.center);
		\draw (22.center) to (13.center);
		\draw (23.center) to (14.center);
		\draw (24.center) to (15.center);
		\draw (25.center) to (16.center);
		\draw (17.center) to (8.center);
		\draw (26.center) to (27.center);
		\draw (36.center) to (27.center);
		\draw (28.center) to (37.center);
		\draw (38.center) to (29.center);
		\draw (39.center) to (30.center);
		\draw (40.center) to (31.center);
		\draw (41.center) to (32.center);
		\draw (42.center) to (33.center);
		\draw (43.center) to (34.center);
		\draw (35.center) to (26.center);
		\draw (44.center) to (45.center);
		\draw (48.center) to (45.center);
		\draw (46.center) to (49.center);
		\draw (50.center) to (47.center);
		\draw (51.center) to (52.center);
		\draw (57.center) to (52.center);
		\draw (53.center) to (58.center);
		\draw (59.center) to (54.center);
		\draw (60.center) to (55.center);
		\draw (61.center) to (56.center);
		\draw (63.center) to (62.center);
		\draw (64.center) to (65.center);
		\draw (74.center) to (65.center);
		\draw (66.center) to (75.center);
		\draw (76.center) to (67.center);
		\draw (77.center) to (68.center);
		\draw (78.center) to (69.center);
		\draw (79.center) to (70.center);
		\draw (80.center) to (71.center);
		\draw (81.center) to (72.center);
		\draw (73.center) to (64.center);
		\draw (82.center) to (83.center);
		\draw (92.center) to (83.center);
		\draw (84.center) to (93.center);
		\draw (94.center) to (85.center);
		\draw (95.center) to (86.center);
		\draw (96.center) to (87.center);
		\draw (97.center) to (88.center);
		\draw (98.center) to (89.center);
		\draw (99.center) to (90.center);
		\draw (91.center) to (82.center);
		\draw (100.center) to (101.center);
		\draw (104.center) to (101.center);
		\draw (102.center) to (105.center);
		\draw (106.center) to (103.center);
		\draw (107.center) to (108.center);
		\draw (113.center) to (108.center);
		\draw (109.center) to (114.center);
		\draw (115.center) to (110.center);
		\draw (116.center) to (111.center);
		\draw (117.center) to (112.center);
		\draw (118.center) to (119.center);
		\draw (124.center) to (119.center);
		\draw (120.center) to (125.center);
		\draw (126.center) to (121.center);
		\draw (127.center) to (122.center);
		\draw (128.center) to (123.center);
		\draw (129.center) to (130.center);
		\draw (135.center) to (130.center);
		\draw (131.center) to (136.center);
		\draw (137.center) to (132.center);
		\draw (138.center) to (133.center);
		\draw (139.center) to (134.center);
		\draw [style={blue_edge}] (31.center) to (87.center);
		\draw [style={blue_edge}] (40.center) to (96.center);
		\draw [style={blue_edge}] (115.center) to (78.center);
	\end{pgfonlayer}
\end{tikzpicture}\]
and the same for $[u_1]$, $[v_1]$ on the other patch. We quotient out a $\overline{Z}$ operator in $[v_1]$ and $[v_2]$ going along the right and left boundaries, like so:
\[\input{z_logical_counterexample2.tikz}\]
leaving a $\overline{Z}$-merged code. This has new $\overline{Z}$ operators, which belong to the equivalence class $[u_1+u_2]$. These operators are of the form:
\[\input{z_logical_counterexample3.tikz}\]
to see that these do belong to $[u_1 + u_2]$, label this operator $t$ and see that $t+u_1+u_2$ is in $[0]$, as it forms a contractible loop. Then $[t] = [-u_1 - u_2] = [u_1 + u_2]$, recalling that we are working over $\F_2$. This new operator $t$ has a weight lower than any of those in the original codes, which one can see from the diagrams.

\section{A merged code with larger logical space}\label{app:merge_bigger}

Take the lift-connected surface (LCS) codes from \cite{ORM} with $l = 1$, $L = 3$. This is a $\llbracket 15,3,3\rrbracket$ code $C_\bullet$, with the parity-check matrices:

\[P_Z = \begin{pmatrix}
1& 0& 0& 0& 0& 0& 1& 1& 0& 0& 0& 0& 1& 0& 0\\
0& 1& 0& 0& 0& 0& 0& 1& 1& 0& 0& 0& 0& 1& 0\\
0& 0& 1& 0& 0& 0& 1& 0& 1& 0& 0& 0& 0& 0& 1\\
0& 0& 0& 1& 0& 0& 0& 0& 0& 1& 1& 0& 1& 0& 1\\
0& 0& 0& 0& 1& 0& 0& 0& 0& 0& 1& 1& 1& 1& 0\\
0& 0& 0& 0& 0& 1& 0& 0& 0& 1& 0& 1& 0& 1& 1
\end{pmatrix}\]

\[P_X = \begin{pmatrix}
1& 0& 0& 1& 1& 0& 0& 0& 0& 0& 0& 0& 1& 0& 0\\
0& 1& 0& 0& 1& 1& 0& 0& 0& 0& 0& 0& 0& 1& 0\\
0& 0& 1& 1& 0& 1& 0& 0& 0& 0& 0& 0& 0& 0& 1\\
0& 0& 0& 0& 0& 0& 1& 0& 0& 1& 1& 0& 1& 0& 1\\
0& 0& 0& 0& 0& 0& 0& 1& 0& 0& 1& 1& 1& 1& 0\\
0& 0& 0& 0& 0& 0& 0& 0& 1& 1& 0& 1& 0& 1& 1
\end{pmatrix}\]

This code has a separated $\overline{Z}$-logical with support on qubits $(2, 9, 14)$, starting from $0$. If we merge two copies of $C_\bullet$ along this logical we obtain a $\llbracket 27, 6, 2\rrbracket$ code, when naively one would expect a code with 5 logical qubits. The additional logical has appeared because, while quotienting the logicals together in the two codes, we have inadvertently increased the size of $\ker (P_X)$, increasing the size of the logical space.

One explanation for why this can occur is because of the complication of bases for our chain complexes. In Definition~\ref{def:op_subcomplex}, we set $\tilde{V_{0}} = \bigcup_{u \in \im(\del^{V_\bullet}_{1})} {\rm supp\ } u$, so we incorporated all basis elements with support in the image of the logical operator; if we were to instead set $V_0 = \im(\del^{V_\bullet}_{1})$ we believe that this occurrence would be impossible, as the only quotient would be on precisely those vectors in the image of the logical operator, not those vectors' basis elements. However, we cannot do this in general while keeping all chain maps basis-preserving.

\section{Separation is gauge-fixability}\label{app:separation_gauge_fixability}

We will prove in this section that for all CSS codes having separated and gauge-fixable operators are equivalent properties.

\begin{lemma}\label{lem:gauge_fixing}
A vector $u \in \ker(P_X)\backslash \im(P_Z^\intercal)$ is gauge-fixable iff for every pair $(e_i, e_j)$ of basis vectors in ${\rm supp\ }(u)$ there is a vector $a$ in $\im(P_X^\intercal)$ such that $u \odot a = e_i+e_j$.
\end{lemma}
\proof
If there are two vectors $v, w$ paired with $u$ such that $v \odot u = e_i$ and $w \odot u = e_j$ then $v+w \odot u = e_i + e_j$. As each basis vector in $u$ must be safely correctable, $u$ always has a vector $a = v+w \in \im(P_X^\intercal)$ such that $a \odot u = e_i + e_j$. Going the other way, there must be at least one paired vector $v$ with $u$ such that $u \cdot v = 1$ -- we don't assume that $|u \odot v| = 1$, just that the dot product is 1, i.e. they have an odd number of intersecting basis vectors. This can be reduced to an intersection of 1 by applying a vector in $\im(P_X^\intercal)$ corresponding to a pair $(e_i, e_j)$ to each of the pairs of intersecting basis vectors apart from the last one. This single basis vector can then be moved around $u$ by further applications of vectors in $\im(P_X^\intercal)$.
\endproof
We find this equivalent definition of gauge-fixing more helpful in practice, as it requires only data about the $X$ stabilisers, rather than paired logical operators, the choice of which depends on the choice of basis of $H_1(C_\bullet)$.

\begin{lemma}\label{lem:when_gauge_fix}
Let $\del_A : A_1 \rightarrow A_0$ be a matrix over $\F_2$. Then for any $v \in \ker(\del_A)$, either for any pair of basis vectors of $A_1$ $(e_i, e_j) \in \supp(v)$ there is a vector $b \in \im(\del_A^\intercal)$ such that $v \odot b = e_i+e_j$ or there is another non-zero vector $u \in \ker(\del_A)$ such that $\supp(u) \subset \supp(v)$.
\end{lemma}
\proof
Define the vector space $S = \{w \odot v : w \in \ker(\del_A)^\perp \}$. Observe that any vector in $S$ must have even Hamming weight, and that 
\[S \cong \{w \odot v : w \in \ker(\del_A\restriction_{\supp\ v})^\perp \} = \ker(\del_A\restriction_{\supp\ v})^\perp,\]
as any vectors in $\ker(\del_A)^\perp$ wholly outside of $\supp(v)$ will not contribute to the Hadamard product, and the isomorphism merely entails chopping off some entries which will always be $0$. Now, $\dim(S) \leq |v|-1$, as $\dim(\ker(\del_A\restriction_{\supp\ v})) \geq 1$ by definition. 

Suppose $\dim(S) = |v|-1$. Then $v$ has no other vectors in $\ker(\del_A)$ contained in its support, as then $\dim(\ker(\del_A\restriction_{\supp\ v})) = 1$, and for any pair of basis vectors of $A_1$ $(e_i, e_j) \in \supp(v)$ there is a vector $b \in \im(\del_A^\intercal)$ such that $v \odot b = e_i+e_j$. To see this, view $S$ as the row space of a matrix:
\[\begin{pmatrix}
w_1\odot v\\
w_2\odot v\\
\vdots\\
w_m \odot v\\
\end{pmatrix}
\]
where $m = |v| - 1$, with some chosen basis of $S$. Then, put the matrix in row echelon form by performing Gaussian elimination:
\[\begin{pmatrix}
1 & * & * & * & \cdots & *\\
0 & 1 & * & * & \cdots & *\\
& & & \vdots & & \\
0 & 0 & 0 & 0 & \cdots & 1\\
\end{pmatrix}
\]
where $*$ values are unknown. As $\dim(S) = |v|-1$, without knowing anything else about the code, there will be exactly 1 row which is indented by 2 from the previous row. But each row must have an even number of 1s in it, including the last row, so the matrix must actually have the row echelon form:
\[\begin{pmatrix}
1 & *  & * & \cdots & * & *\\
0 & 1  & * & \cdots & * & *\\
0 & 0  & 1 & \cdots & * & *\\
& & \vdots & & & \\
0 & 0 & 0 & \cdots & 1 & 1\\
\end{pmatrix}
\]
Then, add rows from the bottom to the top as necessary to give
\[\begin{pmatrix}
1 & 0  & 0 & \cdots & 0 & *\\
0 & 1  & 0 & \cdots & 0 & *\\
0 & 0  & 1 & \cdots & 0 & *\\
& & \vdots & & &  \\
0 & 0 & 0 & \cdots & 1 & 1\\
\end{pmatrix}
\]
i.e. an identity matrix with one column unknown to the right. But once again each row must have an even number of non-zero entries, as $S = \ker(\del_A\restriction_{\supp\ v})^\perp$, so the column to the right must have all entries equal to 1. Therefore, we have $|v|-1$ different pairs, and combinations of these suffice to give any pair in $\supp(v)$.

Now, suppose $\dim(S) < |v|-1$. Then, as $\dim(\ker(\del_A\restriction_{\supp\ v})^\perp) > 1$, there must be another vector in $\ker(\del_A)$ contained in $\supp(v)$.
\endproof

This means that a vector $v$ in $\ker(\del_A)$, with no other vectors in $\ker(\del_A)$ contained in $\supp(v)$, will always have the property that for any pair of basis vectors of $A_1$ $(e_i, e_j) \in \supp(v)$ there is a vector $b \in \im(\del_A^\intercal)$ such that $v \odot b = e_i+e_j$.

This lemma implies that separation and gauge-fixing coincide.

\section{Error-corrected $\overline{Z}$-merge with the Shor code}\label{app:shor_merge}
In this appendix we work through an example explicitly, using the techniques of Section~\ref{sec:practical} to perform a distance 3 error-corrected $\overline{Z}\tens\overline{Z}$ measurement between two copies of the Shor code, for which see Example~\ref{ex:shor}.

Let us say the two copies are labelled $(C_\bullet, C^\bullet)$ and $(D_\bullet, D^\bullet)$, with 
\[C_\bullet = D_\bullet = \begin{tikzcd}\F_2^6 \arrow[r, "\del_2"] & \F_2^9 \arrow[r, "\del_{1}"] & \F_2^2\end{tikzcd}\]
and
\[\del^{C_\bullet}_2 = \del^{D_\bullet}_2 = \begin{pmatrix}
1&1&0&0&0&0\\
1&0&0&0&0&0\\
0&1&0&0&0&0\\
0&0&1&1&0&0\\
0&0&1&0&0&0\\
0&0&0&1&0&0\\
0&0&0&0&1&1\\
0&0&0&0&1&0\\
0&0&0&0&0&1
\end{pmatrix};\quad 
\del^{C_\bullet}_{1} = \del^{D_\bullet}_{1} =\begin{pmatrix} 
1&1&1&1&1&1&0&0&0\\
1&1&1&0&0&0&1&1&1
\end{pmatrix}.\]
We will use the $\overline{Z}$ operator $Z_1\tens Z_4\tens Z_7$, denoted $u = \begin{pmatrix}1&0&0&1&0&0&1&0&0 \end{pmatrix}^\intercal$, with $u \in C_1$ and $u\in D_1$, to glue along.

The logical operator subcomplex $V_\bullet$ is then
\[V_\bullet = \begin{tikzcd}\F_2^3 \arrow[r, "\del_{1}^{V_\bullet}"] &\F_2^2\end{tikzcd}\]
with $\del_{1}^{V_\bullet} = \begin{pmatrix} 1&1&0\\1&0&1\end{pmatrix}$ and all other components of $V_\bullet$ being $0$.

We now make the tensor product chain complex $W_\bullet = (P\tens V)_\bullet$ from Definition~\ref{def:tensor_sandwich}, where $P_\bullet = \begin{tikzcd}P_1 \arrow[r, "\begin{pmatrix}1\\1\end{pmatrix}"] & P_0\end{tikzcd}$. We have
\[W_\bullet = \begin{tikzcd}\F_2^3 \arrow[r, "\del_2^{W_\bullet}"] &\F_2^8 \arrow[r,"\del_{1}^{W_\bullet}"] & \F_2^4\end{tikzcd}\]
with
\[\del_2^{W_\bullet} = \begin{pmatrix}1&0&0\\0&1&0\\0&0&1\\1&0&0\\0&1&0\\0&0&1\\1&1&0\\1&0&1\end{pmatrix};\quad \del_{1}^{W_\bullet} = \begin{pmatrix}1&1&0&0&0&0&1&0\\1&0&1&0&0&0&0&1\\0&0&0&1&1&0&1&0\\0&0&0&1&0&1&0&1\end{pmatrix}\]

For $T_\bullet$ we take the two pushouts from Definition~\ref{def:two_pushouts}. First, we have
\[\begin{tikzcd}V_\bullet \arrow[r, hookrightarrow, "g_\bullet"] \arrow[d, hookrightarrow,"f_\bullet"']& C_\bullet \arrow[d, "q_\bullet"]\\
W_\bullet \arrow[r, "p_\bullet"'] & R_\bullet \arrow[ul, phantom, "\usebox\pushout", very near start]\end{tikzcd}\]
Giving
\[R_\bullet = \begin{tikzcd}\F_2^9 \arrow[r, "\del_2^{R_\bullet}"] &\F_2^{14} \arrow[r, "\del_{1}^{R_\bullet}"] &\F_2^4 \end{tikzcd}\]
with $R_2 = W_2\oplus C_2$, as $V_2 = 0$. The other components of $R_\bullet$ require taking quotients, identifying elements of $W_1$ and $C_1$, and the same for $W_0$ and $C_{0}$. One can then use Definition~\ref{def:pushout} to show that
\[\del^{R_\bullet}_2 = \begin{pmatrix}
1&0&0&0&0&0&0&0&0\\
0&1&0&0&0&0&0&0&0\\
0&0&1&0&0&0&0&0&0\\
1&1&0&0&0&0&0&0&0\\
1&0&1&0&0&0&0&0&0\\
1&0&0&1&1&0&0&0&0\\
0&0&0&1&0&0&0&0&0\\
0&0&0&0&1&0&0&0&0\\
0&1&0&0&0&1&1&0&0\\
0&0&0&0&0&1&0&0&0\\
0&0&0&0&0&0&1&0&0\\
0&0&1&0&0&0&0&1&1\\
0&0&0&0&0&0&0&1&0\\
0&0&0&0&0&0&0&0&1
\end{pmatrix}; \quad \del^{R_\bullet}_{1} = \begin{pmatrix}
0&0&0&1&0&1&1&1&1&1&1&0&0&0\\
0&0&0&0&1&1&1&1&0&0&0&1&1&1\\
1&1&0&1&0&0&0&0&0&0&0&0&0&0\\
1&0&1&0&1&0&0&0&0&0&0&0&0&0
\end{pmatrix}.\]
For the second pushout, that is
\[\begin{tikzcd}
V_\bullet \arrow[d, hookrightarrow]\arrow[r, hookrightarrow] & W_\bullet \arrow[r] & R_\bullet \arrow[d]\\
D_\bullet \arrow[rr] & & T_\bullet\arrow[ul, phantom, "\usebox\pushout", very near start]\end{tikzcd}\]
we then have
\[T_\bullet = \begin{tikzcd}\F_2^{15} \arrow[r, "\del_2^{T_\bullet}"] &\F_2^{20} \arrow[r, "\del_{1}^{T_\bullet}"] &\F_2^{4} \end{tikzcd}.\]
The differentials are somewhat unwieldy, but we include them for completeness:
\[\del_2^{T_\bullet}= 
\begin{pmatrix}
1&1&0&0&0&0&0&0&0&0&0&0&0&0&0\\
1&0&1&0&0&0&0&0&0&0&0&0&0&0&0\\
1&0&0&1&1&0&0&0&0&0&0&0&0&0&0\\
0&0&0&1&0&0&0&0&0&0&0&0&0&0&0\\
0&0&0&0&1&0&0&0&0&0&0&0&0&0&0\\
0&1&0&0&0&1&1&0&0&0&0&0&0&0&0\\
0&0&0&0&0&1&0&0&0&0&0&0&0&0&0\\
0&0&0&0&0&0&1&0&0&0&0&0&0&0&0\\
0&0&1&0&0&0&0&1&1&0&0&0&0&0&0\\
0&0&0&0&0&0&0&1&0&0&0&0&0&0&0\\
0&0&0&0&0&0&0&0&1&0&0&0&0&0&0\\
1&0&0&0&0&0&0&0&0&1&1&0&0&0&0\\
0&0&0&0&0&0&0&0&0&1&0&0&0&0&0\\
0&0&0&0&0&0&0&0&0&0&1&0&0&0&0\\
0&1&0&0&0&0&0&0&0&0&0&1&1&0&0\\
0&0&0&0&0&0&0&0&0&0&0&1&0&0&0\\
0&0&0&0&0&0&0&0&0&0&0&0&1&0&0\\
0&0&1&0&0&0&0&0&0&0&0&0&0&1&1\\
0&0&0&0&0&0&0&0&0&0&0&0&0&1&0\\
0&0&0&0&0&0&0&0&0&0&0&0&0&0&1
\end{pmatrix}
\]

\[\del_{1}^{T_\bullet}= 
\begin{pmatrix}
1&0&1&1&1&1&1&1&0&0&0&0&0&0&0&0&0&0&0&0\\
0&1&1&1&1&0&0&0&1&1&1&0&0&0&0&0&0&0&0&0\\
1&0&0&0&0&0&0&0&0&0&0&1&1&1&1&1&1&0&0&0\\
0&1&0&0&0&0&0&0&0&0&0&1&1&1&0&0&0&1&1&1
\end{pmatrix}
\]

One can check the various properties of this code. For example, ${\rm rank}\del_{1}^{T_\bullet} = 4$ and ${\rm rank}\del_{2}^{T_\bullet} = 15$. Thus $\dim H_1(T_\bullet) = \dim T_1 - 4 - 15 = 1$, and so the code $(T_\bullet, T^\bullet)$ encodes one logical qubit.

We can compare this with $(C\oplus D)_\bullet$ from before the merge:
\[(C\oplus D)_\bullet = \begin{tikzcd}\F_2^{12} \arrow[r, "\del_2^{(C\oplus D)_\bullet}"] &\F_2^{18} \arrow[r, "\del_{1}^{(C\oplus D)_\bullet}"] &\F_2^{4} \end{tikzcd}\]
where the differentials are easy to see from those of $C_\bullet$ and $D_\bullet$, with $\del_2^{(C\oplus D)_\bullet} = \del_2^{C_\bullet}\oplus \del_2^{D_\bullet}$ etc. Evidently, $((C\oplus D)_\bullet,(C\oplus D)^\bullet)$ encodes 2 logical qubits. As expected, there are 2 new qubits, and 3 new $Z$-measurements in $(T_\bullet, T^\bullet)$. Each of the 2 new qubits participates in 2 of the new $Z$-measurements (and no other $Z$-measurements). We can check that $d^Z_{T} \geq 3$, i.e. the code has distance bounded below.

For the error-corrected $\overline{Z}\tens\overline{Z}$ measurement, we therefore start with the code $((C\oplus D)_\bullet,(C\oplus D)^\bullet)$. Recall that this has $d=3$. We then initialise the 2 new qubits in the $\ket{+}$ state and measure 3 rounds of the stabilisers specified by $\del_2^{T_\bullet}$ and $\del_{1}^{T_\bullet}$. As the 2 new qubits each participate in 2 of the new $Z$-measurements, the product of the outcomes is insensitive to initialisation errors. We apply the gauge-fixing operators from Example~\ref{ex:shor_fixing} to correct for the 3 new $Z$-measurements which may output the -1 measurement outcome. We end up with the code $(T_\bullet, T^\bullet)$.
\bibliographystyle{eptcs}
\end{document}